\documentclass[twocolumn,english,superscriptaddress]{revtex4-2}
\usepackage{lmodern}
\usepackage{lmodern}

\usepackage[T1]{fontenc}
\usepackage[latin9]{inputenc}
\usepackage{babel}
\usepackage{array}
\usepackage{multirow}
\usepackage{amsmath}
\usepackage{amssymb}
\usepackage{graphicx}
\usepackage{wasysym}
\PassOptionsToPackage{normalem}{ulem}
\usepackage{ulem}
\usepackage[unicode=true,pdfusetitle,
 bookmarks=true,bookmarksnumbered=false,bookmarksopen=false,
 breaklinks=false,pdfborder={0 0 1},backref=false,colorlinks=false]
 {hyperref}

\makeatletter

\providecommand{\tabularnewline}{\\}
\newcommand{\lyxdot}{.}

\usepackage{babel}
\newenvironment{cellvarwidth}[1][t]
    {\begin{varwidth}[#1]{\linewidth}}
    {\@finalstrut\@arstrutbox\end{varwidth}}
\usepackage{varwidth}
\usepackage{array}

\makeatother

\begin{document}
\title{The non-equilibrium Marshak wave problem in non-homogeneous media}
\author{Nitay Derei}
\affiliation{Racah Institute of Physics, The Hebrew University, 9190401 Jerusalem,
Israel}
\author{Shmuel Balberg}
\affiliation{Racah Institute of Physics, The Hebrew University, 9190401 Jerusalem,
Israel}
\author{Shay I. Heizler}
\affiliation{Racah Institute of Physics, The Hebrew University, 9190401 Jerusalem,
Israel}
\author{Elad Steinberg}
\affiliation{Racah Institute of Physics, The Hebrew University, 9190401 Jerusalem,
Israel}
\author{Ryan G. McClarren}
\affiliation{Department of Aerospace and Mechanical Engineering, University of
Notre Dame, Fitzpatrick Hall, Notre Dame, IN 46556, USA}
\author{Menahem Krief}
\email{menahem.krief@mail.huji.ac.il}

\affiliation{Racah Institute of Physics, The Hebrew University, 9190401 Jerusalem,
Israel}
\affiliation{Department of Aerospace and Mechanical Engineering, University of
Notre Dame, Fitzpatrick Hall, Notre Dame, IN 46556, USA}
\begin{abstract}
We derive a family of similarity solutions to the nonlinear non-equilibrium
Marshak wave problem for an inhomogeneous planar medium which is coupled
to a time dependent radiation driving source. We employ the non-equilibrium
gray diffusion approximation in the supersonic regime. The solutions
constitute a generalization of the non-equilibrium nonlinear solutions
that were developed recently for homogeneous media. Self-similar solutions
are constructed for a power law time dependent surface temperature,
a spatial power law density profile and a material model with power
law temperature and density dependent opacities and specific energy
density. The extension of the problem to non-homogeneous media enables
the existence of similarity solutions for a general power law specific
material energy. It is shown that the solutions exist for specific
values of the temporal temperature drive and spatial density exponents,
which depend on the material exponents. We also illustrate how the
similarity solutions take various qualitatively different forms which
are analyzed with respect to various parameters. Based on the solutions,
we define a set of non-trivial benchmarks for supersonic non-equilibrium
radiative heat transfer. The similarity solutions are compared to
gray diffusion simulations as well as to detailed implicit Monte-Carlo
and discrete-ordinate transport simulations in the optically-thick
regime, showing a great agreement, which highlights the benefit of
these solutions as a code verification test problem. 
\end{abstract}
\maketitle

\section{Introduction}

The theory of radiation hydrodynamics is at the heart of various high
energy density systems, such as inertial confinement fusion and astrophysical
phenomena \cite{lindl2004physics,falize2011similarity,robey2001experimental,hurricane2014fuel,cohen2020key,brutman2024primary,steinberg2024stream}.
Analytic solutions for the equations of radiation hydrodynamics are
an important and practical aspect of the analysis and design of high
energy density experiments \cite{sigel1988x,lindl1995development,cohen2018modeling,cohen2020key,heizler2021radiation,malka2022supersonic,courtois2024characterization,liao2024inverse}
and are frequently used for the verification of computer simulations
\cite{calder2002validating,reinicke1991point,lowrie2008radiative,mcclarren2011benchmarks,mcclarren2021two,mcclarren2011solutions,mcclarren2008analytic,bennett2021self,krief2021analytic,giron2021solutions,giron2023solutions,krief2023piston,krief2024self,krief2024unified,heizler2024accurate}.

The theory of Marshak waves which was developed in the seminal work
\cite{marshak1958effect} and was further generalized in Refs. \cite{petschek1960penetration,zeldovich1967physics,chow1967propagation,pert1977class,pakula1985self,kaiser1989x,reinicke1991point,shestakov1999time,hammer2003consistent,garnier2006self,saillard2010principles,lane2013new,shussman2015full,heizler2016self,cohen2018modeling,krief2024unified},
is a fundamental phenomena that describes the nonlinear propagation
of radiation and the subsequent thermalization of a material that
is illuminated by an intense radiation energy source. In most cases,
radiative heat conduction plays a pivotal role in the process. At
high temperatures and for non-opaque materials, the radiative heat
wave may propagate faster than the speed of sound, resulting in a
supersonic Marshak wave \cite{hammer2003consistent,garnier2006self,smith2010solutions,shussman2015full,hristov2016integral,hristov2018heat,malka2022supersonic,krief2024self,krief2024unified},
for which the material motion is negligible. 

Originally, the Marshak wave problem was addressed under the assumption
of local thermodynamic equilibrium between the radiation field and
the heated material. This scenario is only valid for systems which
are optically thick with respect to the emission-absorption process.
Pomraning and subsequently Su and Olson, in their seminal works \cite{pomraning1979non,bingjing1996benchmark},
derived a widely used \cite{timmes2005automated,hayes2006simulating,krumholz2007equations,gittings2008rage,kamm2008enhanced,zhang2012castro,skinner2013two,rider2016robust,ramsey2015radiation,menon2022vettam}
solution for a non-equilibrium linear Marshak wave problem in the diffusion limit of radiative transfer, assuming a temperature
independent opacity, a material energy density that depends on temperature
as $\sim T^{4}$ and a spatially homogeneous media. This solution
was further extended, yet in the linear regime, in Refs. \cite{ganapol1983non,ganapol2002two,brunner2006development,miller2008some,ghosh2014analytical}.
However, opacities of real materials usually depend strongly on temperature,
so that nonlinear conduction prevails in most high-energy-density
systems. Recently, in Ref. \cite{krief2024self}, a family of new
solutions to the non-equilibrium Marshak wave problem was developed
for nonlinear conduction scenarios. These solutions were developed
by employing a dimensional analysis, that results in solutions which
are self-similar under the assumptions of a general power law temperature
dependent opacities, a material energy density that varies as $\sim T^{4}$
(as in the Pomraning-Su-Olson linear solution) and a spatially homogeneous
media. Since the material energy density of most materials vary as
$\sim T^{\beta}$ where usually $\beta\neq4$, the nonlinear solutions
in Ref. \cite{krief2024self} cannot be applied for a general (and
more realistic) material energy density. In fact, these nonlinear
solutions are not self-similar for $\beta\neq4$.

In this work, we develop new self-similar solutions to the non-equilibrium
nonlinear Marshak wave problem, which are essentially a generalization
of the solutions of Ref. \cite{krief2024self}, to non-homogeneous
media. The self-similarity of the solutions is enabled, by introducing
a more general setup, for which the material has a non-homogeneous
power law spatial density profile of the form $\rho\left(x\right)=\rho_{0}x^{-\omega}$.
The time dependent surface temperature drive is of the form $T_{s}\left(t\right)=T_{0}t^{\tau}$,
and the material model obey a temperature-density power law, with
total and absorption opacities of the form $k_{t}\left(T,\rho\right)=\frac{1}{\mathcal{G}}T^{-\alpha}\rho^{1+\lambda}$
and $k_{a}\left(T,\rho\right)=\frac{1}{\mathcal{G}'}T^{-\alpha'}\rho^{1+\lambda'}$,
and a material energy density $u\left(T,\rho\right)=\mathcal{F}T^{\beta}\rho^{1-\mu}$
. The resulting solutions allow self-similarity for $\beta\neq4$
if and only if $\omega\neq0$, and are therefore, a direct generalization
of the solutions of Ref. \cite{krief2024self}, which assumes $\beta=4$
and $\omega=0$. It is shown that this new family of nonlinear self-similar
solutions exists for specific values of the temporal exponent $\tau$
and the spatial exponent $\omega$, which are related to the the material
model exponents $\alpha,\lambda,\alpha',\lambda',\beta,\mu$. The
properties and behavior of the solutions with respect to various parameters
are analyzed in detail. Finally, the generalized solutions are used
to define a set of non-trivial benchmarks for supersonic non-equilibrium
radiative heat transfer. The solutions are compared to detailed numerical
stochastic and deterministic radiation transport simulations as well
as gray diffusion simulations.

\begin{figure}[t]
\begin{centering}
\includegraphics[width=1\columnwidth]{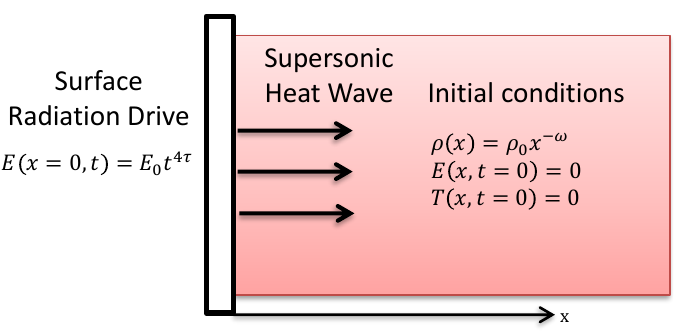} 
\par\end{centering}
\caption{A description of the generalized non-equilibrium Marshak wave problem.
A radiation temperature drive which obeys a temporal power law is
applied at the surface ($x=0$) of a planar medium which is initially
cold with zero radiation field and has an inhomogeneous material density
which obeys a spatial power law. A nonlinear Marshak wave propagates
into the medium and is conducted supersonically (with no material
motion). \label{fig:problemdiagram}}
\end{figure}

\section{Statement of the problem\label{sec:Statement-of-the}}

Supersonic heat conduction by radiation is a common scenario in high
energy density flows, in which radiative heat conduction dominates
and hydrodynamic motion is negligible, and the material density remains
constant in time. The non-equilibrium 1-group (gray) supersonic radiative
transfer problem in planar slab symmetry for the radiation energy
density $E\left(x,t\right)$ and the material energy density $u\left(x,t\right)$,
is formulated by the following coupled equations (two-temperature
approximation) \cite{pomraning1982radiation,pomraning2005equations,mihalas1999foundations,brunner2002forms,krumholz2007equations,heizler2012asymptotic,krief2024self}:

\begin{equation}
\frac{\partial E}{\partial t}+\frac{\partial F}{\partial x}=ck_{a}\left(U-E\right),\label{eq:main_eq}
\end{equation}
\begin{equation}
\frac{\partial u}{\partial t}=ck_{a}\left(E-U\right),\label{eq:main_mat}
\end{equation}
where $c$ is the speed of light, $k_{a}$ the radiation absorption
macroscopic cross section (which is also referred to as the absorption
coefficient or absorption opacity), and $U=aT^{4}$ with $T$ the
material temperature and $a=\frac{8\pi^{5}k_{B}^{4}}{15h^{3}c^{3}}$
the radiation constant. The effective radiation temperature $T_{r}$
is related to the radiation energy density by $E=aT_{r}^{4}$. For
optically thick media, the diffusion approximation of radiative transfer
is applicable, and the radiation energy flux obeys Fick's law: 
\begin{equation}
F=-D\frac{\partial E}{\partial x},\label{eq:fick}
\end{equation}
where the radiation diffusion coefficient is given by:

\begin{equation}
D=\frac{c}{3k_{t}},\label{eq:diffusion_coeff}
\end{equation}
where $k_{t}=\rho\kappa_{R}$ is the total (absorption+scattering)
macroscopic transport cross section, which we also refer to as the
total opacity coefficient, where $\kappa_{R}$ the Rosseland mean
opacity and $\rho$ is the (time independent) material mass density.

We now define a setup for a Marshak wave problem, which is also described
schematically in Fig. \ref{fig:problemdiagram}. We assume a semi-infinite
planar medium which is initially cold with no radiation field,
\begin{equation}
E\left(x,t=0\right)=U\left(x,t=0\right)=0,\label{eq:init_cond}
\end{equation}
with a (time independent) material mass density profile which is given
by a spatial power law
\begin{equation}
\rho\left(x\right)=\rho_{0}x^{-\omega}.\label{eq:rho_omega}
\end{equation}
We note that for a planar system to have a finite mass over a finite
distance from the origin, we must require $\omega<1$. A Marshak wave
is driven by imposing a surface radiation temperature at $x=0$, which
obeys a temporal power law of the form: 
\begin{equation}
T_{r}\left(x=0,t\right)\equiv T_{s}\left(t\right)=T_{0}t^{\tau},\label{eq:Tbound}
\end{equation}
so that the radiation energy density at the system left boundary is:

\begin{equation}
E\left(x=0,t\right)=E_{0}t^{4\tau},\ E_{0}=aT_{0}^{4}.\label{eq:bc}
\end{equation}
As was discussed in Ref. \cite{krief2024self} (and in Refs. \cite{rosen2005fundamentals,cohen2018modeling,cohen2020key,heizler2021radiation,krief2024unified}
for Marshak wave in thermodynamic equilibrium), this boundary
condition of an imposed surface temperature is different than the
common Marshak (or Milne) boundary condition, which represents the
incoming flux from a heat bath. We will derive below in section \ref{subsec:Marshak-boundary-condition}
a relation between the surface radiation temperature and the heat
bath temperature, which will allow to define the same problem by a
Marshak boundary condition with a prescribed bath temperature as a
function of time.

We assume a power law temperature and density dependence of the total
opacity

\begin{equation}
k_{t}\left(T,\rho\right)=\frac{1}{\mathcal{G}}T^{-\alpha}\rho^{1+\lambda},\label{eq:ross_opac_powerlaw}
\end{equation}
the absorption opacity 
\begin{equation}
k_{a}\left(T,\rho\right)=\frac{1}{\mathcal{G}'}T^{-\alpha'}\rho^{1+\lambda'},\label{eq:planck_opac_powerlaw}
\end{equation}
and the material energy density (energy per unit volume)
\begin{equation}
u\left(T,\rho\right)=\mathcal{F}T^{\beta}\rho^{1-\mu}.\label{eq:eos}
\end{equation}
This power law material model, which is characterized by the positive
dimensional constants $\mathcal{G}$, $\mathcal{G}'$, $\mathcal{F}$
and the non-negative exponents $\alpha,$ $\lambda,$ $\alpha',$
$\lambda',$ $\beta,$ and $\mu\leq1$, serves as a good approximation
for the properties of many materials and is commonly employed in the
analysis of high energy density phenomena \cite{pakula1985self,hammer2003consistent,garnier2006self,smith2010solutions,shussman2015full,hristov2018heat,krief2021analytic,malka2022supersonic,krief2024self,krief2024unified}.
Values of the exponents for several materials that were used in experiments
of supersonic and subsonic Marshak waves are given in table \ref{tab:materialparams},
adopted from Refs. \cite{cohen2020key,heizler2021radiation,farmer2024high}
(and references therein), for which photon scattering is negligible
($k_{t}=k_{a}$, so that $\alpha=\alpha'$, $\lambda=\lambda'$).

The problem defined by Eqs. \eqref{eq:main_eq}-\eqref{eq:eos} is
a generalization of the problem defined in Ref. \cite{krief2024self},
where it was assumed that $\omega=0$ and $\beta=4$, to the more
general case of an inhomogeneous density profile, $\omega\neq0$,
and realistic materials for which usually $\beta\neq4$ (as evident
from table \ref{tab:materialparams}).

Using the density profile \eqref{eq:rho_omega} and the material model
defined in Eqs. \eqref{eq:ross_opac_powerlaw}-\eqref{eq:eos}, Eqs.
\eqref{eq:main_eq}-\eqref{eq:main_mat} are written in closed form
as a set of nonlinear coupled partial differential equations for $E$
and $U$:

\begin{align}
\frac{\partial E}{\partial t} & =K\frac{\partial}{\partial x}\left(x^{\omega\left(1+\lambda\right)}U^{\frac{\alpha}{4}}\frac{\partial E}{\partial x}\right)\label{eq:Eform}\\
 & +Mx^{-\omega\left(1+\lambda'\right)}U^{-\frac{\alpha'}{4}}\left(U-E\right),\nonumber 
\end{align}

\begin{equation}
\frac{\partial U}{\partial t}=Px^{-\omega\left(\lambda'+\mu\right)}U^{-\frac{\alpha'}{4}}\left(E-U\right),\label{eq:Uform}
\end{equation}
where we have defined the dimensional constants: 
\begin{equation}
K=\frac{c\mathcal{G}}{3}\rho_{0}^{-1-\lambda}a^{-\frac{\alpha}{4}},\label{eq:KDEF}
\end{equation}
\begin{equation}
M=\frac{c}{\mathcal{G}'}\rho_{0}^{1+\lambda'}a^{\frac{\alpha'}{4}},\label{eq:MDEF}
\end{equation}
\begin{equation}
P=\frac{4ca^{\frac{\alpha'+\beta}{4}}\rho_{0}^{\lambda'+\mu}}{\beta\mathcal{G}'\mathcal{F}}.\label{eq:PDEF}
\end{equation}

\section{A Self-Similar solution\label{sec:Self-Similar-solution}}

\begin{figure*}[t]
\begin{centering}
\includegraphics[width=1\textwidth]{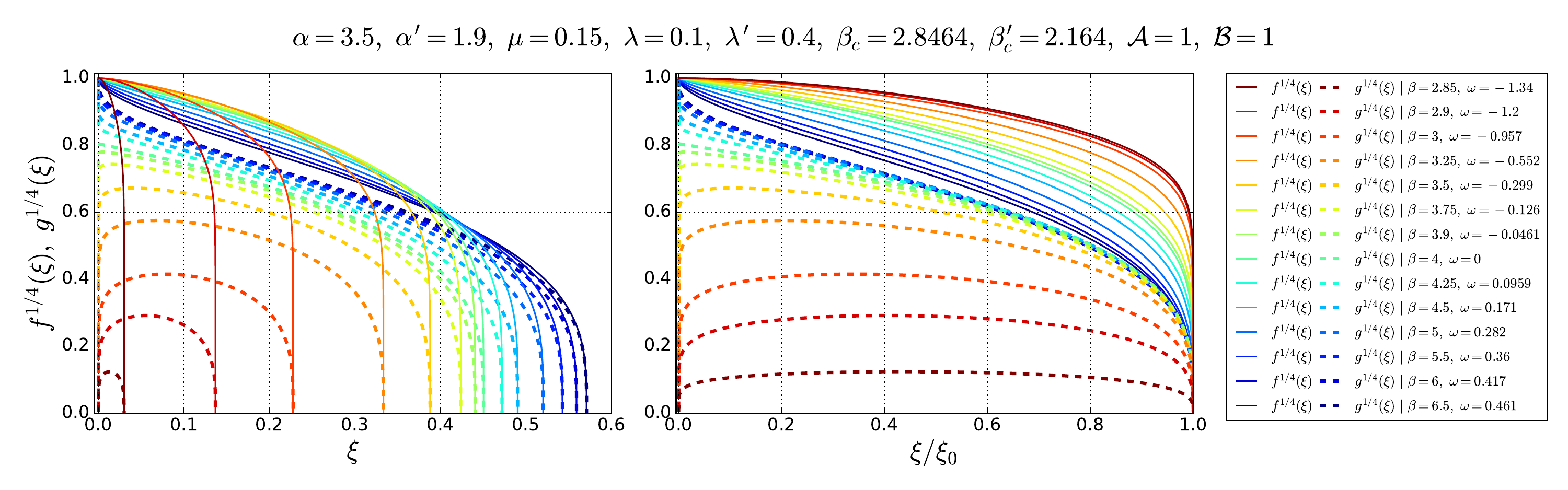}
\par\end{centering}
\caption{The radiation $f^{1/4}\left(\xi\right)$ (solid lines) and material
$g^{1/4}\left(\xi\right)$ (dashed lines) temperature similarity profiles,
for varying values of $\beta$, as listed in the legend (along with
the resulting values of $\omega$). The parameters $\mathcal{A},\ \mathcal{B}$,
the material exponents and the resulting $\beta_{c},\beta_{c}'$ {[}Eqs.
\eqref{eq:betac}-\eqref{eq:betac_p}{]}, are listed in the title.
The profiles are shown as a function of $\xi$ (left figure) and of
$\xi/\xi_{0}$ (right figure), in order to display the variation of
the profiles independently of the front coordinate $\xi_{0}$. Three
types of solutions, as discussed in Sec. \ref{subsec:The-solution-near},
are evident: (i) for $\beta_{c}<\beta<4$, the density vanishes at
the origin ($\omega<0$) and therefore, so does the material temperature,
(ii) for $\beta=4$, the density is constant ($\omega=0$) and finite,
which results in a material temperature that is finite at origin and
lower than the radiation temperature and (iii) for $\beta>4$, the
density diverges at the origin ($\omega>0$), so that the material
and radiation temperatures are equal at the origin. \label{fig:profiles_beta}}
\end{figure*}
\begin{figure*}[t]
\begin{centering}
\includegraphics[width=1\textwidth]{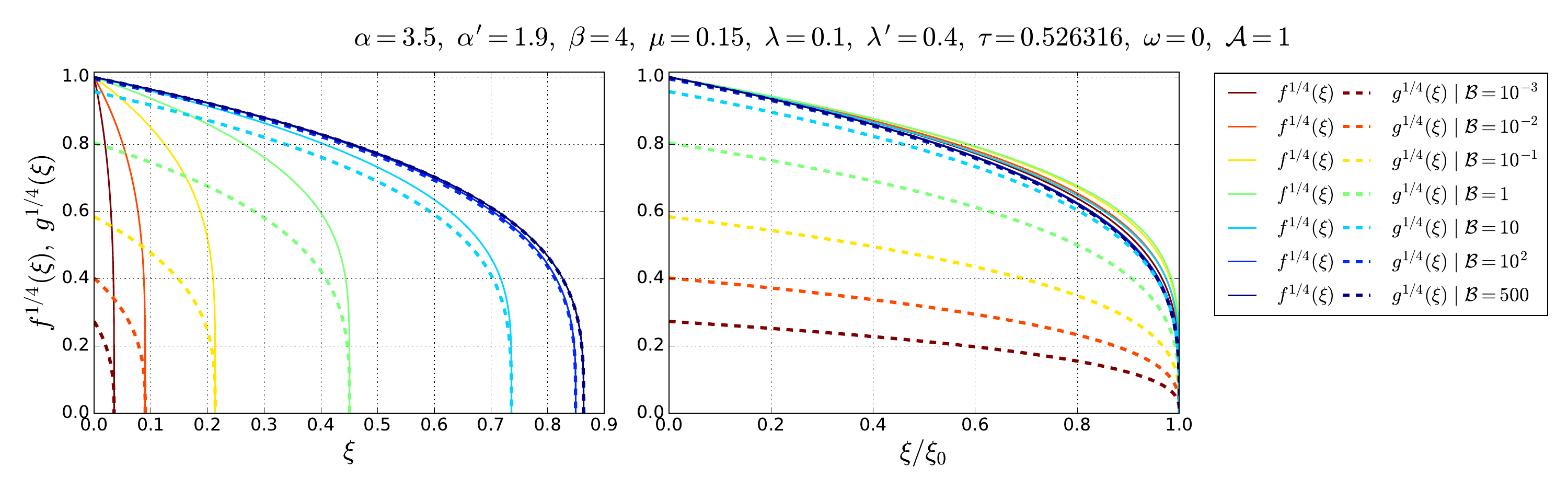}
\par\end{centering}
\caption{Same as Fig. \ref{fig:profiles_beta}, but for a material model with
a fixed $\beta=4$ (as listed in the title), for $\mathcal{A}=1$
and varying values of $\mathcal{B}$ which are listed in the legend.
The choice $\beta=4$ results in $\omega=0$ (a homogeneous density
profile), which according to the analysis in Sec. \ref{subsec:The-solution-near},
results in a finite value of the material temperature at the origin.
It is evident that the material and radiation temperature profiles
become closer when $\mathcal{B}$ is increased, and that thermodynamic
equilibrium limit $f^{1/4}\left(\xi\right)\approx g^{1/4}\left(\xi\right)$
is reached when $\mathcal{B}\gg1$. As discussed in the text, it is
also evident that $\xi_{0}$ increases with $\mathcal{B}$. \label{fig:profiles_omega_0}}
\end{figure*}

\begin{figure*}[t]
\begin{centering}
\includegraphics[width=1\textwidth]{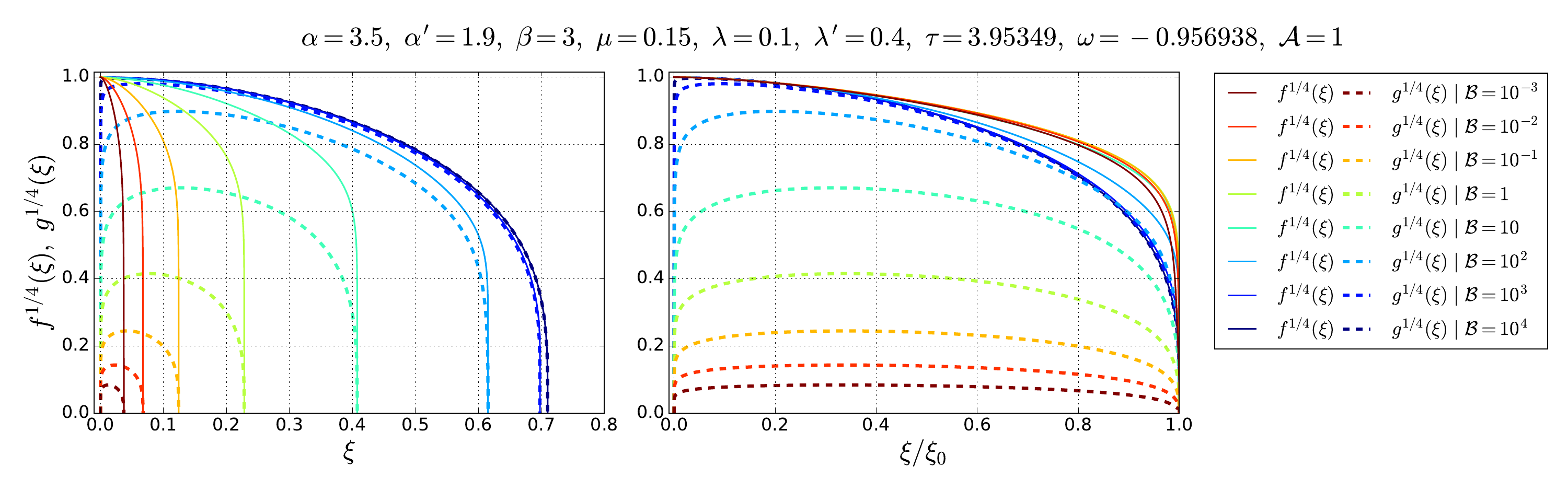}
\par\end{centering}
\caption{Same as Fig. \ref{fig:profiles_omega_0}, but for a choice of parameters
for which $\omega<0$ (an increasing density profile). According to
the analysis in Sec. \ref{subsec:The-solution-near}, this case results
in $g\left(\xi\to0\right)=0$, for any value of $\mathcal{B}$. It
is evident as $\mathcal{B}\rightarrow\infty$, the material and radiation
temperatures approach each other, while maintaining a zero material
temperature at the origin, which results in increasingly larger slopes
near the origin.\label{fig:profiles_omega_neg}}
\end{figure*}

\begin{figure*}[t]
\begin{centering}
\includegraphics[width=1\textwidth]{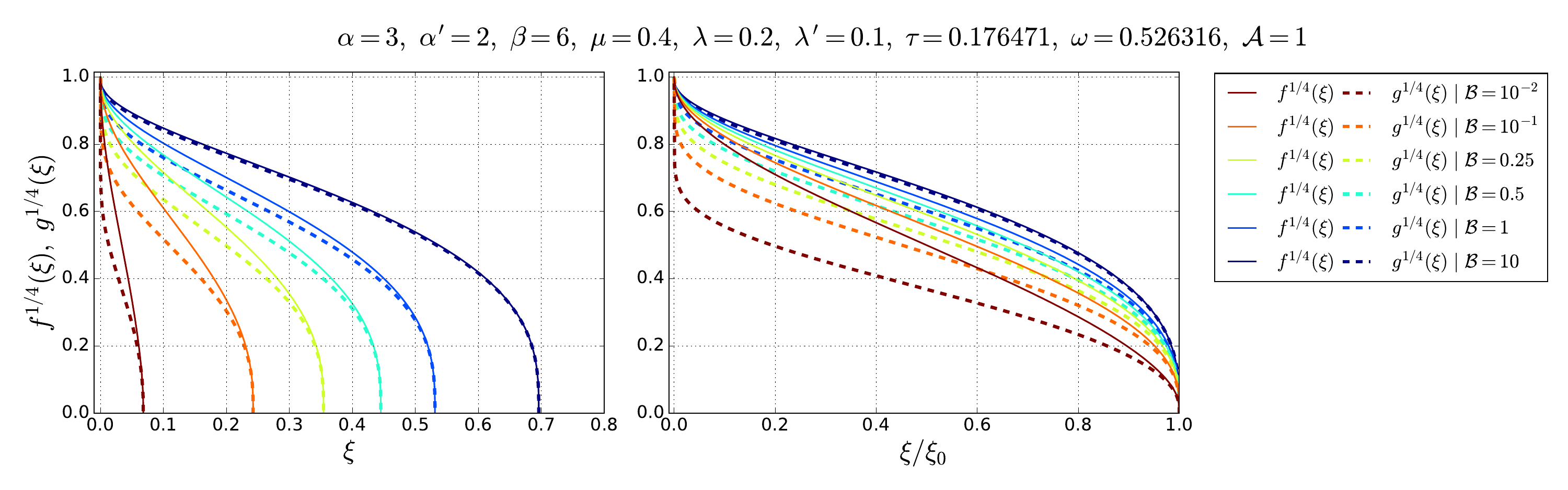}
\par\end{centering}
\caption{Same as Fig. \ref{fig:profiles_omega_neg}, but for a choice of parameters
for which $\omega>0$ (a decreasing density profile), that results
in $g\left(\xi\to0\right)=1$, which must hold for any value of $\mathcal{B}$.
It is evident as $\mathcal{B}\rightarrow0$, the material temperature
becomes increasingly lower than the radiation temperature, while maintaining
equilibrium at the origin, $g\left(\xi\to0\right)=f\left(\xi\to0\right)=1$.
This results in increasingly steep negative slopes near the origin.
\label{fig:profiles_omega_pos}}
\end{figure*}

\begin{figure*}[t]
\begin{centering}
\includegraphics[width=1\textwidth]{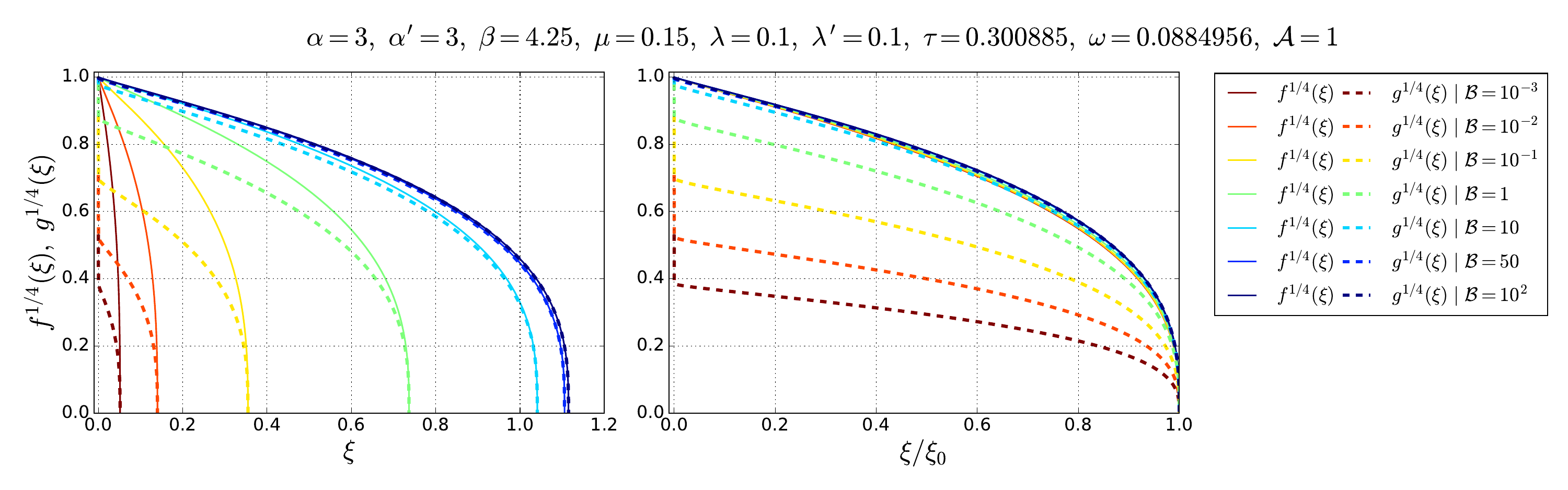}
\par\end{centering}
\caption{Same as Fig. \ref{fig:profiles_omega_pos}, but for a choice of parameters
with a smaller $\omega>0$, which results in an extremely sharp drop
of the material temperature near the origin.\label{fig:profiles_omega_pos_small}}
\end{figure*}

\begin{figure*}[t]
\begin{centering}
\includegraphics[width=1\textwidth]{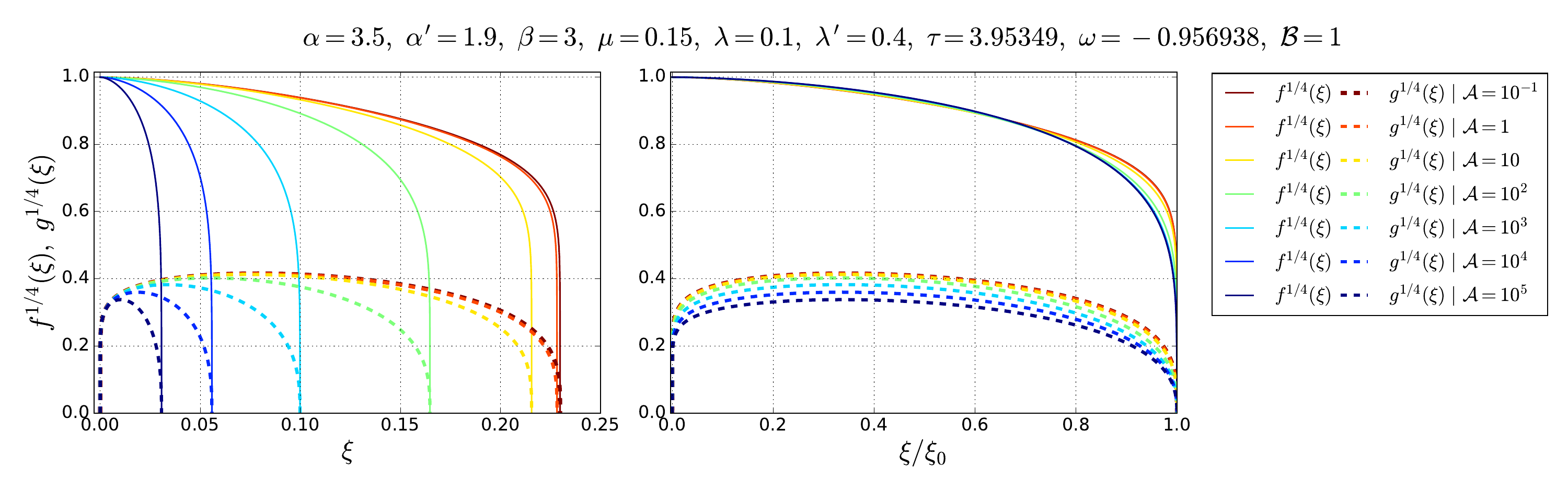}
\par\end{centering}
\caption{Same as Fig. \ref{fig:profiles_omega_0}, but for varying values of
$\mathcal{A}$ (as listed in the legend) and a constant value
$\mathcal{B}=1$. The choice of parameters is the same as in Fig.
\ref{fig:profiles_omega_neg}. It is evident that when $\mathcal{A}\lesssim1$
it has a small effect on $\xi_{0}$ and the profile shapes. For larger
values of $\mathcal{A}$, the value of $\xi_{0}$ and $g\left(\xi\right)$
decrease with respect to $\mathcal{A}$. \label{fig:profiles_omega_neg_A}}
\end{figure*}

\begin{figure}[h]
\begin{centering}
\includegraphics[width=1.01\columnwidth]{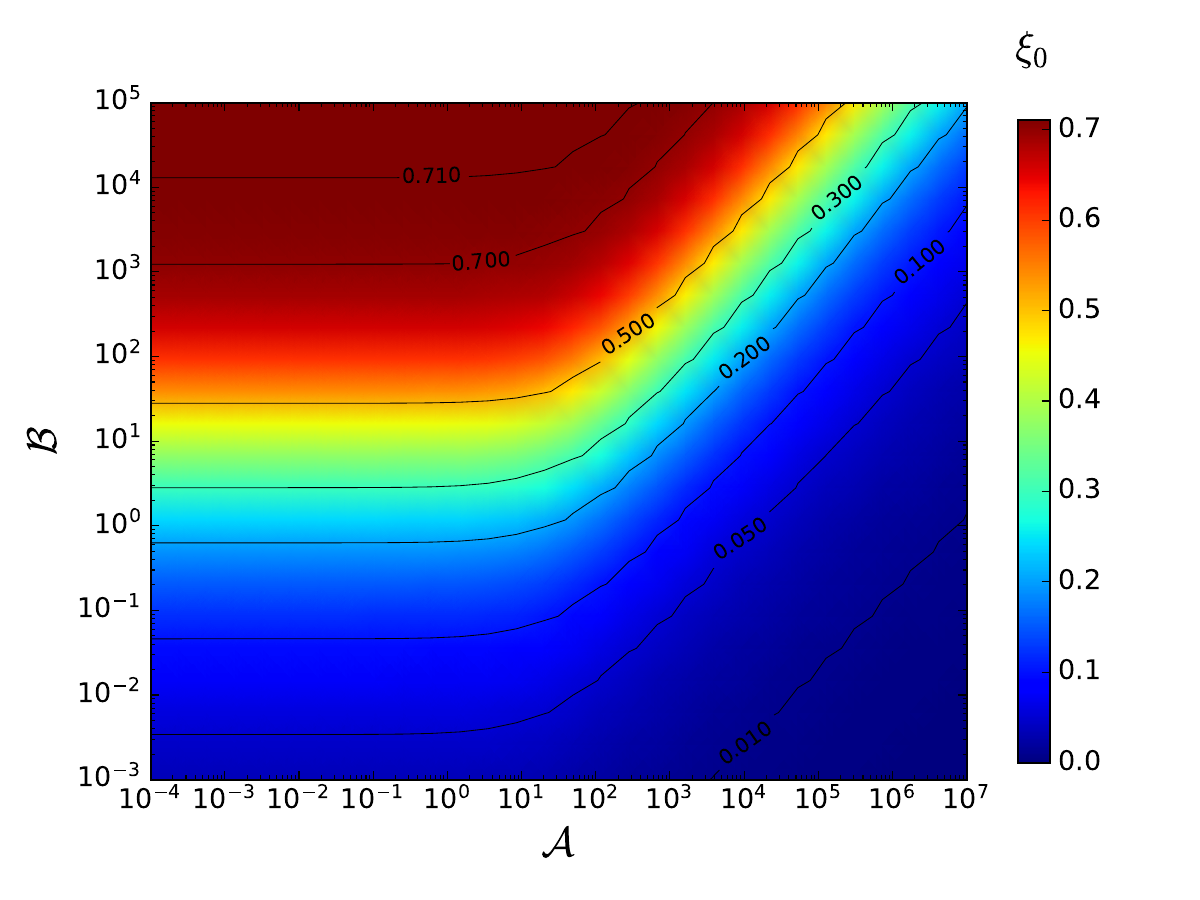}
\par\end{centering}
\begin{centering}
\includegraphics[width=1.01\columnwidth]{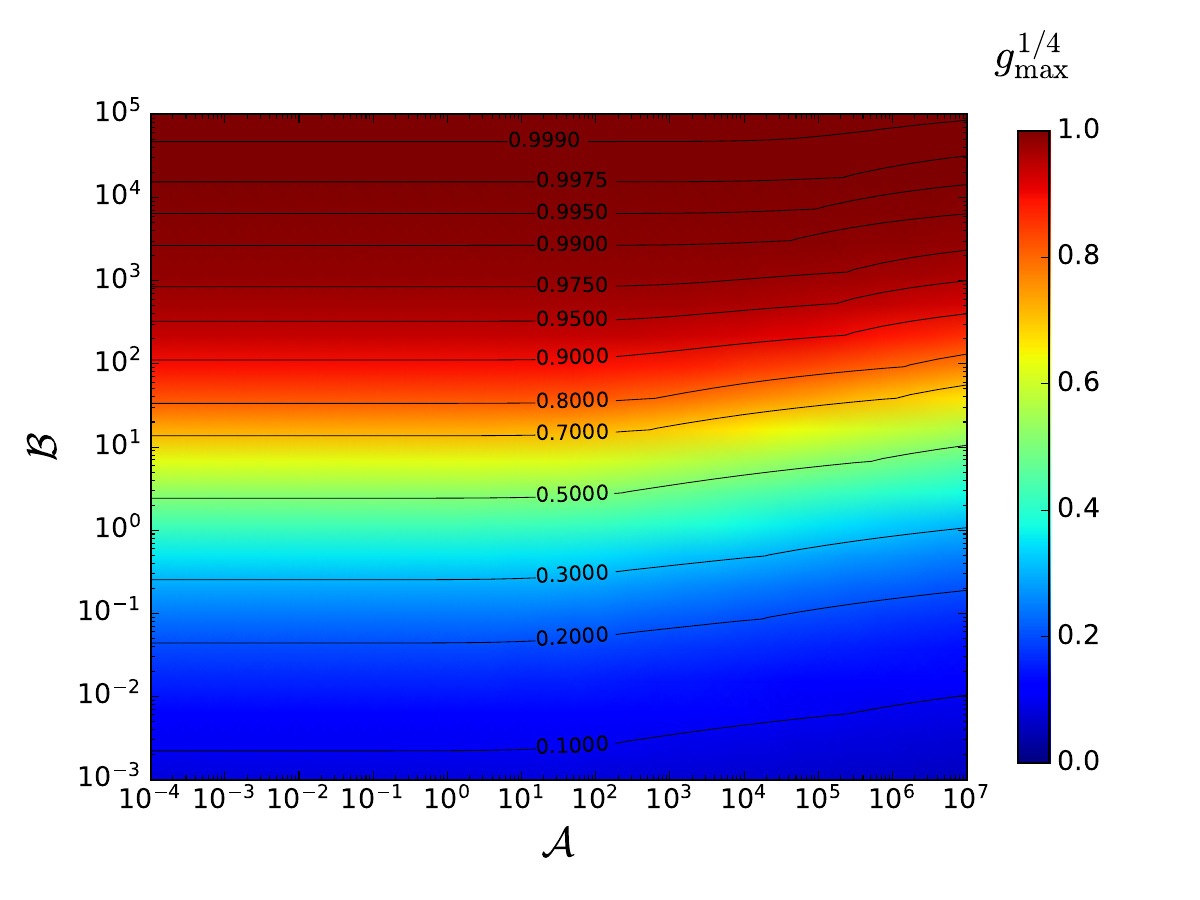}
\par\end{centering}
\caption{Color maps of the heat front coordinate $\xi_{0}$ (upper figure)
and the maximal value of the material temperature profile $g_{\text{max}}^{1/4}=\protect\underset{0\protect\leq\xi\protect\leq\xi_{0}}{\text{max}}\left[g^{1/4}\left(\xi\right)\right]$,
as a function of $\mathcal{A}$ and $\mathcal{B}$. The material exponents
are the same as in Figs. \ref{fig:profiles_omega_neg},\ref{fig:profiles_omega_neg_A}
(namely, $\alpha=3.9,\,\alpha'=1.9,\,\beta=3,\,\mu=0.15,\,\lambda=0.1$
and $\lambda'=0.4$, for which $\omega<0$). It is evident that $\xi_{0}$
and $g_{\text{max}}^{1/4}$ are increasing functions of $\mathcal{B}$
(in agreement with Fig. \ref{fig:profiles_omega_neg}), and that thermodynamic
equilibrium is reached for $\mathcal{B}\gg1$ (as $g_{\text{max}}^{1/4}$
approaches unity). It is also evident that $\mathcal{A}$ has an effect
on $\xi_{0}$ and $g_{\text{max}}^{1/4}$, only for large values $\mathcal{A}\apprge10$,
for which $\xi_{0}$ and $g_{\text{max}}^{1/4}$ are decreasing functions
of $\mathcal{A}$ (in agreement with Fig. \ref{fig:profiles_omega_neg_A}).
\label{fig:meshplots}}
\end{figure}

In Appendix \ref{sec:Dimensional-analysis} it is shown in detail,
using the method of dimensional analysis \cite{buckingham1914physically,zeldovich1967physics,barenblatt1996scaling,shussman2015full,krief2021analytic},
that the solution of the problem defined by the nonlinear gray diffusion
model in Eqs. \eqref{eq:Eform}-\eqref{eq:Uform} with the initial
and boundary conditions \eqref{eq:init_cond},\eqref{eq:bc}, is self-similar,
if and only if the surface temperature exponent and the spatial density
exponent obey the following relations in terms of the various material
exponents: 
\begin{align}
\tau & =\frac{1-\mu}{\left(\beta-4\right)\left(1+\lambda'\right)+\alpha'\left(1-\mu\right)},\label{eq:tau_ss}\\
\omega & =\frac{2\left(\beta-4\right)}{\left(\beta-4\right)\left(2+\lambda+\lambda'\right)+\left(\alpha+\alpha'\right)\left(1-\mu\right)}.\label{eq:omega_ss}
\end{align}
It is evident that $\omega=0$ if and only if $\beta=4$, in which
case $\tau=\frac{1}{\alpha'}$, in agreement with Ref. \cite{krief2024self}.
Conversely, $\beta\neq4$ results in $\omega\neq0$, and the problem
has a self-similar solution only for an inhomogeneous density profile.
As shown in Appendix \ref{sec:Dimensional-analysis}, when Eqs. \eqref{eq:tau_ss}-\eqref{eq:omega_ss}
hold, the problem defines two dimensionless constants: 
\begin{align}
\mathcal{A} & =E_{0}^{-\frac{\alpha+\alpha'}{4\left(\alpha\tau+1\right)}}K^{\frac{\alpha'\tau-1}{\alpha\tau+1}}M,\label{eq:adef}\\
\mathcal{B} & =E_{0}^{\frac{4-\beta-\alpha-\alpha'}{4\left(\alpha\tau+1\right)}}K^{\frac{\left(\alpha'+\beta-4\right)\tau-1}{\alpha\tau+1}}P,\label{eq:bdef}
\end{align}
and the Marshak wave problem has a self-similar solution which is
expressed in terms dimensionless similarity profiles $f\left(\xi\right)$,
$g\left(\xi\right)$ as: 
\begin{equation}
E\left(x,t\right)=E_{0}t^{4\tau}f\left(\xi\right),\label{eq:Ess}
\end{equation}
\begin{equation}
U\left(x,t\right)=E_{0}t^{4\tau}g\left(\xi\right),\label{eq:Uss}
\end{equation}
where the dimensionless similarity coordinate is given by: 
\begin{equation}
\xi=\frac{x}{t^{\delta}\left(KE_{0}^{\frac{\alpha}{4}}\right)^{\frac{1}{2-\omega\left(1+\lambda\right)}}},\label{eq:xsi_def}
\end{equation}
and the similarity exponent is:{\small{}
\begin{equation}
\delta=\frac{1+\alpha\tau}{2-\omega\left(\lambda+1\right)}=\frac{1}{2}\left(1+\frac{\alpha\left(1-\mu\right)+\left(\beta-4\right)\left(1+\lambda\right)}{\alpha'\left(1-\mu\right)+\left(\beta-4\right)\left(1+\lambda'\right)}\right).\label{eq:delta_def}
\end{equation}
}The radiation and material temperature profiles are therefore given
by 
\begin{equation}
T_{r}\left(x,t\right)=T_{0}t^{\tau}f^{1/4}\left(\xi\right),\label{eq:Trss}
\end{equation}
\begin{equation}
T\left(x,t\right)=T_{0}t^{\tau}g^{1/4}\left(\xi\right).\label{eq:Tmss}
\end{equation}
We see that when the absorption and total opacity have the same temperature
and density dependence ($\alpha=\alpha'$ and $\lambda=\lambda'$
or $\alpha=\alpha'$ and $\beta=4$), we acquire $\delta=1$, in agreement
with Ref. \cite{krief2024self}. As shown in Appendix \ref{sec:Dimensional-analysis},
by plugging the self-similar representation \eqref{eq:Ess}-\eqref{eq:Uss}
into the gray diffusion system \eqref{eq:Eform}-\eqref{eq:Uform},
all dimensional quantities are factored out, and the following (dimensionless)
second order ordinary differential equations (ODE) system for the
similarity profiles is obtained:{\small{}
\begin{align}
4\tau & f\left(\xi\right)-\delta\xi f'\left(\xi\right)=\xi^{\omega\left(1+\lambda\right)}\Bigg(\frac{1}{\xi}\omega\left(\lambda+1\right)g^{\frac{\alpha}{4}}\left(\xi\right)f'\left(\xi\right)\nonumber \\
 & +g^{\frac{\alpha}{4}-1}\left(\xi\right)\left[\frac{\alpha}{4}f'\left(\xi\right)g'\left(\xi\right)+g\left(\xi\right)f''\left(\xi\right)\right]\Bigg)\nonumber \\
 & -\mathcal{A}\xi^{-\omega\left(1+\lambda'\right)}g^{-\frac{\alpha'}{4}}\left(\xi\right)\left(f\left(\xi\right)-g\left(\xi\right)\right),\label{eq:f_ode}
\end{align}
\begin{equation}
4\tau g\left(\xi\right)-\delta\xi g'\left(\xi\right)=\mathcal{B}\xi^{-\omega\left(\lambda'+\mu\right)}g^{1-\frac{\alpha'+\beta}{4}}\left(\xi\right)\left(f\left(\xi\right)-g\left(\xi\right)\right).\label{eq:g_ode}
\end{equation}
}The surface radiation temperature boundary condition {[}Eq. \eqref{eq:bc}{]},
is written in terms of the radiation energy similarity profile as:
\begin{equation}
f\left(0\right)=1.\label{eq:f0_bc}
\end{equation}
It is evident that the dimensionless problem defined by Eqs. \eqref{eq:f_ode}-\eqref{eq:f0_bc}
depends only on the material exponents $\alpha,\alpha',\lambda,\lambda',\beta,\mu$
and the dimensionless constants $\mathcal{A},\mathcal{B}$. We also
see that if $\beta=4$ we get $\mathcal{B}=\epsilon\mathcal{A}$ where
$\epsilon=\frac{a}{\mathcal{F}}$, and the dimensionless problem is
reduced to the homogeneous solution given in Ref. \cite{krief2024self}.

Nonlinear conduction is characterized by a steep heat front, that
is, there exists a finite heat front coordinate, $\xi_{0}$, such
that $f\left(\xi\right)=g\left(\xi\right)=0$ for $\xi\geq\xi_{0}$.
According to Eq. \eqref{eq:xsi_def}, the heat front propagates in
time according to a temporal power law: 
\begin{equation}
x_{F}\left(t\right)=\xi_{0}t^{\delta}\left(KE_{0}^{\frac{\alpha}{4}}\right)^{\frac{1}{2-\omega\left(1+\lambda\right)}}.\label{eq:xheat}
\end{equation}
As shown in many works on nonlinear Marshak waves in thermodynamic
equilibrium, for which $T_{r}\equiv T$ \cite{marshak1958effect,petschek1960penetration,castor2004radiation,mihalas1999foundations,nelson2009semi,lane2013new,shussman2015full,garnier2006self,krief2024self,krief2024unified}
and more recently in non-equilibrium as well \cite{krief2024self},
it is customary to calculate the value of $\xi_{0}$ via iterations
of a ``shooting method''. This is done by integrating Eqs. \eqref{eq:f_ode}-\eqref{eq:g_ode}
starting from a trial value of $\xi_{0}$ towards the origin, $\xi=0$,
resulting in a numerical value for $f\left(0\right)$. The trial $\xi_{0}$
is adjusted until the surface radiation temperature boundary condition,
$f\left(0\right)=1$, is satisfied. To the best of our knowledge,
analytical solutions to Eqs. \eqref{eq:f_ode}-\eqref{eq:g_ode} exist
only when $\beta=4$ ($\omega=0$) and $\alpha=\alpha'$, as was found
in Ref. \cite{krief2024self}, otherwise, the similarity profiles
and $\xi_{0}$ must be obtained numerically. Numerical results for
the similarity profiles are shown in Figs. \ref{fig:profiles_beta}-\ref{fig:profiles_omega_neg_A},
which display the different characteristics of the solutions for cases
with $\omega=0$, $\omega>0$ and $\omega<0$, as well as the behavior
of the solutions for varying values of $\mathcal{A}$ and $\mathcal{B}$.
These characteristics of the solutions are discussed in detail below
in Sec. \ref{subsec:The-solution-near}. 

\subsection{Range of validity}

The similarity solution is invalid when $\omega\geq1$ (in which case
there is an infinite mass over any finite distance from the origin),
or when $\delta\leq0$ (the wave does not propagate outwards). In
order to further analyze the solution's properties and range of validity,
we first consider the case with $\mu<1$, for which it is customary
to write $\tau$, $\omega$ and $\delta$ {[}Eqs. \eqref{eq:tau_ss}-\eqref{eq:omega_ss},
\eqref{eq:delta_def}{]}, as functions of $\beta$: 
\begin{equation}
\tau\left(\beta\right)=\left(\frac{1-\mu}{1+\lambda'}\right)\frac{1}{\beta-\beta_{c}},\label{eq:tau_beta}
\end{equation}
\begin{align}
\omega\left(\beta\right) & =\left(\frac{1}{1+\overline{\lambda}}\right)\frac{\beta-4}{\beta-\beta_{c}'},\label{eq:omega_beta}
\end{align}
\begin{equation}
\delta\left(\beta\right)=\left(\frac{1+\overline{\lambda}}{1+\lambda'}\right)\frac{\beta-\beta_{c}'}{\beta-\beta_{c}},\label{eq:delta_beta}
\end{equation}
where we have defined two critical values of $\beta$:
\begin{equation}
\beta_{c}=4-\frac{\alpha'\left(1-\mu\right)}{1+\lambda'},\label{eq:betac}
\end{equation}
\begin{equation}
\beta_{c}'=4-\frac{\overline{\alpha}\left(1-\mu\right)}{1+\overline{\lambda}}.\label{eq:betac_p}
\end{equation}
and the average opacity temperature and density exponents:
\begin{equation}
\overline{\alpha}=\frac{1}{2}\left(\alpha+\alpha'\right),\ \ \overline{\lambda}=\frac{1}{2}\left(\lambda+\lambda'\right).
\end{equation}
We see that $\tau$ diverges at $\beta=\beta_{c}<4$ while $\omega$
diverges at $\beta=\beta_{c}'<4$. We also see that $\tau<0$ for
$\beta<\beta_{c}$, $\tau>0$ for $\beta>\beta_{c}$ and there is
no solution with $\tau=0$. These cases correspond, respectively,
to a decreasing, increasing and constant temperature drive {[}see
Eq. \eqref{eq:Tbound}{]}. Similarly, $\omega<0$ for $\beta_{c}'<\beta<4,$
while $\omega=0$ for $\beta=4$ and $\omega>0$ otherwise; cases
which correspond, respectively, to an increasing, constant and decreasing
spatial density profiles {[}see Eq. \eqref{eq:rho_omega}{]}. In addition,
$\omega<\frac{1}{1+\overline{\lambda}}<1$ for $\beta>\beta_{c}'$.
We also see that $\delta<0$ for $\text{\ensuremath{\min\left(\beta_{c},\beta_{c}'\right)}}<\beta<\ensuremath{\text{max}\left(\beta_{c},\beta_{c}'\right)}$.
As a result, solutions with $\tau<0$ exist only for $\omega>0$,
while on the other hand, solutions with $\tau>0$ exist for negative,
positive and zero $\omega$. In addition, for $0\leq\beta<\ensuremath{\min\left(\beta_{c},\beta_{c}'\right)}$,
for which $\omega>0$ and $\tau<0$, the solution is valid provided
that $\omega<1$ as well. In summary, the solution is always invalid
for $\text{\ensuremath{\min\left(\beta_{c},\beta_{c}'\right)}}\leq\beta\leq\ensuremath{\text{max}\left(\beta_{c},\beta_{c}'\right)}$
and also for values of $\beta$ in the range $0\leq\beta<\ensuremath{\min\left(\beta_{c},\beta_{c}'\right)}$
for which $\omega\left(\beta\right)>1$.

\begin{figure*}[t]
\begin{centering}
\includegraphics[width=0.95\textwidth]{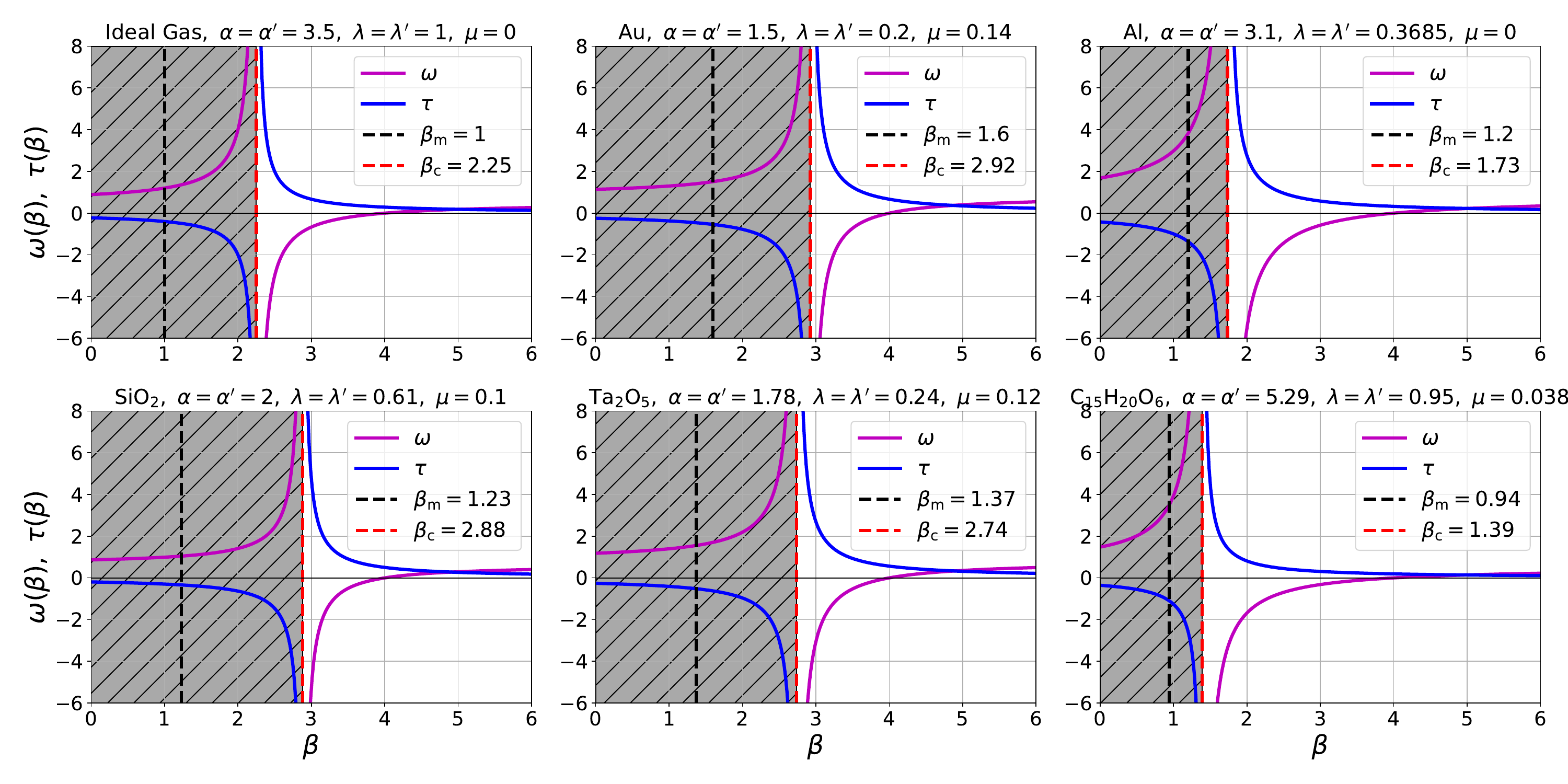}
\par\end{centering}
\caption{The temperature drive exponent $\tau$ (blue) and spatial density
exponent $\omega$ (magenta), for which the solution of the Marshak
wave problem is self-similar {[}Eqs. \eqref{eq:tau_ss}-\eqref{eq:omega_ss}
or \eqref{eq:tau_beta}-\eqref{eq:omega_beta}{]}, as a function of
the material energy density power $\beta$. The results are shown
for $\alpha=\alpha'$, $\lambda=\lambda'$ using the parameters $\alpha,\lambda,\mu$
of various materials (given in Table \ref{tab:materialparams}). Vertical
lines are shown for the critical value $\beta=\beta_{c}=\beta_{c}'$
{[}Eqs. \eqref{eq:betac}-\eqref{eq:betac_p}{]}, and $\beta=\beta_{m}$,
the actual temperature exponent of each material (see Table \ref{tab:materialparams}).
It is evident that $\omega>1$ for $\beta<\beta_{c}$, which means
that similarity solutions do not exist for $\beta<\beta_{c}$ (marked
area). Since for all materials shown, $\beta_{c}>\beta_{m}$, the
solution is not self-similar for $\beta=\beta_{m}$. It is also evident
that $\tau>0$ for $\beta>\beta_{c}$ and $\tau<0$ for $\beta<\beta_{c}$,
which means that similarity solutions must have an increasing temperature
drive ($\tau>0$). Finally, for the shown materials $\beta_{c}<4$,
and since $\omega$ changes sign at $\beta=4$, self-similar solutions
exist for both increasing ($\omega<0$) and decreasing ($\omega>0$)
spatial density profiles.\label{fig:tau_omega_profiles}}
\end{figure*}

\begin{figure*}[t]
\begin{centering}
\includegraphics[width=0.33\textwidth]{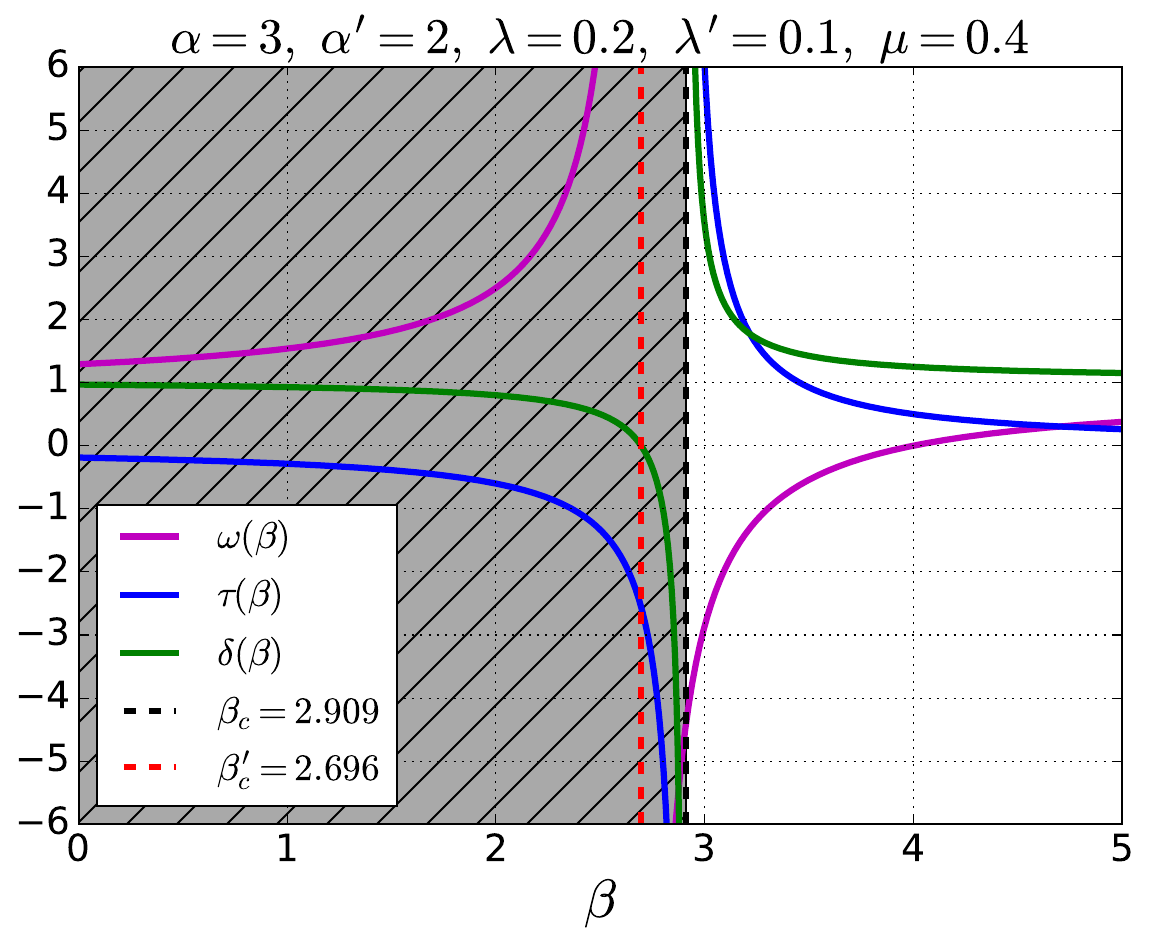}\includegraphics[width=0.33\textwidth]{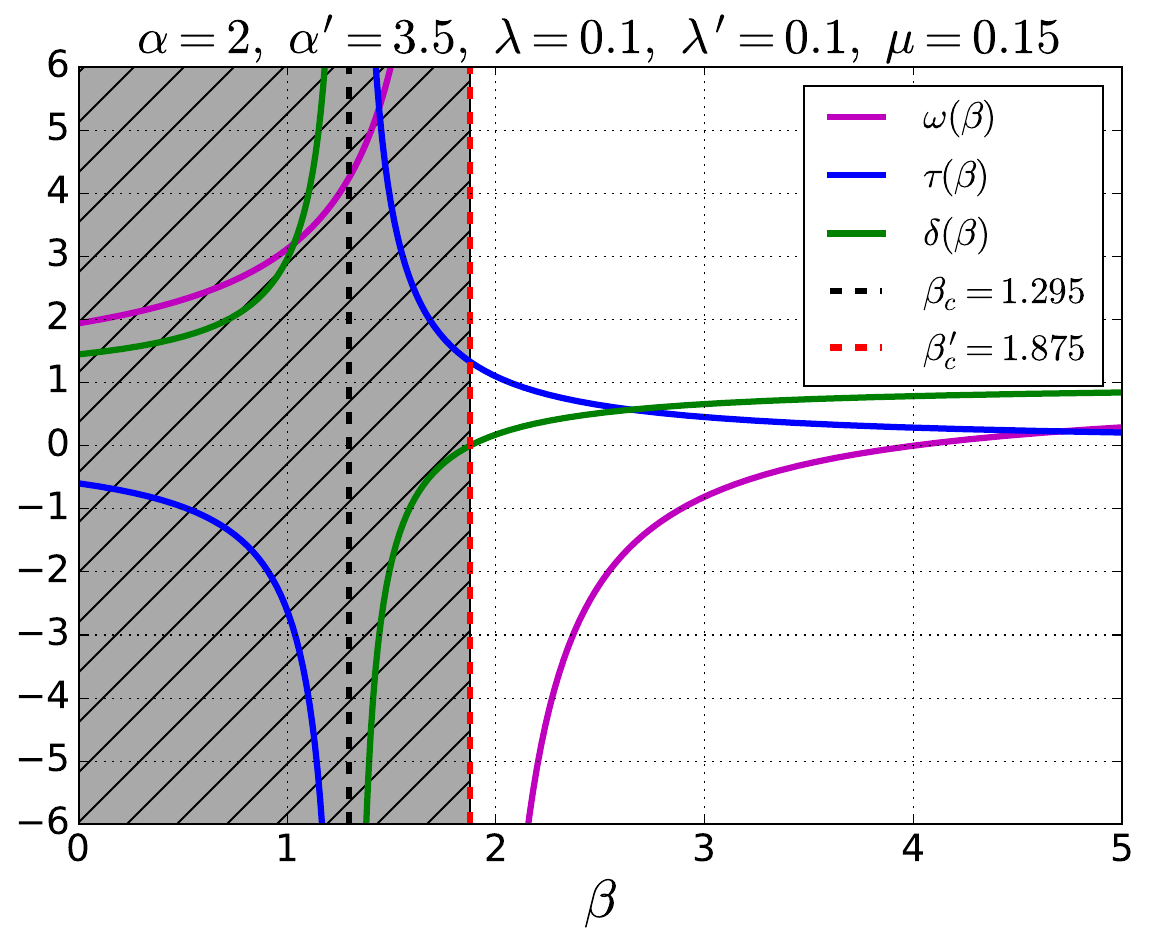}\includegraphics[width=0.33\textwidth]{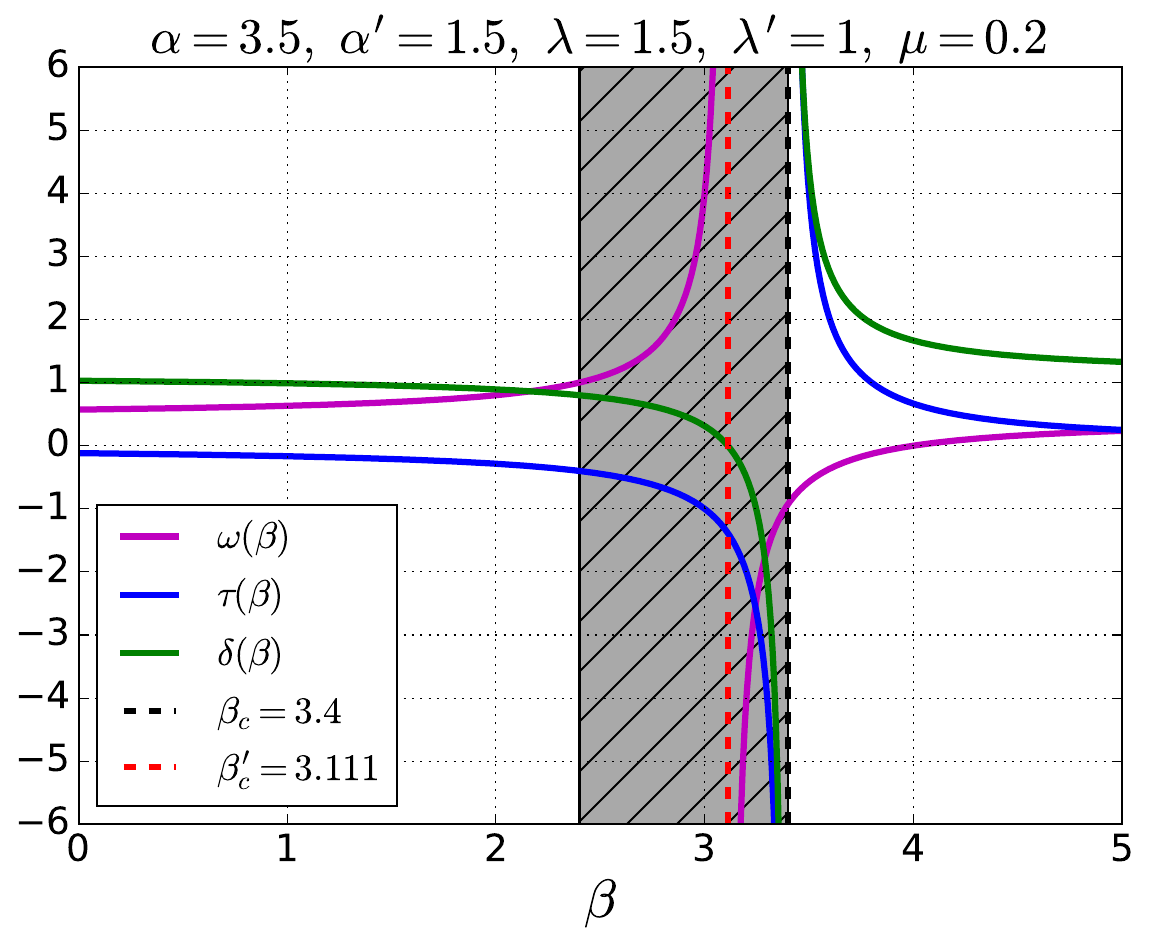}
\par\end{centering}
\begin{centering}
\includegraphics[width=0.33\textwidth]{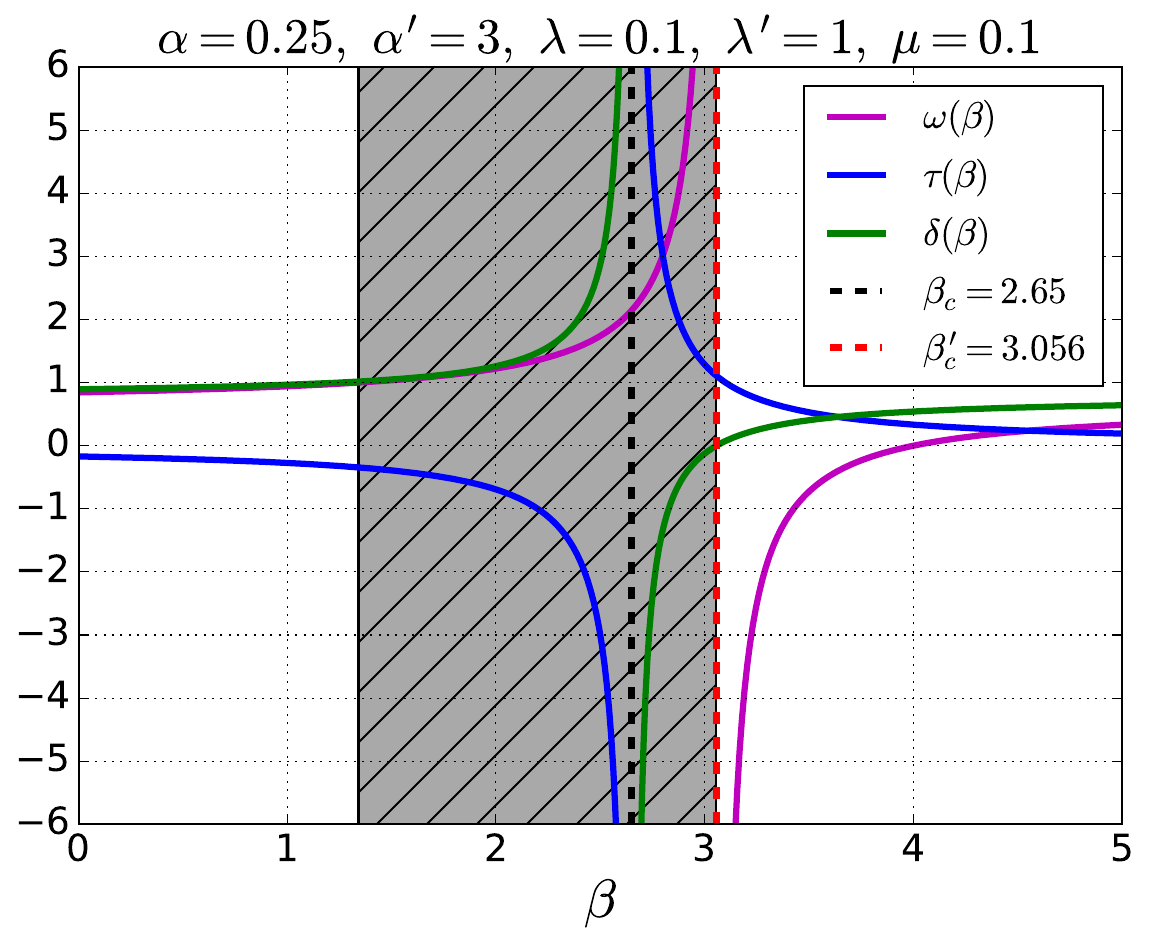}\includegraphics[width=0.33\textwidth]{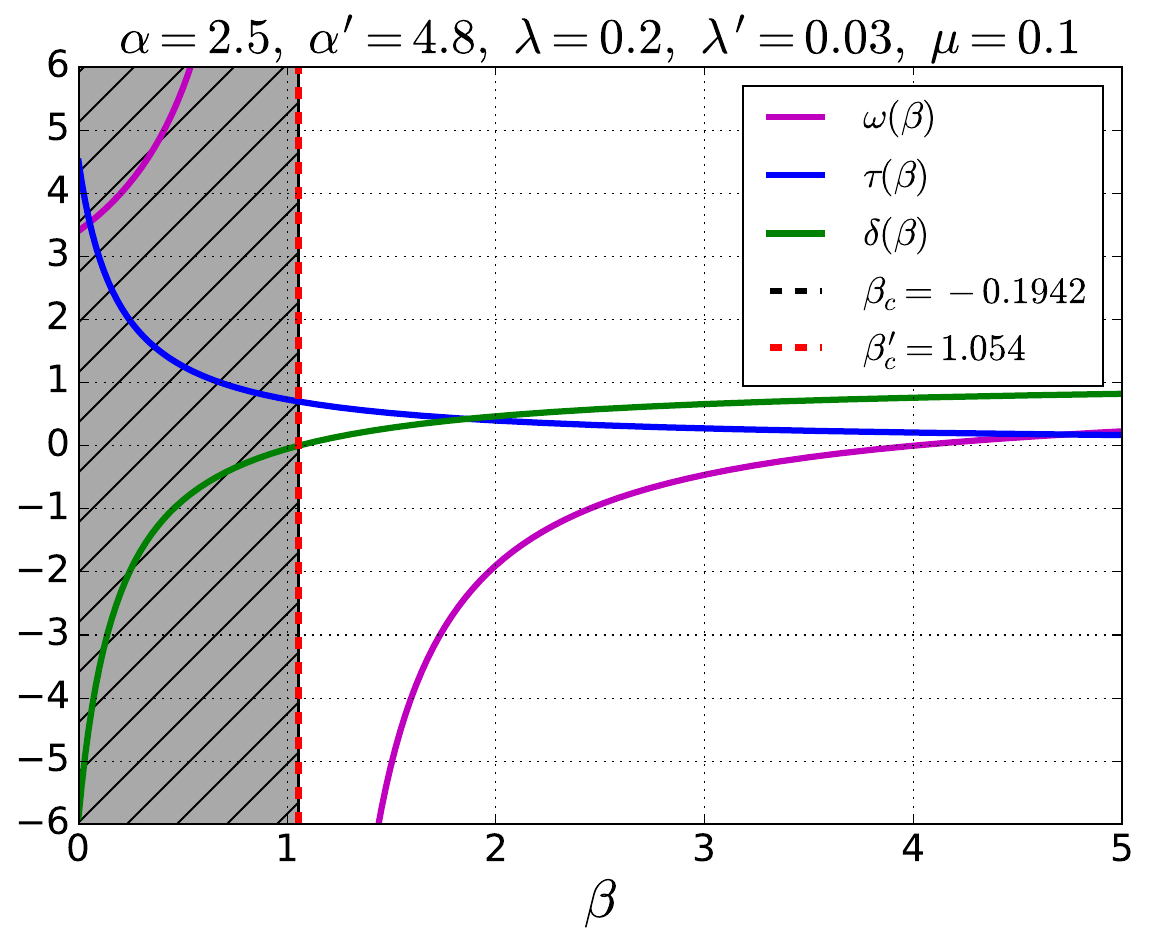}\includegraphics[width=0.33\textwidth]{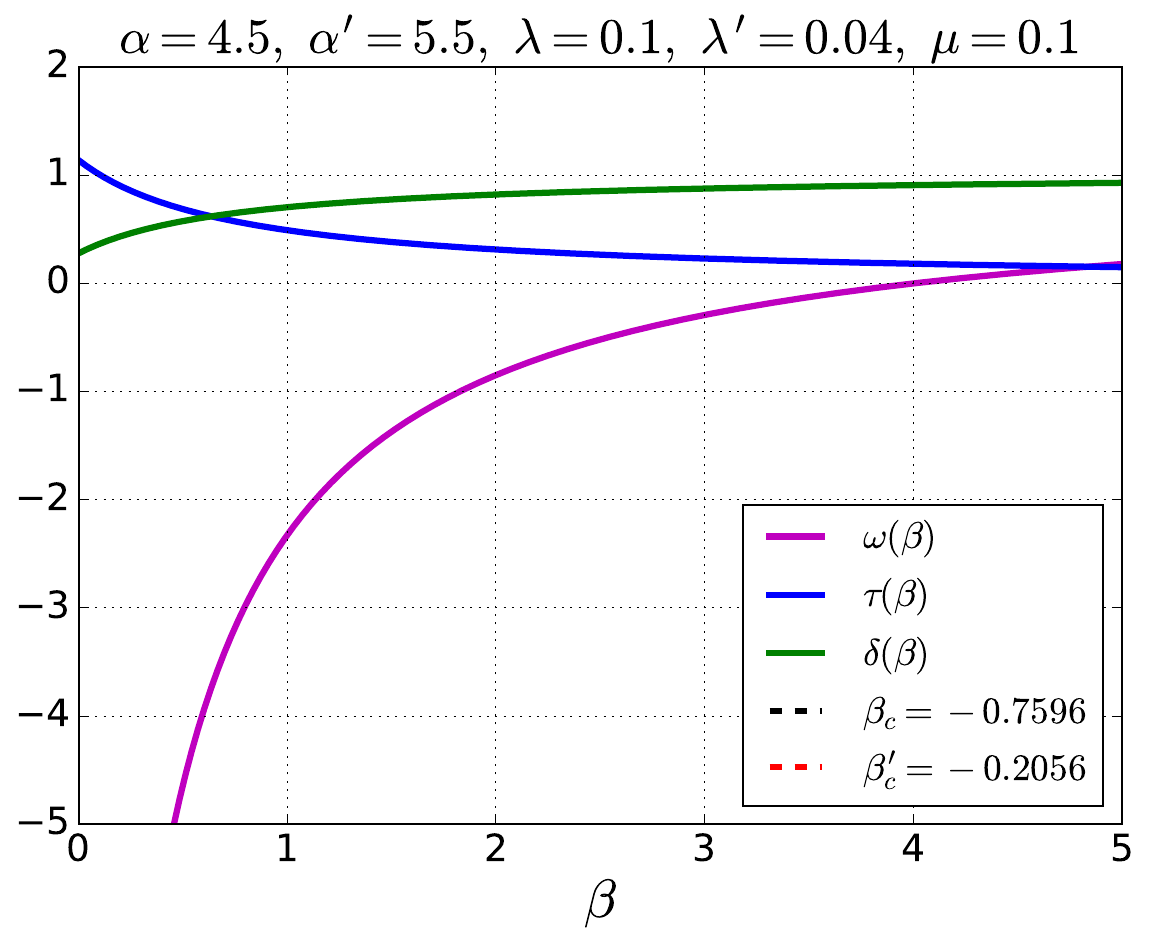}
\par\end{centering}
\caption{Similar plot to Fig. \ref{fig:tau_omega_profiles}, but for cases
for which $\alpha\protect\neq\alpha'$ or $\lambda\protect\neq\lambda'$.
The similarity exponent $\delta\left(\beta\right)$ {[}Eq. \eqref{eq:delta_def}
or \eqref{eq:delta_beta}{]} is also plotted. Vertical lines are shown
for the critical values $\beta=\beta_{c}$ {[}Eq. \ref{eq:betac}{]}
and $\beta=\beta_{c}'$ {[}Eq. \ref{eq:betac_p}{]}, with the values
listed in the legend. The range of $\beta$ for which the self-similar
solution is invalid (marked in gray), is where $\delta\left(\beta\right)\protect\leq0$
or $\omega\left(\beta\right)\protect\geq1$, as discussed in the text.
Six different cases are shown, with different characteristics and
ranges of validity. \label{fig:tau_omega_profiles_betac}}
\end{figure*}

To illustrate these properties, we first analyze the more simple case
in which the total and absorption opacities have the same temperature
and density dependence, that is, when $\alpha=\alpha'$ and $\lambda=\lambda'$.
In that case $\tau$ and $\omega$ diverge at the same value of $\beta=\beta_{c}=\beta_{c}'$,
while $\delta\equiv1$ and is independent of $\beta$. In Fig. \ref{fig:tau_omega_profiles}
the resulting $\tau\left(\beta\right)$ and $\omega\left(\beta\right)$
are plotted as a function of $\beta$, for several material models
(assuming $\alpha=\alpha'$, $\lambda=\lambda'$). In all of those
cases it turns out that $\omega>1$ for $\beta<\beta_{c}$, so that
the similarity solution is invalid for $\beta\leq\beta_{c}$. Conversely,
since $\underset{\beta\rightarrow\infty}{\lim}\omega\rightarrow\frac{1}{1+\lambda}<1$,
the solution is valid for any $\beta>\beta_{c}$. It also happens
that for each of the chosen materials (given in Table \ref{tab:materialparams}),
$\beta_{c}$ is larger than the value of the material's $\beta$,
and therefore, there are no self-similar solutions for these given
materials. In addition, since $\tau>0$ if and only if $\beta>\beta_{c}$,
the similarity solutions exist in those cases only for a strictly
increasing radiation temperature drive ($\tau>0$). In Fig. \ref{fig:tau_omega_profiles_betac}
we show similar plots for $\tau\left(\beta\right)$, $\omega\left(\beta\right)$
and $\delta\left(\beta\right)$ for more general cases of $\alpha,\alpha',\lambda,\lambda',\mu$,
with $\alpha\neq\alpha'$ or $\lambda\neq\lambda'$. Various different
characteristics are evident. Some cases have valid solutions only
for $\beta>\beta_{c}$ or $\beta>\beta_{c}'$ for which $\tau\left(\beta\right)>0$.
Other cases have valid solutions in a range $0\leq\beta<\beta^{*}<\text{min\ensuremath{\left(\beta_{c},\beta_{c}'\right)}}$
for which $\tau\left(\beta\right)<0$. Interestingly, some cases have
$\beta_{c},\beta_{c}'<0$, so that the solutions are valid for any
$\beta$, and have $\tau\left(\beta\right)>0$. It is also evident
that there exists solutions with $0<\delta<1$ (decelerating heat
front) as well as $\delta>1$ (accelerating heat front). 

Finally, for $\mu=1$ (a density independent material energy density)
and $\beta\neq4$, Eqs. \eqref{eq:tau_ss}-\eqref{eq:omega_ss}, \eqref{eq:delta_def}
give the following $\beta$ independent exponents:
\begin{align}
\tau & =0,\label{eq:tau_mu1}
\end{align}
\begin{equation}
\omega=\frac{1}{1+\overline{\lambda}},\label{eq:omega_mu1}
\end{equation}
{\small{}
\begin{equation}
\delta=\frac{1+\overline{\lambda}}{1+\lambda'}.\label{eq:delta_mu1}
\end{equation}
}{\small\par}

In summary, it was shown that depending on the material exponents,
$\alpha,\alpha',\lambda,\lambda',\beta,\mu$, the generalized solutions,
when they are valid, can have a temporally increasing ($\tau>0$),
decreasing ($\tau<0$) and constant ($\tau=0$) surface temperature
drive.

\begin{table}
\begin{centering}
\begin{tabular}{ccccc}
 &  &  &  & \tabularnewline
\hline 
\hline 
Material & $\alpha$ & $\beta$ & $\lambda$ & $\mu$\tabularnewline
\hline 
$\mathrm{Au}$ & $1.5$ & $1.6$ & $0.2$ & $0.14$\tabularnewline
$\mathrm{Al}$ & $3.1$ & $1.2$ & $0.368$ & $0$\tabularnewline
$\mathrm{SiO_{2}}$ & $2$ & $1.23$ & $0.61$ & $0.1$\tabularnewline
$\mathrm{Ta_{2}O_{5}}$ & $1.78$ & $1.37$ & $0.24$ & $0.12$\tabularnewline
$\mathrm{C_{15}H_{20}O_{6}}$ & $5.29$ & $0.94$ & $0.95$ & $0.038$\tabularnewline
Ideal Gas & $3.5$ & $1$ & $1$ & $0$\tabularnewline
\hline 
\hline 
 &  &  &  & \tabularnewline
\end{tabular}
\par\end{centering}
\caption{Values of material temperature and density exponents $\alpha,\beta,\lambda,\mu$
for various materials (adopted from Refs. \cite{cohen2020key,heizler2021radiation,farmer2024high}
and references therein). In the last row, the values are listed for
a material with an ideal gas equation of state a pure free-free absorption
opacity. \label{tab:materialparams}}
\end{table}

\subsection{The solution near the origin\label{subsec:The-solution-near}}

The behavior of the solution near the system's boundary, can be analyzed
without having to solve the full coupled ODE system \eqref{eq:f_ode}-\eqref{eq:g_ode}.
We expand the solution to first order near $\xi\rightarrow0$:

\begin{align}
f\left(\xi\right) & \approx1+f'\left(0\right)\xi+O\left(\xi^{2}\right),\label{eq:expansion_origing_1}\\
g\left(\xi\right) & \approx g_{0}+g'\left(0\right)\xi+O\left(\xi^{2}\right).\label{eq:expansion_origing_2}
\end{align}
where $g_{0}=g\left(\xi=0\right)$. By substituting the expansion
\eqref{eq:expansion_origing_1}-\eqref{eq:expansion_origing_2} into
the matter equation \eqref{eq:g_ode} and keeping the zero order terms,
we find:

\begin{equation}
\begin{cases}
g_{0}=1 & \omega>0\\
g_{0}\left(1+\frac{4\tau}{\mathcal{B}}g_{0}^{\frac{\alpha'+\beta-4}{4}}\right)=1 & \omega=0\\
g_{0}=0 & \omega<0
\end{cases}\label{eq:g0_eq_general}
\end{equation}
These three qualitatively different forms of the solution near the
origin are displayed in Figs. \ref{fig:profiles_beta}-\ref{fig:profiles_omega_neg_A}.
For $\omega=0$ (so that $\beta=4$ and $\tau=1/\alpha'$), $g_{0}$
obeys a nonlinear equation, as was found in Ref. \cite{krief2024self},
whose solution may attain, as a function of $\mathcal{B}$ and $\alpha'$,
any value in the range $0<g_{0}<1$, that is, the dimensionless material
temperature at the origin is finite and can vary continuously between
0 (highly non-LTE) and 1 (the LTE limit), as shown in Fig. \ref{fig:profiles_omega_0}.
On the other hand, we see that for $\omega>0$ (that is, $\beta>4$),
$g_{0}=f\left(0\right)=1$, that is, the material and radiation are
always in equilibrium at the origin. This is a result of the density
$\rho\propto x^{-\omega}$ diverging at the origin, leading to a divergent
absorption coefficient $k_{a}\propto\rho^{\lambda'+1}\propto x^{-\omega\left(\lambda'+1\right)}$,
which results in an infinitely strong coupling and an immediate equilibration
at $\xi\rightarrow0$, as shown in Figs. \ref{fig:profiles_beta},
\ref{fig:profiles_omega_pos} and \ref{fig:profiles_omega_pos_small}.
Conversely, when $\omega<0$ (that is, $\beta<4$), the density and
consequently the absorption coefficient vanish at the origin, which
leads to no coupling between the radiation and material, which results
in a cold material at $\xi\rightarrow0$, so that $g_{0}=0$. This
results in a material temperature profile which is not monotonic,
as shown in Figs. \ref{fig:profiles_omega_neg} and \ref{fig:profiles_omega_neg_A}.
These three types of solutions are shown in Fig. \ref{fig:profiles_omega_0},
where the temperature profiles are displayed for a varying value of
$\beta$.

\subsection{The thermodynamic equilibrium limit\label{subsec:The-LTE-limit}}

Using Eqs. \eqref{eq:Uform} and \eqref{eq:bdef}, it can be inferred
that the dimensionless constant parameter $\mathcal{B}$ quantifies
the material coupling to the radiation field by the emission absorption
process, since:

\begin{equation}
\frac{\partial U}{\partial t}\propto\mathcal{B}x^{-\omega\left(\lambda'+\mu\right)}U^{-\frac{\alpha'}{4}}\left(E-U\right),\label{eq:ueq}
\end{equation}
which shows that the $\mathcal{B}$ determines the equilibration rate.
Equivalently, Eq. \eqref{eq:g_ode} shows that when $\mathcal{B}\gg1$
we have $g\left(\xi\right)\approx f\left(\xi\right)$, that is, the
material and radiation temperature profiles approach a common form,
that is, a state of thermodynamic equilibrium is reached locally.
This fact is demonstrated in Figs. \ref{fig:profiles_omega_0}-\ref{fig:profiles_omega_pos_small},
where it is evident that as $\mathcal{B}$ increases, $g^{1/4}\left(\xi\right)$
and $f^{1/4}\left(\xi\right)$ converge to the same profile. As discussed
in the previous section, when $\omega<0$, the material temperature
must vanish at the origin for any value of $\mathcal{A}$ and $\mathcal{B}$.
However, since $f\left(0\right)=1$, when $\mathcal{B}\rightarrow\infty$
the thermodynamic equilibrium limit must be reached, which
results in a very sharp increase of $g\left(\xi\right)$ near the
origin from 0 to 1, as shown in Fig. \ref{fig:profiles_omega_neg}.
Conversely, when $\omega>0$, we must have $g_{0}=1$, as equilibrium
is always reached at the origin, independently of the values of $\mathcal{A}$
and $\mathcal{B}$. However, for $\mathcal{B}\lesssim1$ a significant
state of non-equilibrium occurs for $\xi>0$, which results in a sharp
decrease of $g\left(\xi\right)$ near the origin from a value of 1
to a lower finite value, as shown in Figs. \ref{fig:profiles_omega_pos}-\ref{fig:profiles_omega_pos_small}.

In Fig. \ref{fig:meshplots} the heat front coordinate $\xi_{0}$
and the maximal value of the dimensionless material temperature, $g_{\text{max}}^{1/4}=\underset{0\leq\xi\leq\xi_{0}}{\text{max}}\left[g^{1/4}\left(\xi\right)\right]$,
are shown as functions of $\mathcal{A}$ and $\mathcal{B}$, for a
case with $\omega<0$. From the discussion above, when the equilibrium
limit is reached, the value of $g_{\text{max}}^{1/4}$ should reach
unity, for any value of $\omega$. It is evident from Fig. \ref{fig:meshplots}
that $g_{\text{max}}^{1/4}$ depends weakly on $\mathcal{A}$, and
approaches unity as the value of $\mathcal{B}$ increases. 

Finally, it is evident from Figs. \ref{fig:profiles_omega_0}-\ref{fig:profiles_omega_pos_small},
as well as from Fig. \ref{fig:meshplots}, that the front coordinate
$\xi_{0}$ is an increasing function of $\mathcal{B}$. This can be
expected, since larger values of $\mathcal{B}$ give rise to a larger
coupling between the radiation and material, leading to a higher material
temperature, which results in a smaller total opacity {[}see Eq. \ref{eq:ross_opac_powerlaw}{]},
that leads to a faster heat propagation. Figs. \ref{fig:profiles_omega_neg_A}
and \ref{fig:meshplots} show that for moderate values $\mathcal{A}\lesssim1$,
$\xi_{0}$ and the profile shapes depends very weakly on $\mathcal{A}$.
For larger values of $\mathcal{A}$, the material temperature profiles
$g^{1/4}\left(\xi\right)$ decrease with respect to $\mathcal{A}$,
resulting in a smaller values of $\xi_{0}$.

\subsection{Marshak boundary condition\label{subsec:Marshak-boundary-condition}}

\begin{figure}[h]
\begin{centering}
\includegraphics[width=0.99\columnwidth]{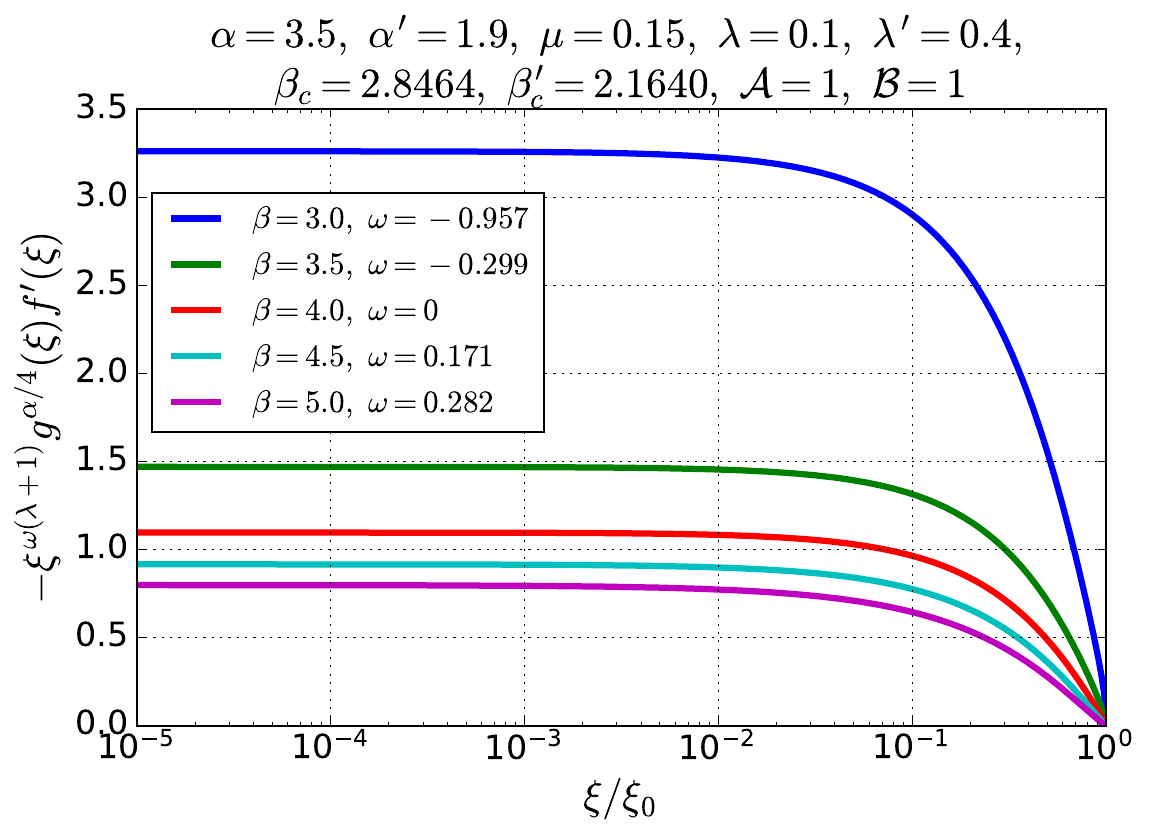}
\par\end{centering}
\caption{The flux similarity profile {[}see Eq. \ref{eq:self_similar_flux}{]},
for various values of $\omega$ (as listen in the legend), and for
a choice of parameters which is listed in the title (same as in Fig.
\ref{fig:profiles_beta}). \label{fig:selfsimflux}}
\end{figure}

Non-equilibrium Marshak waves \cite{pomraning1979non,bingjing1996benchmark,krief2024self},
as well as many other non-equilibrium heat wave benchmarks \cite{fleck1971implicit,larsen1988grey,olson2000diffusion,chang2007deterministic,morel2007linear,densmore2012hybrid,steinberg2022multi,steinberg2023frequency,mclean2022multi,yee2017stable,brunner2023family,zhang2023fully,liu2023implicit,li2024unified},
are specified in terms of a given incoming radiative flux, rather
than a temperature boundary condition (as in Eq. \eqref{eq:bc}).
The latter boundary condition applies more naturally in the diffusion
approximation, while the former is more natural to use in the solution
of the radiation transport equation, which has the angular surface
flux as a boundary condition (as will be discussed below in section
\ref{subsec:Transport-setup}). Nevertheless, these two different
boundary conditions can be related \cite{rosen2005fundamentals,cohen2018modeling,cohen2020key,heizler2021radiation,krief2024self,krief2024unified}.

The incoming flux boundary condition, which is known as the Marshak
boundary condition \cite{pomraning1979non,bingjing1996benchmark,rosen2005fundamentals,krief2024self}
is given by:

\begin{equation}
\frac{4}{c}F_{\text{inc}}\left(t\right)=E\left(x=0,t\right)+\frac{2}{c}F\left(x=0,t\right),\label{eq:maesh_bc_def}
\end{equation}
where $F_{\text{inc}}\left(t\right)$ is the incoming flux at $x=0$.
The incoming flux to a medium which is coupled to a heat bath at temperature
$T_{\text{bath}}\left(t\right)$, is given by $F_{\text{inc}}\left(t\right)=\frac{ac}{4}T_{\text{bath}}^{4}\left(t\right)$.
The Marshak boundary condition \eqref{eq:maesh_bc_def} results from
the diffusion limit of the exact Milne boundary condition of radiation
transport (see section \ref{subsec:Transport-setup} below). Using
the surface radiation temperature, $E\left(x=0,t\right)=aT_{s}^{4}\left(t\right)$,
in the Marshak boundary condition \eqref{eq:maesh_bc_def} gives:

\begin{equation}
T_{\text{bath}}\left(t\right)=\left(T_{s}^{4}\left(t\right)+\frac{2}{ac}F\left(x=0,t\right)\right)^{\frac{1}{4}},\label{eq:Tbath_marsh_bc_F}
\end{equation}
which is a relation between the bath temperature, the surface radiation
temperature and the net surface flux. By using Eqs. \eqref{eq:fick}-\eqref{eq:diffusion_coeff},
the radiation flux can be written in a self-similar form:

\begin{align}
F\left(x,t\right) & =K^{\frac{1}{2-\omega\left(1+\lambda\right)}}E_{0}^{1+\frac{\alpha}{4\left(2-\omega\left(1+\lambda\right)\right)}}t^{4\tau+\delta-1}\mathcal{S}\left(\xi\right),\label{eq:heat_flux}
\end{align}
where the (dimensionless) similarity flux profile is given by:
\begin{equation}
\mathcal{S}\left(\xi\right)=-\xi^{\omega\left(1+\lambda\right)}g^{\frac{\alpha}{4}}\left(\xi\right)f'\left(\xi\right).\label{eq:self_similar_flux}
\end{equation}
The radiation surface temperature is given by Eq. \eqref{eq:Tbound},
so that the bath temperature, according to Eq. \eqref{eq:Tbath_marsh_bc_F},
can be written explicitly as a function of time:
\begin{equation}
T_{\text{bath}}\left(t\right)=\left(1+Bt^{\delta-1}\right)^{\frac{1}{4}}T_{0}t^{\tau},\label{eq:Tbath_marsh_bc}
\end{equation}
where we have defined the bath coefficient: 
\begin{align}
B & =\frac{2}{c}\left(\frac{c\mathcal{G}}{3}\rho_{0}^{-1-\lambda}T_{0}^{\alpha}\right)^{\frac{1}{2-\omega\left(1+\lambda\right)}}\mathcal{S}\left(0\right),\label{eq:Bdef}
\end{align}
where the dimensionless radiation flux at the origin is: 
\begin{equation}
\mathcal{S}\left(0\right)=-\lim_{\xi\to0}\xi^{\omega\left(1+\lambda\right)}g^{\frac{\alpha}{4}}\left(\xi\right)f'\left(\xi\right).\label{eq:selfsimflux}
\end{equation}
We note that $\mathcal{S}\left(0\right)$ does not diverge for both
cases of diverging and vanishing density at the origin. This is demonstrated
in Fig. \ref{fig:selfsimflux}, where $\mathcal{S}\left(\xi\right)$
is displayed for various values of $\omega$. We also note that Eq.
\eqref{eq:Tbath_marsh_bc} shows that only for $\delta=1$ the bath
temperature is given by a temporal power law, which has the same temporal
power $\tau$ of the surface temperature (in agreement with Ref. \cite{krief2024self}).
Finally, we note that since $B\propto\mathcal{G}^{\frac{1}{2-\omega\left(\lambda+1\right)}}$,
when the total opacity increases, the bath temperature $T_{\text{bath}}\left(t\right)$
becomes closer to the surface temperature $T_{s}\left(t\right)$.

\section{Comparison with simulations\label{sec:simulations}}

\subsection{Transport setup\label{subsec:Transport-setup}}

We now construct a setup for gray transport calculations of the gray
diffusion problem defined in Sec. \ref{sec:Statement-of-the}. The
one dimensional, one group (gray) transport equation for the radiation
intensity field $I\left(x,\mu,t\right)$ in slab symmetry is \cite{su1997analytical,olson2000diffusion,pomraning2005equations,steinberg2022multi,castor2004radiation,mihalas1999foundations,krief2024self}

\begin{align}
\left(\frac{1}{c}\frac{\partial}{\partial t}+\mu\frac{\partial}{\partial x}\right)I\left(x,\mu,t\right)+ & \left(k_{a}+k_{s}\right)I\left(x,\mu,t\right)\nonumber \\
=\frac{ac}{4\pi}k_{a}T^{4}\left(x,t\right)+\frac{1}{2}k_{s} & \int_{-1}^{1}d\mu'I\left(x,\mu',t\right),\label{eq:Treq}
\end{align}
where $\mu$ is the directional angle cosine, $k_{a}=k_{a}\left(T,\rho\right)$
and $k_{s}=k_{s}\left(T,\rho\right)$, are, respectively, the absorption
and elastic scattering macroscopic cross sections, which
are functions of the local material temperature and density. The radiation
field is coupled to the material via the material energy equation:

\begin{align}
\frac{\partial u\left(T,\rho\right)}{\partial t} & =k_{a}\left[2\pi\int_{-1}^{1}d\mu'I\left(x,\mu',t\right)-acT^{4}\left(x,t\right)\right].\label{eq:tr_mat}
\end{align}
where $u\left(T,\rho\right)$ is the material energy density. The
radiation energy density is given by: 
\begin{equation}
E\left(x,t\right)=\frac{2\pi}{c}\int_{-1}^{1}d\mu'I\left(x,\mu',t\right),
\end{equation}
and the radiation temperature is $T_{r}\left(x,t\right)=\left(E\left(x,t\right)/a\right)^{1/4}$.
The diffusion limit holds for optically thick problems. In that case,
the transport problem \eqref{eq:Treq}-\eqref{eq:tr_mat} can be approximated
by the gray diffusion problem defined by equations \eqref{eq:main_eq}-\eqref{eq:fick},
with the total opacity $k_{t}=k_{s}+k_{a}$. Therefore, a transport
setup of the diffusion problem defined in Sec. \ref{sec:Statement-of-the}
should have the following effective elastic scattering opacity:
\begin{align}
k_{s}\left(T,\rho\right) & =k_{t}\left(T,\rho\right)-k_{a}\left(T,\rho\right)\nonumber \\
 & =\frac{1}{\mathcal{G}}T^{-\alpha}\rho^{1+\lambda}-\frac{1}{\mathcal{G}'}T^{-\alpha'}\rho^{1+\lambda'}.\label{eq:scatt}
\end{align}
We note that unless $\alpha=\alpha'$ and $\lambda=\lambda'$, this
form the scattering opacity is not a good model for real materials,
but is used here to construct a gray transport problem which is mathematically
equivalent, in the optically thick limit, to a gray diffusion problem
whose total and absorption opacities are power laws with respect to
temperature and density. Moreover, since the scattering opacity must
be positive, the gray transport problem is well defined only if $k_{t}\left(T,\rho\right)\geq k_{a}\left(T,\rho\right)$,
which must hold for the relevant temperatures and densities in the
problem. This constraint does not have to hold for the gray diffusion
problem, which is well defined for any functions $k_{t}$, $k_{a}$.
We note that due to the complex nature of opacity spectra of mid or
high-Z hot dense plasmas, the total (Rosseland) opacity can be lower
than the absorption (Planck) opacity \cite{pomraning2005equations,mihalas1999foundations,castor2004radiation,krief2018new,krief2018star},
in which case an equivalent gray transport problem cannot be defined.

The boundary condition for the transport problem is defined via the
incident radiation field at $x=0$ for incoming directions $\mu>0$,
of a a black body radiation bath
\begin{equation}
I\left(x=0,\mu,t\right)=\frac{ac}{4\pi}T_{\text{bath}}^{4}\left(t\right),\label{eq:Ibath_tr}
\end{equation}
where the time dependent bath temperature drive is taken from the
Marshak (Milne) boundary condition using the radiation temperature
and flux which are taken from the gray diffusion solution {[}Eq. \eqref{eq:Tbath_marsh_bc}{]},
as detailed in Sec. \ref{subsec:Marshak-boundary-condition}. The
Marshak boundary condition {[}Eq. \eqref{eq:maesh_bc_def}{]} is derived
by the diffusion limit approximation of the exact transport boundary
condition given by Eq. \eqref{eq:Ibath_tr}.

Since the diffusion limit is applicable for optically thick problems,
we expect gray transport results to have a good agreement with diffusion
simulations and the self-similar solutions in this limit. Nevertheless,
for optically thick problems with low absorption but high photon scattering
($k_{a}\ll k_{t}$), such that $\mathcal{B}\lesssim1$, we expect
transport results to agree well with diffusion, while the radiation
and material are significantly out of equilibrium. Several examples
of this scenario will be shown below.

\subsection{Test cases\label{subsec:Test-cases}}

Based on the self-similar solutions, we define six benchmarks for
which we specify in detail the setups for gray diffusion and transport
computer simulations. We conducted gray diffusion simulations and
stochastic implicit Monte-Carlo (IMC) \cite{fleck1971implicit,mcclarren2009modified,brunner2006comparison,cleveland2014mitigating,noebauer2012monte,wollaber2016four,noebauer2019monte}
and deterministic discrete-ordinates ($S_{N}$) transport simulations.
The diffusion simulations were performed without the application of
flux limiters. The $S_{N}$ simulations were performed using a numerical
method that is detailed in Ref. \cite{mcclarren2022data}, while
the IMC simulations employed the novel numerical scheme that is described
in Refs. \cite{steinberg2022new,steinberg2022multi,steinberg2023frequency}.

All cases are run until the final time $t=1\text{ns}$, and the radiation
surface temperature drive is increased to a final temperature of $1\text{keV}$:
\begin{equation}
T_{s}\left(t\right)=\left(\frac{t}{\text{ns}}\right)^{\tau}\ \text{keV},\label{eq:Ts}
\end{equation}
so that $T_{0}=\text{keV}/\text{ns}^{\tau}$ for all cases. In addition,
we take in all cases a spatial density profile:
\begin{equation}
\rho\left(x\right)=\left(\frac{x}{\text{cm}}\right)^{-\omega}\ \text{g}/\text{cm}^{3}.
\end{equation}
so that $\rho_{0}=1\text{g}\cdot\text{cm}^{\omega-3}$. The temperature
profiles are plotted at the final time and also at the times when
the heat front reaches $20\%$ and $60\%$ of the final front position. 

In tests 1 and 2 we define a material model with $\beta<4$, so that
$\omega<0$ and that the material temperature is zero at the origin,
increases until reaching a maximum and then decreases again towards
the front. In cases 3-5 the material models have $\beta>4$ so that
$\omega>0$, and the radiation and material temperatures are equal
at the origin. Tests 1-4 have $\mathcal{B}\lesssim1$ so that a significant
deviation from equilibrium occurs. Test 5 has $\mathcal{B}\gg1$ which
results in the thermodynamic equilibrium limit, for which
the radiation and material temperatures are very close throughout
the heat wave. Finally, Test 6, which also has $\mathcal{B}\gg1$,
but since the density is extremely small in a wide range near the
origin, a state of equilibrium is not reached in that region. 

All cases were defined such that the total optical depth is large,
so that the diffusion limit is applicable, and transport simulations
results should agree with the gray diffusion self-similar solutions
and simulations. 

Since exact closed form analytical solutions of Eqs. \eqref{eq:f_ode}-\eqref{eq:g_ode}
for the temperature profiles do no exist, tabulated exact numerical
solution profiles for all cases are given in table \ref{tab:tests_table}.
In addition, we give simple and closed form but approximate fitted
analytic profiles in table \ref{tab:test_fits}, which are accurate
to about $0.5\%$.

\begin{table*}[t]
\centering{}{\small{}}%
\begin{tabular}{|c|c|c|c|c|c|c|c|c|c|c|c|c|}
\cline{2-13} \cline{3-13} \cline{4-13} \cline{5-13} \cline{6-13} \cline{7-13} \cline{8-13} \cline{9-13} \cline{10-13} \cline{11-13} \cline{12-13} \cline{13-13} 
\multicolumn{1}{c|}{} & \multicolumn{2}{c|}{{\small{}Test 1}} & \multicolumn{2}{c|}{{\small{}Test 2}} & \multicolumn{2}{c|}{{\small{}Test 3}} & \multicolumn{2}{c|}{{\small{}Test 4}} & \multicolumn{2}{c|}{{\small{}Test 5}} & \multicolumn{2}{c|}{{\small{}Test 6}}\tabularnewline
\hline 
{\small{}$\xi_{0}$} & \multicolumn{2}{c|}{{\small{}$1.274605$}} & \multicolumn{2}{c|}{{\small{}$0.615503$}} & \multicolumn{2}{c|}{{\small{}$0.314115$}} & \multicolumn{2}{c|}{{\small{}$0.484638$}} & \multicolumn{2}{c|}{{\small{}$0.530730$}} & \multicolumn{2}{c|}{{\small{}$1.198678$}}\tabularnewline
\hline 
\hline 
{\small{}$\xi/\xi_{0}$} & {\small{}$f^{1/4}\left(\xi\right)$} & {\small{}$g^{1/4}\left(\xi\right)$} & {\small{}$f^{1/4}\left(\xi\right)$} & {\small{}$g^{1/4}\left(\xi\right)$} & {\small{}$f^{1/4}\left(\xi\right)$} & {\small{}$g^{1/4}\left(\xi\right)$} & {\small{}$f^{1/4}\left(\xi\right)$} & {\small{}$g^{1/4}\left(\xi\right)$} & {\small{}$f^{1/4}\left(\xi\right)$} & {\small{}$g^{1/4}\left(\xi\right)$} & {\small{}$f^{1/4}\left(\xi\right)$} & {\small{}$g^{1/4}\left(\xi\right)$}\tabularnewline
\hline 
\hline 
{\small{}0} & {\small{}1} & {\small{}0} & {\small{}1} & {\small{}0} & {\small{}1} & {\small{}1} & {\small{}1} & {\small{}1} & {\small{}1} & {\small{}1} & {\small{}1} & {\small{}0}\tabularnewline
\hline 
{\small{}$10^{-6}$} & {\small{}1} & {\small{}0.3463} & {\small{}1} & {\small{}0.010182} & {\small{}0.99987} & {\small{}0.97389} & {\small{}0.99795} & {\small{}0.98549} & {\small{}0.99968} & {\small{}0.99951} & {\small{}1} & {\small{}0.0025247}\tabularnewline
\hline 
{\small{}$10^{-5}$} & {\small{}1} & {\small{}0.40853} & {\small{}1} & {\small{}0.019658} & {\small{}0.99959} & {\small{}0.95101} & {\small{}0.9955} & {\small{}0.97292} & {\small{}0.99899} & {\small{}0.99872} & {\small{}1} & {\small{}0.0079838}\tabularnewline
\hline 
{\small{}0.0001} & {\small{}1} & {\small{}0.48063} & {\small{}1} & {\small{}0.037954} & {\small{}0.99868} & {\small{}0.91732} & {\small{}0.99012} & {\small{}0.95124} & {\small{}0.9968} & {\small{}0.99637} & {\small{}1} & {\small{}0.025247}\tabularnewline
\hline 
{\small{}0.0005} & {\small{}1} & {\small{}0.53693} & {\small{}1} & {\small{}0.060113} & {\small{}0.99692} & {\small{}0.88791} & {\small{}0.98262} & {\small{}0.92829} & {\small{}0.9928} & {\small{}0.99221} & {\small{}1} & {\small{}0.056454}\tabularnewline
\hline 
{\small{}0.001} & {\small{}1} & {\small{}0.56257} & {\small{}1} & {\small{}0.073278} & {\small{}0.99553} & {\small{}0.87386} & {\small{}0.97769} & {\small{}0.91585} & {\small{}0.98977} & {\small{}0.98909} & {\small{}1} & {\small{}0.079837}\tabularnewline
\hline 
{\small{}0.005} & {\small{}0.99999} & {\small{}0.62477} & {\small{}1} & {\small{}0.11605} & {\small{}0.98909} & {\small{}0.83765} & {\small{}0.9593} & {\small{}0.87914} & {\small{}0.97658} & {\small{}0.97567} & {\small{}1} & {\small{}0.17848}\tabularnewline
\hline 
{\small{}0.01} & {\small{}0.99995} & {\small{}0.65239} & {\small{}1} & {\small{}0.14146} & {\small{}0.98378} & {\small{}0.82} & {\small{}0.94659} & {\small{}0.85883} & {\small{}0.9663} & {\small{}0.96526} & {\small{}1} & {\small{}0.25225}\tabularnewline
\hline 
{\small{}0.05} & {\small{}0.99837} & {\small{}0.71601} & {\small{}0.99998} & {\small{}0.22394} & {\small{}0.95741} & {\small{}0.76988} & {\small{}0.89478} & {\small{}0.79375} & {\small{}0.91888} & {\small{}0.91757} & {\small{}1} & {\small{}0.55298}\tabularnewline
\hline 
{\small{}0.1} & {\small{}0.99298} & {\small{}0.73924} & {\small{}0.99983} & {\small{}0.27272} & {\small{}0.93334} & {\small{}0.74045} & {\small{}0.85459} & {\small{}0.75214} & {\small{}0.87876} & {\small{}0.87737} & {\small{}1} & {\small{}0.74168}\tabularnewline
\hline 
{\small{}0.15} & {\small{}0.98375} & {\small{}0.74747} & {\small{}0.99931} & {\small{}0.30574} & {\small{}0.91201} & {\small{}0.71883} & {\small{}0.8218} & {\small{}0.72084} & {\small{}0.84492} & {\small{}0.84351} & {\small{}1} & {\small{}0.84704}\tabularnewline
\hline 
{\small{}0.2} & {\small{}0.97077} & {\small{}0.74769} & {\small{}0.99815} & {\small{}0.33111} & {\small{}0.89177} & {\small{}0.70038} & {\small{}0.79232} & {\small{}0.69399} & {\small{}0.81402} & {\small{}0.81262} & {\small{}0.99998} & {\small{}0.90706}\tabularnewline
\hline 
{\small{}0.25} & {\small{}0.95413} & {\small{}0.74215} & {\small{}0.9961} & {\small{}0.35158} & {\small{}0.87192} & {\small{}0.68356} & {\small{}0.76462} & {\small{}0.66952} & {\small{}0.78473} & {\small{}0.78336} & {\small{}0.99992} & {\small{}0.94156}\tabularnewline
\hline 
{\small{}0.3} & {\small{}0.93389} & {\small{}0.73185} & {\small{}0.99286} & {\small{}0.36838} & {\small{}0.85207} & {\small{}0.66761} & {\small{}0.73788} & {\small{}0.64642} & {\small{}0.75633} & {\small{}0.755} & {\small{}0.99977} & {\small{}0.96171}\tabularnewline
\hline 
{\small{}0.35} & {\small{}0.91009} & {\small{}0.71732} & {\small{}0.98818} & {\small{}0.38214} & {\small{}0.83193} & {\small{}0.65208} & {\small{}0.71159} & {\small{}0.62408} & {\small{}0.72837} & {\small{}0.72709} & {\small{}0.9994} & {\small{}0.97363}\tabularnewline
\hline 
{\small{}0.4} & {\small{}0.88273} & {\small{}0.69885} & {\small{}0.98178} & {\small{}0.39317} & {\small{}0.81128} & {\small{}0.63665} & {\small{}0.68537} & {\small{}0.60208} & {\small{}0.70049} & {\small{}0.69926} & {\small{}0.99862} & {\small{}0.98055}\tabularnewline
\hline 
{\small{}0.45} & {\small{}0.85176} & {\small{}0.67658} & {\small{}0.97339} & {\small{}0.4016} & {\small{}0.78988} & {\small{}0.62109} & {\small{}0.65893} & {\small{}0.5801} & {\small{}0.6724} & {\small{}0.67123} & {\small{}0.99713} & {\small{}0.98406}\tabularnewline
\hline 
{\small{}0.5} & {\small{}0.81706} & {\small{}0.65054} & {\small{}0.96271} & {\small{}0.40745} & {\small{}0.76751} & {\small{}0.60515} & {\small{}0.63196} & {\small{}0.55784} & {\small{}0.64383} & {\small{}0.64274} & {\small{}0.99452} & {\small{}0.9848}\tabularnewline
\hline 
{\small{}0.55} & {\small{}0.77847} & {\small{}0.62067} & {\small{}0.94943} & {\small{}0.41063} & {\small{}0.7439} & {\small{}0.58862} & {\small{}0.60419} & {\small{}0.53503} & {\small{}0.61451} & {\small{}0.61348} & {\small{}0.9902} & {\small{}0.9828}\tabularnewline
\hline 
{\small{}0.6} & {\small{}0.73573} & {\small{}0.58681} & {\small{}0.93315} & {\small{}0.41097} & {\small{}0.71871} & {\small{}0.57122} & {\small{}0.57528} & {\small{}0.51135} & {\small{}0.5841} & {\small{}0.58315} & {\small{}0.9834} & {\small{}0.97764}\tabularnewline
\hline 
{\small{}0.65} & {\small{}0.68847} & {\small{}0.54869} & {\small{}0.9134} & {\small{}0.40817} & {\small{}0.69153} & {\small{}0.55264} & {\small{}0.54484} & {\small{}0.48644} & {\small{}0.55223} & {\small{}0.55136} & {\small{}0.9731} & {\small{}0.96855}\tabularnewline
\hline 
{\small{}0.7} & {\small{}0.63614} & {\small{}0.50588} & {\small{}0.88956} & {\small{}0.40178} & {\small{}0.66178} & {\small{}0.53243} & {\small{}0.51236} & {\small{}0.45982} & {\small{}0.51839} & {\small{}0.5176} & {\small{}0.95799} & {\small{}0.95434}\tabularnewline
\hline 
{\small{}0.75} & {\small{}0.57793} & {\small{}0.45773} & {\small{}0.86074} & {\small{}0.39109} & {\small{}0.62862} & {\small{}0.50997} & {\small{}0.4771} & {\small{}0.43082} & {\small{}0.48185} & {\small{}0.48116} & {\small{}0.93621} & {\small{}0.93324}\tabularnewline
\hline 
{\small{}0.8} & {\small{}0.51255} & {\small{}0.4032} & {\small{}0.82555} & {\small{}0.37496} & {\small{}0.59073} & {\small{}0.48427} & {\small{}0.43795} & {\small{}0.39843} & {\small{}0.44152} & {\small{}0.44092} & {\small{}0.90506} & {\small{}0.90262}\tabularnewline
\hline 
{\small{}0.85} & {\small{}0.43783} & {\small{}0.34055} & {\small{}0.78155} & {\small{}0.3514} & {\small{}0.54581} & {\small{}0.45357} & {\small{}0.39301} & {\small{}0.36086} & {\small{}0.39551} & {\small{}0.39501} & {\small{}0.86008} & {\small{}0.85806}\tabularnewline
\hline 
{\small{}0.9} & {\small{}0.34949} & {\small{}0.2665} & {\small{}0.72362} & {\small{}0.31631} & {\small{}0.4891} & {\small{}0.4142} & {\small{}0.33844} & {\small{}0.31453} & {\small{}0.33999} & {\small{}0.33961} & {\small{}0.79256} & {\small{}0.79086}\tabularnewline
\hline 
{\small{}0.95} & {\small{}0.23673} & {\small{}0.17304} & {\small{}0.6366} & {\small{}0.2585} & {\small{}0.40721} & {\small{}0.35543} & {\small{}0.26372} & {\small{}0.24948} & {\small{}0.26449} & {\small{}0.26424} & {\small{}0.6789} & {\small{}0.67742}\tabularnewline
\hline 
{\small{}0.973} & {\small{}0.16711} & {\small{}0.11688} & {\small{}0.57037} & {\small{}0.21332} & {\small{}0.34744} & {\small{}0.31057} & {\small{}0.21224} & {\small{}0.20336} & {\small{}0.21273} & {\small{}0.21256} & {\small{}0.58759} & {\small{}0.58611}\tabularnewline
\hline 
{\small{}0.99} & {\small{}0.095194} & {\small{}0.06148} & {\small{}0.48037} & {\small{}0.15471} & {\small{}0.27093} & {\small{}0.25003} & {\small{}0.15048} & {\small{}0.14639} & {\small{}0.15078} & {\small{}0.15068} & {\small{}0.46451} & {\small{}0.46284}\tabularnewline
\hline 
{\small{}0.996} & {\small{}0.056619} & {\small{}0.033764} & {\small{}0.41126} & {\small{}0.11438} & {\small{}0.21683} & {\small{}0.20467} & {\small{}0.11006} & {\small{}0.10809} & {\small{}0.11027} & {\small{}0.11022} & {\small{}0.3751} & {\small{}0.37309}\tabularnewline
\hline 
{\small{}0.998} & {\small{}0.038217} & {\small{}0.0214} & {\small{}0.366} & {\small{}0.090878} & {\small{}0.18385} & {\small{}0.17586} & {\small{}0.087024} & {\small{}0.085901} & {\small{}0.087202} & {\small{}0.087161} & {\small{}0.32016} & {\small{}0.31776}\tabularnewline
\hline 
{\small{}0.999} & {\small{}0.025796} & {\small{}0.013543} & {\small{}0.32583} & {\small{}0.072157} & {\small{}0.15626} & {\small{}0.15105} & {\small{}0.068887} & {\small{}0.068248} & {\small{}0.069032} & {\small{}0.069003} & {\small{}0.27415} & {\small{}0.27125}\tabularnewline
\hline 
{\small{}0.9999} & {\small{}0.006992} & {\small{}0.0029419} & {\small{}0.22149} & {\small{}0.033435} & {\small{}0.092131} & {\small{}0.090927} & {\small{}0.031834} & {\small{}0.031738} & {\small{}0.031904} & {\small{}0.031895} & {\small{}0.16769} & {\small{}0.16188}\tabularnewline
\hline 
{\small{}0.99999} & {\small{}0.0018935} & {\small{}0.00063494} & {\small{}0.14959} & {\small{}0.015319} & {\small{}0.054835} & {\small{}0.054569} & {\small{}0.014756} & {\small{}0.014741} & {\small{}0.01479} & {\small{}0.014788} & {\small{}0.10621} & {\small{}0.094976}\tabularnewline
\hline 
{\small{}0.999999} & {\small{}0.00050668} & {\small{}0.00013483} & {\small{}0.095918} & {\small{}0.006454} & {\small{}0.032663} & {\small{}0.032606} & {\small{}0.0068493} & {\small{}0.0068472} & {\small{}0.0068734} & {\small{}0.0068726} & {\small{}0.06881} & {\small{}0.052622}\tabularnewline
\hline 
\end{tabular}\caption{Radiation and material temperature profiles, as a function of $\xi/\xi_{0}=x/x_{F}\left(t\right)$,
obtained from numerical solutions of the ODE system \eqref{eq:f_ode}-\eqref{eq:g_ode},
for the test cases defined in Sec. \ref{subsec:Test-cases}. The numerical
values of $\xi_{0}$ for each case are also listed. In tests 1 and
2 we have $\omega<0$, so that $g\left(\xi=0\right)=0$, and $g\left(\xi\right)$
is increasing until reaching a maximum and then decreasing again towards
the front. In cases 3-5 we have $\omega>0$, so that $g\left(\xi=0\right)=1$,
and $g\left(\xi\right)$ is strictly decreasing. Test 5 has $\mathcal{B}\gg1$
which results in the thermodynamic equilibrium limit, as $g\left(\xi\right)\approx f\left(\xi\right)$.
Test 6 also has $\mathcal{B}\gg1$, but since the density is extremely
small in the inner region $\xi/\xi_{0}\protect\leq0.3$, equilibrium
is not reached in that region. \label{tab:tests_table}}
\end{table*}

\begin{table*}[t]
\begin{centering}
\begin{tabular}{|>{\centering}V{\linewidth}|c|c|c|}
\cline{2-4} \cline{3-4} \cline{4-4} 
\multicolumn{1}{c|}{} & Fitted temperatures profiles & Average Error & Maximal Error\tabularnewline
\hline 
\multirow{2}{*}{Test 1} & \begin{cellvarwidth}$\vphantom{|}$

$f^{1/4}\left(\xi\right)=\begin{cases}
1-0.75567\left(\frac{\xi}{\xi_{0}}\right)^{2.0416} & \xi<0.75\xi_{0}\\
1.2527\left(1-\frac{\xi}{\xi_{0}}\right)^{0.55623} & \xi\geq0.75\xi_{0}
\end{cases}$

$\vphantom{|}$\end{cellvarwidth} & $0.15\%$ & $0.44\%$\tabularnewline
\cline{2-4} \cline{3-4} \cline{4-4} 
 & \begin{cellvarwidth}$\vphantom{|}$

$g^{1/4}\left(\xi\right)=\begin{cases}
-0.15937\left(\frac{\xi}{\xi_{0}}\right)^{0.2194}+\left(\frac{\xi}{\xi_{0}}\right)^{0.074842} & \xi<0.05\xi_{0}\\
\left(0.63674+0.55611\left(\frac{\xi}{\xi_{0}}\right)^{0.56101}\right)\left(1-\frac{\xi}{\xi_{0}}\right)^{0.63964} & \xi\geq0.05\xi_{0}
\end{cases}$

$\vphantom{|}$\end{cellvarwidth} & $0.067\%$ & $0.45\%$\tabularnewline
\hline 
\multirow{2}{*}{Test 2} & \begin{cellvarwidth}$\vphantom{|}$

$f^{1/4}\left(\xi\right)=\begin{cases}
1-0.35369\left(\frac{\xi}{\xi_{0}}\right)^{3.2539} & \xi<0.75\xi_{0}\\
1.1098\left(1-\frac{\xi}{\xi_{0}}\right)^{0.1834} & \xi\geq0.75\xi_{0}
\end{cases}$

$\vphantom{|}$\end{cellvarwidth} & $0.077\%$ & $0.66\%$\tabularnewline
\cline{2-4} \cline{3-4} \cline{4-4} 
 & \begin{cellvarwidth}$\vphantom{|}$

$g^{1/4}\left(\xi\right)=\begin{cases}
0.52694\left(\frac{\xi}{\xi_{0}}\right)^{0.28559}-\left(\frac{\xi}{\xi_{0}}\right)^{3.5311} & \xi<0.1\xi_{0}\\
\left(0.15073+0.55903\left(\frac{\xi}{\xi_{0}}\right)^{0.62589}\right)\left(1-\frac{\xi}{\xi_{0}}\right)^{0.32965} & \xi\geq0.1\xi_{0}
\end{cases}$

$\vphantom{|}$\end{cellvarwidth} & $0.15\%$ & $0.45\%$\tabularnewline
\hline 
\multirow{2}{*}{Test 3} & \begin{cellvarwidth}$\vphantom{|}$

$f^{1/4}\left(\xi\right)=\begin{cases}
1-0.31578\left(\frac{\xi}{\xi_{0}}\right)^{0.66822} & \xi<0.2\xi_{0}\\
\left(0.95143-0.11368\left(\frac{\xi}{\xi_{0}}\right)^{1.4227}\right)\left(1-\frac{\xi}{\xi_{0}}\right)^{0.24206} & \xi\geq0.2\xi_{0}
\end{cases}$

$\vphantom{|}$\end{cellvarwidth} & $0.18\%$ & $0.57\%$\tabularnewline
\cline{2-4} \cline{3-4} \cline{4-4} 
 & \begin{cellvarwidth}$\vphantom{|}$

$g^{1/4}\left(\xi\right)=\begin{cases}
1+0.75349\left(\frac{\xi}{\xi_{0}}\right)^{0.37109}-\left(\frac{\xi}{\xi_{0}}\right)^{0.24703} & \xi<0.05\xi_{0}\\
\left(1.2663-0.58658\left(\frac{\xi}{\xi_{0}}\right)^{0.061645}\right)\left(1-\frac{\xi}{\xi_{0}}\right)^{0.21797} & \xi\geq0.05\xi_{0}
\end{cases}$

$\vphantom{|}$\end{cellvarwidth} & $0.058\%$ & $0.57\%$\tabularnewline
\hline 
\multirow{2}{*}{Test 4} & \begin{cellvarwidth}$\vphantom{|}$

$f^{1/4}\left(\xi\right)=\begin{cases}
1-0.44613\left(\frac{\xi}{\xi_{0}}\right)^{0.47963} & \xi<0.2\xi_{0}\\
\left(0.93272-0.19194\left(\frac{\xi}{\xi_{0}}\right)^{0.56565}\right)\left(1-\frac{\xi}{\xi_{0}}\right)^{0.34578} & \xi\geq0.2\xi_{0}
\end{cases}$

$\vphantom{|}$\end{cellvarwidth} & $0.11\%$ & $0.63\%$\tabularnewline
\cline{2-4} \cline{3-4} \cline{4-4} 
 & \begin{cellvarwidth}$\vphantom{|}$

$g^{1/4}\left(\xi\right)=\begin{cases}
1+0.60577\left(\frac{\xi}{\xi_{0}}\right)^{0.3164}-\left(\frac{\xi}{\xi_{0}}\right)^{0.27358} & \xi<0.05\xi_{0}\\
\left(1.149-0.48665\left(\frac{\xi}{\xi_{0}}\right)^{0.11809}\right)\left(1-\frac{\xi}{\xi_{0}}\right)^{0.32793} & \xi\geq0.05\xi_{0}
\end{cases}$

$\vphantom{|}$\end{cellvarwidth} & $0.052\%$ & $0.22\%$\tabularnewline
\hline 
Test 5 & \begin{cellvarwidth}$\vphantom{|}$

$f^{1/4}\left(\xi\right)=g^{1/4}\left(\xi\right)=\begin{cases}
1-0.43238\left(\frac{\xi}{\xi_{0}}\right)^{0.55116} & \xi<0.1\xi_{0}\\
\left(0.97888-0.23325\left(\frac{\xi}{\xi_{0}}\right)^{0.53204}\right)\left(1-\left(\frac{\xi}{\xi_{0}}\right)\right)^{0.34725} & \xi\geq0.1\xi_{0}
\end{cases}$

$\vphantom{|}$\end{cellvarwidth} & $0.12\%$ & $0.54\%$\tabularnewline
\hline 
\multirow{2}{*}{Test 6} & \begin{cellvarwidth}$\vphantom{|}$

$f^{1/4}\left(\xi\right)=\begin{cases}
1-0.38037\left(\frac{\xi}{\xi_{0}}\right)^{6.1993} & \xi<0.85\xi_{0}\\
\left(1.2945-0.089492\left(\frac{\xi}{\xi_{0}}\right)^{53.199}\right)\left(1-\frac{\xi}{\xi_{0}}\right)^{0.21389} & \xi\geq0.85\xi_{0}
\end{cases}$

$\vphantom{|}$\end{cellvarwidth} & $0.033\%$ & $0.30\%$\tabularnewline
\cline{2-4} \cline{3-4} \cline{4-4} 
 & \begin{cellvarwidth}$\vphantom{|}$

{\footnotesize{}$g^{1/4}\left(\xi\right)=\begin{cases}
0.3831\frac{\xi}{\xi_{0}}-9.5148\left(\frac{\xi}{\xi_{0}}\right)^{2}+12.852\left(\frac{\xi}{\xi_{0}}\right)^{3}+2.4707\left(\frac{\xi}{\xi_{0}}\right)^{0.49788} & \xi<0.3\xi_{0}\\
0.66687+2.0778\frac{\xi}{\xi_{0}}-5.2893\left(\frac{\xi}{\xi_{0}}\right)^{2}+6.4522\left(\frac{\xi}{\xi_{0}}\right)^{3}-3.2832\left(\frac{\xi}{\xi_{0}}\right)^{4}, & 0.3\xi_{0}\leq\xi<0.8\xi_{0}\\
\left(-586.76+588.16\left(\frac{\xi}{\xi_{0}}\right)^{0.00058395}\right)\left(1-\frac{\xi}{\xi_{0}}\right)^{0.23844} & \xi\geq0.8\xi_{0}
\end{cases}$}{\footnotesize\par}

$\vphantom{|}$\end{cellvarwidth} & $0.07\%$ & $0.77\%$\tabularnewline
\hline 
\end{tabular}
\par\end{centering}
\centering{}\caption{Approximate analytic radiation and material temperature similarity
profiles, as functions of $\xi/\xi_{0}=x/x_{F}\left(t\right)$, for
the test cases defined in Sec. \ref{subsec:Test-cases} (the exact
numerical profiles are tabulated in table \ref{tab:tests_table}).
The maximal and average relative errors (in the range $10^{-6}\protect\leq\xi/\xi_{0}\protect\leq0.95$),
relative to the exact profiles, are also listed. Test 5 is near thermodynamic
equilibrium, with a difference of 0.1\% between the exact material
and radiation temperatures. \label{tab:test_fits}}
\end{table*}

\subsubsection*{TEST 1}

\begin{figure}[t]
\begin{centering}
\includegraphics[width=1\columnwidth]{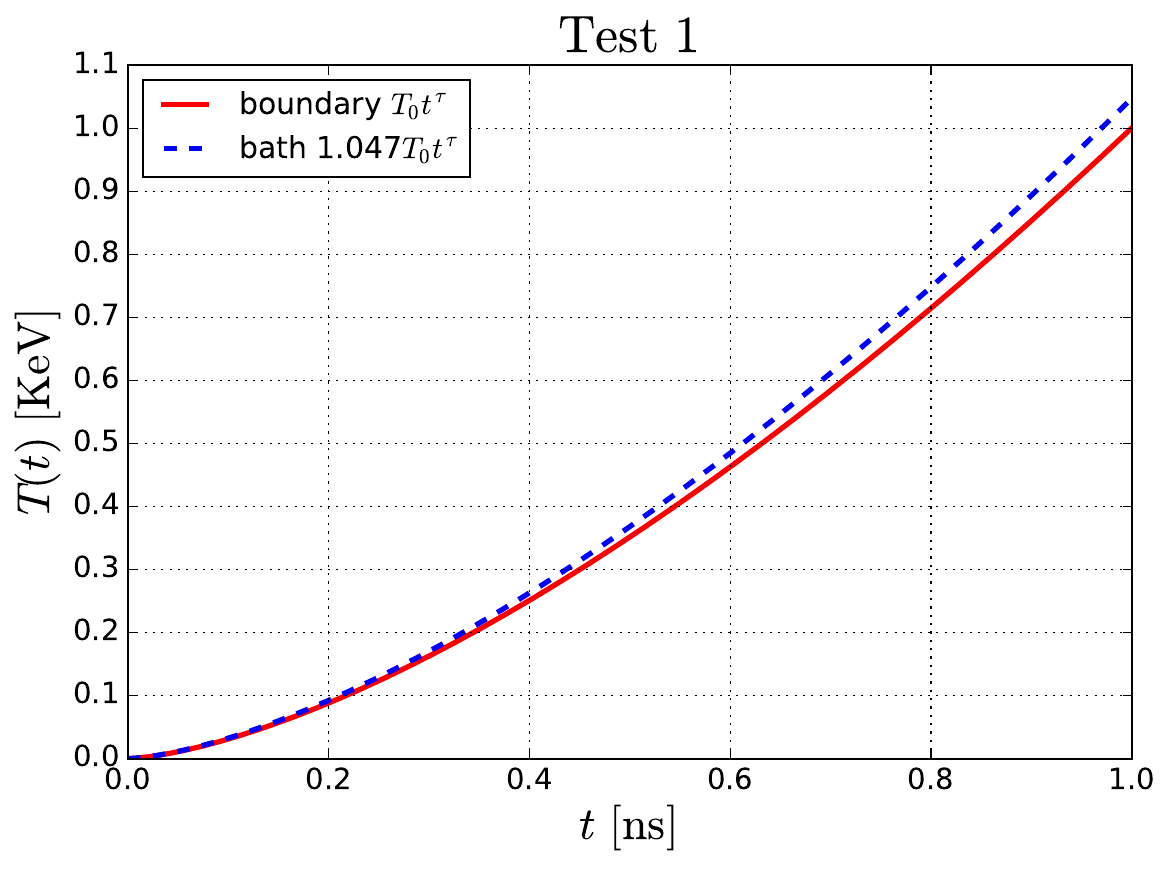}
\par\end{centering}
\caption{A comparison between the surface and bath drive temperatures for test
1.\label{fig:simulation_1_Tbath}}
\end{figure}

\begin{figure}[t]
\begin{centering}
\includegraphics[width=1\columnwidth]{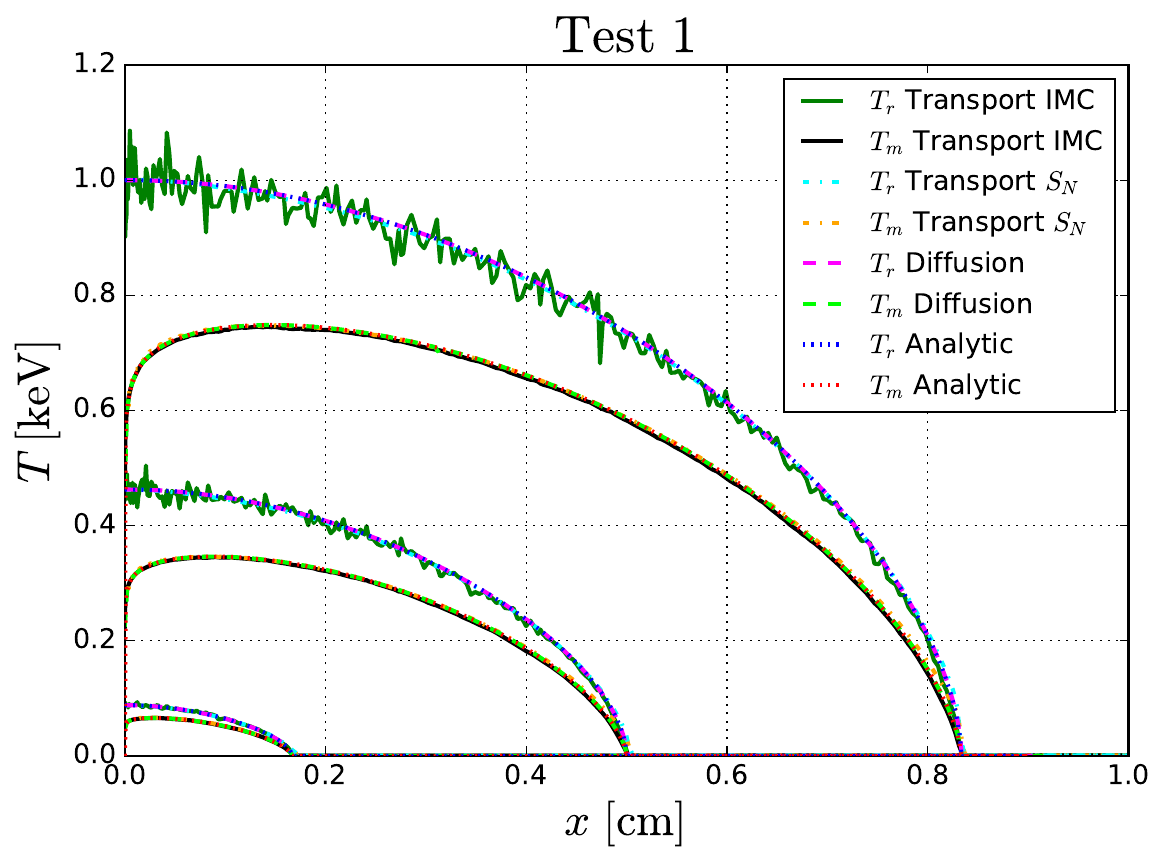}
\par\end{centering}
\caption{Radiation and material temperature profiles for Test 1. Results are
shown at times $t=0.2,\ 0.6$ and 1ns, as obtained from the gray diffusion
self-similar solution, a gray diffusion simulation and from Implicit-Monte-Carlo
(IMC) and discrete ordinates ($S_{N}$) transport simulations.\label{fig:simulation_1}}
\end{figure}

We take $\alpha=\alpha'=1.5$, $\lambda=\lambda'=0.2$, $\mathcal{G}=0.025\frac{\text{cm}\left(\text{g}/\text{cm}^{3}\right)^{1.2}}{\text{keV}^{1.5}}$,
$\mathcal{G}'=10\frac{\text{cm\ensuremath{\left(\text{g}/\text{cm}^{3}\right)^{1.2}}}}{\text{keV}^{1.5}}$,
so that the total and absorption opacities are:
\begin{align*}
k_{t}\left(T,\rho\right) & =40\left(\frac{T}{\text{keV}}\right)^{-1.5}\left(\frac{\rho}{\text{g}/\text{cm}^{3}}\right)^{1.2}\ \text{cm}^{-1},\\
k_{a}\left(T,\rho\right) & =0.1\left(\frac{T}{\text{keV}}\right)^{-1.5}\left(\frac{\rho}{\text{g}/\text{cm}^{3}}\right)^{1.2}\ \text{cm}^{-1}.
\end{align*}
For transport simulations, the scattering opacity {[}Eq. \eqref{eq:scatt}{]}
is given by: 
\begin{align*}
k_{s}\left(T,\rho\right) & =39.9\left(\frac{T}{\text{keV}}\right)^{-1.5}\left(\frac{\rho}{\text{g}/\text{cm}^{3}}\right)^{1.2}\ \text{cm}^{-1},
\end{align*}
which has the form of a power-law since the absorption and total opacities
have the same temperature and density exponents. For the material
energy density we take $\mu=0.14$, so that the critical exponent
is $\beta_{c}=\beta_{c}'=2.925$ {[}see Eqs. \eqref{eq:betac}-\eqref{eq:betac_p}{]}.
We set $\beta=3.4$ and $\mathcal{F}=10^{14}\frac{\text{keV}^{-3.4}}{\left(\text{g}/\text{cm}^{3}\right)^{0.86}}\frac{\text{erg}}{\text{cm}^{3}}$,
so that the material energy density is given by:
\[
u\left(T,\rho\right)=10^{14}\left(\frac{T}{\text{keV}}\right)^{3.4}\left(\frac{\rho}{\text{g}/\text{cm}^{3}}\right)^{0.86}\ \text{\ensuremath{\frac{\text{erg}}{\text{cm}^{3}}}}.
\]
For the exponents of this material model, using Eqs. \eqref{eq:tau_ss}-\eqref{eq:omega_ss},
a self similar solution exists for a surface temperature temporal
exponent $\tau=\frac{86}{57}\approx1.50877$ and a spatial density
exponent $\omega=-\frac{20}{19}\approx-1.05263$. Therefore, the surface
temperature and the material density profile are:
\[
T_{s}\left(t\right)=\left(\frac{t}{\text{ns}}\right)^{\frac{86}{57}}\ \text{keV},
\]

\[
\rho\left(x\right)=\left(\frac{x}{\text{cm}}\right)^{\frac{20}{19}}\ \text{g}/\text{cm}^{3}.
\]
 Using Eqs. \eqref{eq:adef}-\eqref{eq:bdef}, we find the dimensionless
constants of the problem: $\mathcal{A}=1.75246$ and $\mathcal{B}=4.15619$.
The similarity exponent is $\delta=1$, since the absorption and total
opacities have the same exponents {[}see Eq. \eqref{eq:delta_def}{]}
and the wave travels at constant speed. The numerical solution of
the similarity equations \eqref{eq:f_ode}-\eqref{eq:g_ode} gives
the heat front coordinate $\xi_{0}=1.2746051$, so that the heat front
position, according to Eq. \eqref{eq:xheat} is:
\[
x_{F}\left(t\right)=0.8332614\left(\frac{t}{\text{ns}}\right)\ \text{cm}.
\]
The resulting dimensionless flux at the origin is $\mathcal{S}\left(0\right)=4.62922$,
so that the bath temperature is {[}Eq. \eqref{eq:Tbath_marsh_bc}{]}:
\[
T_{\text{bath}}\left(t\right)=1.0470478\left(\frac{t}{\text{ns}}\right)^{\frac{86}{57}}\ \text{keV},
\]
which is used in transport simulations via the incoming bath radiation
flux {[}Eq. \eqref{eq:Ibath_tr}{]} or in diffusion simulations via
the Marshak boundary condition {[}Eq. \eqref{eq:maesh_bc_def}{]}.
We note that the bath temperature has a power law form since $\delta=1$.
Diffusion simulations can be run, alternatively, using the surface
temperature boundary condition {[}Eq. \eqref{eq:Tbound}{]}. A comparison
of the surface and bath temperatures as a function of time is displayed
in Fig. \ref{fig:simulation_1_Tbath}.

The radiation and material temperature profiles are given by the self-similar
solution {[}Eqs. \eqref{eq:Trss}-\eqref{eq:Tmss}{]}: 
\[
T_{r}\left(x,t\right)=\left(\frac{t}{\text{ns}}\right)^{\frac{86}{57}}f^{1/4}\left(\xi_{0}x/x_{F}\left(t\right)\right)\ \text{keV},
\]
\[
T\left(x,t\right)=\left(\frac{t}{\text{ns}}\right)^{\frac{86}{57}}g^{1/4}\left(\xi_{0}x/x_{F}\left(t\right)\right)\ \text{keV}.
\]
As discussed in Sec. \ref{sec:Self-Similar-solution}, there is no
analytical solution for the temperature profiles. The resulting numerical
solution is tabulated in table \ref{tab:tests_table}. Simple closed
form approximate expressions for the temperature profiles are given
in table \eqref{tab:test_fits}, with an accuracy which is better
than $0.5$\%. The profiles in tables \ref{tab:tests_table} and \ref{tab:test_fits}
are given as functions of $\xi/\xi_{0}=x/x_{F}\left(t\right)$. As
discussed in Sec. \ref{subsec:The-solution-near}, since we have $\beta<4$,
the spatial density increases in space, and the material temperature
is reduced to zero towards the origin.

Since $\mathcal{B}$ is not much larger than unity, we expect a significant
deviation from equilibrium. This is seen in Fig. \ref{fig:simulation_1},
where radiation and material temperature profiles of the self-similar
gray diffusion solution are compared to the results of numerical gray
diffusion and transport simulations. 

\subsubsection*{TEST 2}

\begin{figure}[t]
\begin{centering}
\includegraphics[width=1\columnwidth]{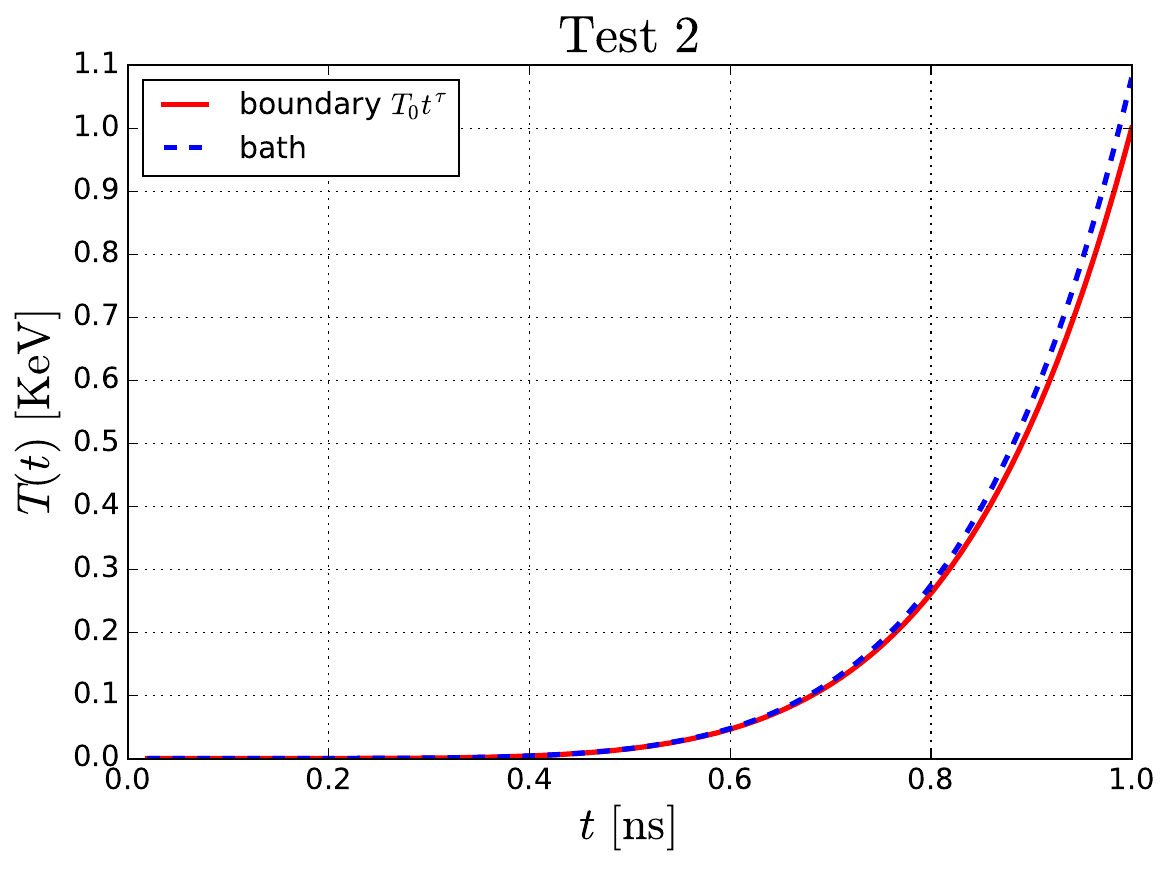}
\par\end{centering}
\caption{A comparison between the surface and bath drive temperatures for test
2.\label{fig:simulation_2-Tbath}}
\end{figure}

\begin{figure}[t]
\begin{centering}
\includegraphics[width=1\columnwidth]{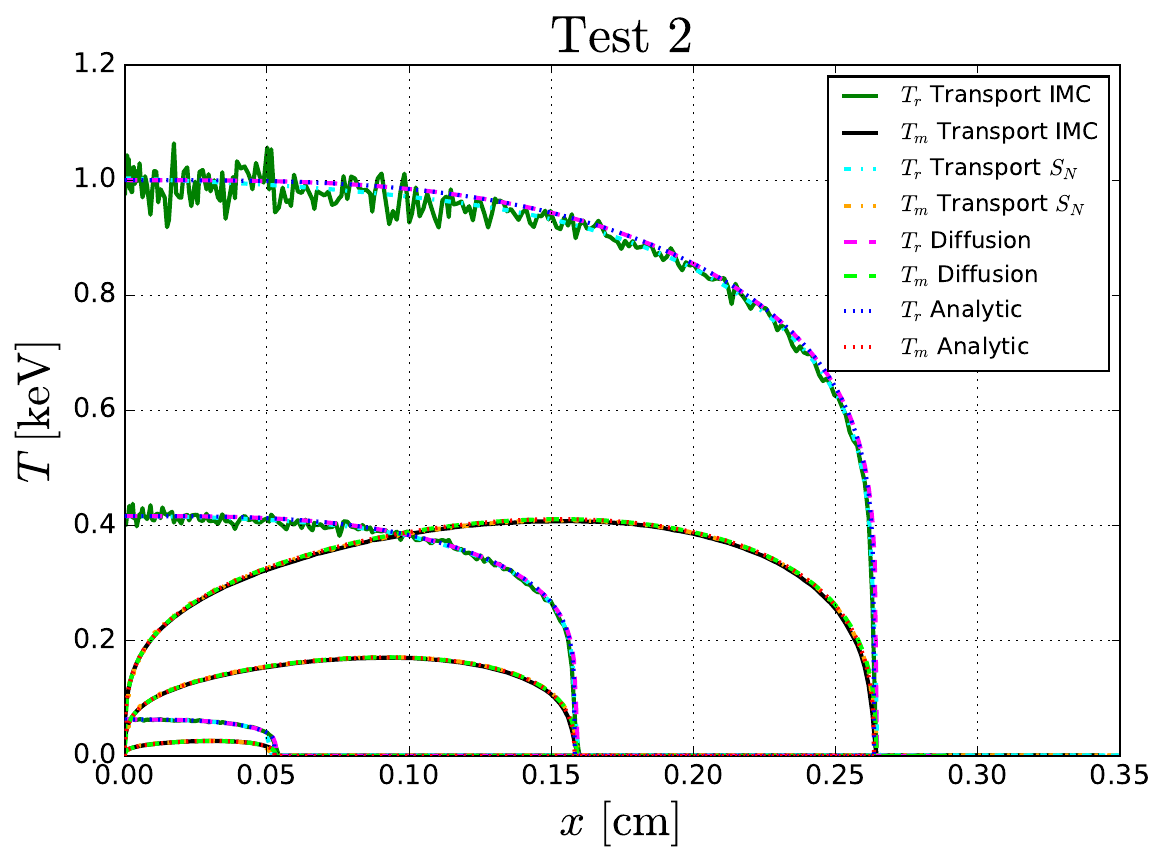}
\par\end{centering}
\caption{Radiation and material temperature profiles for Test 2. Results are
shown at times $t=0.631385,\ 0.864201$ and 1ns, as obtained from
a gray diffusion simulation and from Implicit-Monte-Carlo (IMC) and
discrete ordinates ($S_{N}$) transport simulations. \label{fig:simulation_2}}
\end{figure}

We define, as in test 1, another case with $\omega<0$, but with a
sharper heat front. We take $\alpha=3$, $\alpha'=2$, $\lambda=0.2$,
$\lambda'=0.1$, $\mathcal{G}=10^{-3}\frac{\text{cm}\left(\text{g}/\text{cm}^{3}\right)^{1.2}}{\text{keV}^{3}}$,
$\mathcal{G}'=10\frac{\text{cm\ensuremath{\left(\text{g}/\text{cm}^{3}\right)^{1.1}}}}{\text{keV}^{2}}$,
so that the total and absorption opacities are:
\begin{align*}
k_{t}\left(T,\rho\right) & =10^{3}\left(\frac{T}{\text{keV}}\right)^{-3}\left(\frac{\rho}{\text{g}/\text{cm}^{3}}\right)^{1.2}\ \text{cm}^{-1},\\
k_{a}\left(T,\rho\right) & =0.1\left(\frac{T}{\text{keV}}\right)^{-2}\left(\frac{\rho}{\text{g}/\text{cm}^{3}}\right)^{1.1}\ \text{cm}^{-1}.
\end{align*}
For transport simulations, the scattering opacity {[}Eq. \eqref{eq:scatt}{]}
is given by: 
\begin{align*}
k_{s}\left(T,\rho\right) & =10^{3}\left(\frac{T}{\text{keV}}\right)^{-3}\left(\frac{\rho}{\text{g}/\text{cm}^{3}}\right)^{1.2}\\
 & -0.1\left(\frac{T}{\text{keV}}\right)^{-2}\left(\frac{\rho}{\text{g}/\text{cm}^{3}}\right)^{1.1}\ \text{cm}^{-1},
\end{align*}
which is not a power-law form as in test 1, since the absorption and
total opacities have different temperature and density exponents.
We also take $\mu=0.4$, so that $\beta_{c}=2.909$ and $\beta_{c}'=2.69565$.
We set $\beta=3$ and $\mathcal{F}=10^{14}\frac{\text{keV}^{-3}}{\left(\text{g}/\text{cm}^{3}\right)^{0.6}}\frac{\text{erg}}{\text{cm}^{3}}$,
so that the material energy density is given by:

\[
u\left(T,\rho\right)=10^{14}\left(\frac{T}{\text{keV}}\right)^{3}\left(\frac{\rho}{\text{g}/\text{cm}^{3}}\right)^{0.6}\ \text{\ensuremath{\frac{\text{erg}}{\text{cm}^{3}}}}.
\]
For the exponents of this material model we get $\tau=6$ and $\omega=-\frac{20}{7}\approx-2.8571$,
so that the surface temperature and density profile are:
\[
T_{s}\left(t\right)=\left(\frac{t}{\text{ns}}\right)^{6}\ \text{keV},
\]

\[
\rho\left(x\right)=\left(\frac{x}{\text{cm}}\right)^{\frac{20}{7}}\ \text{g}/\text{cm}^{3}.
\]
The dimensionless constants of the problem are $\mathcal{A}=0.20833$
and $\mathcal{B}=1.63201$. As in the previous case, $\mathcal{B}$
is not large and we expect a significant deviation from equilibrium.
From Eq. \eqref{eq:delta_def} we find the similarity exponent $\delta=\frac{7}{2}$,
so that the the wave front accelerates over time. The resulting numerical
heat front similarity coordinate is $\xi_{0}=0.615503394$, so that
the heat front position is:
\[
x_{F}\left(t\right)=0.26348387\left(\frac{t}{\text{ns}}\right)^{\frac{7}{2}}\ \text{cm}.
\]
The resulting numerical dimensionless flux at the origin is $\mathcal{S}\left(0\right)=12.5696$,
so that the bath temperature is {[}Eq. \eqref{eq:Tbath_marsh_bc}{]}:
\[
T_{\text{bath}}\left(t\right)=\left(1+0.358968\left(\frac{t}{\text{ns}}\right)^{\frac{5}{2}}\right)^{\frac{1}{4}}\left(\frac{t}{\text{ns}}\right)^{6}\ \text{keV},
\]
which is not in a power law form, since $\delta\neq1$. A comparison
of the surface and bath temperatures as a function of time is displayed
in Fig. \ref{fig:simulation_2-Tbath}.

The radiation and material temperature profiles are given by the self-similar
solution {[}Eqs. \eqref{eq:Trss}-\eqref{eq:Tmss}{]}: 
\[
T_{r}\left(x,t\right)=\left(\frac{t}{\text{ns}}\right)^{6}f^{1/4}\left(\xi_{0}x/x_{F}\left(t\right)\right)\ \text{keV},
\]
\[
T\left(x,t\right)=\left(\frac{t}{\text{ns}}\right)^{6}g^{1/4}\left(\xi_{0}x/x_{F}\left(t\right)\right)\ \text{keV}.
\]
 The resulting numerical profiles are tabulated in table \ref{tab:tests_table}
and approximate closed form expressions are given in table \eqref{tab:test_fits}
as a function of $\xi/\xi_{0}=x/x_{F}\left(t\right)$. As in the previous
case, since we have $\beta<4$, the material temperature is decreased
to zero towards the origin.

The temperature profiles are displayed in Fig. \ref{fig:simulation_2},
showing again a great agreement between the various simulations and
analytic gray diffusion solution. 

\subsubsection*{TEST 3}

\begin{figure}[t]
\begin{centering}
\includegraphics[width=1\columnwidth]{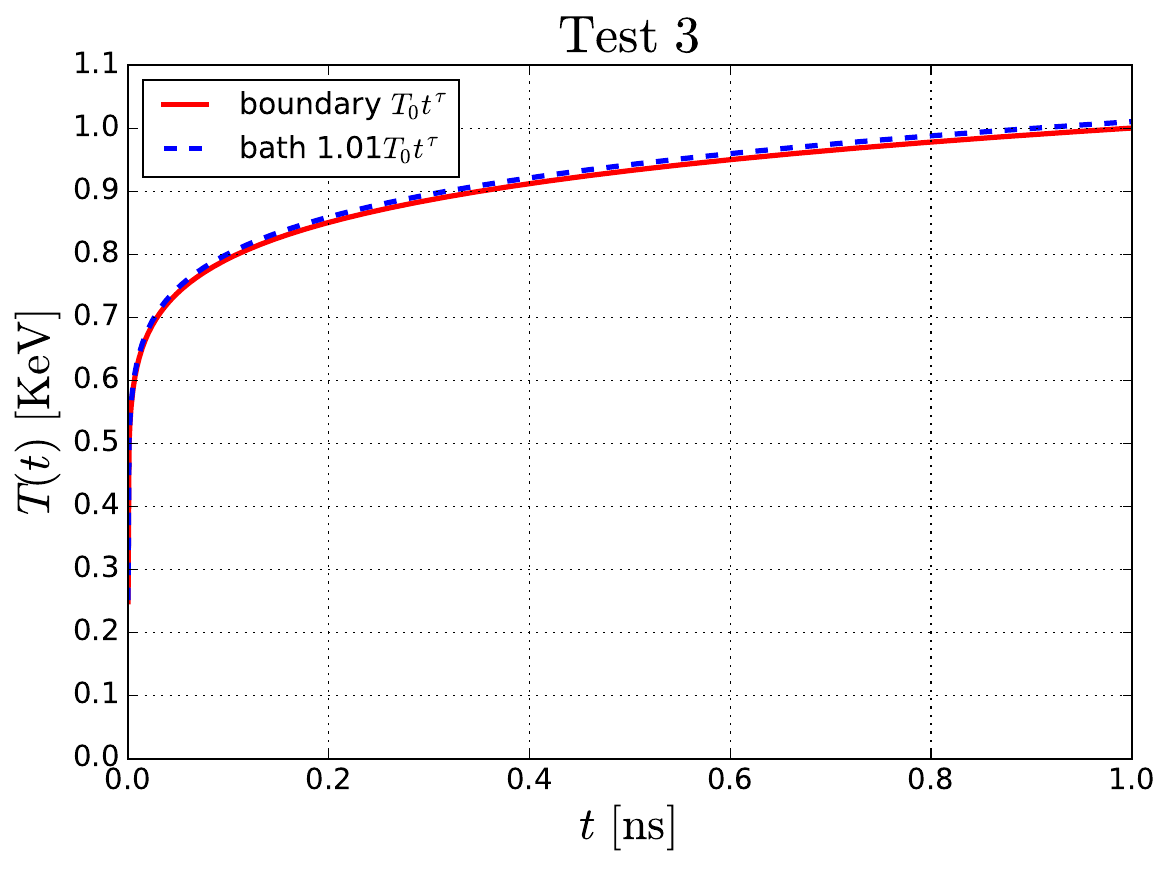}
\par\end{centering}
\caption{A comparison between the surface and bath drive temperatures for test
3. \label{fig:simulation_3_Tbath}}
\end{figure}

\begin{figure}[t]
\begin{centering}
\includegraphics[width=1\columnwidth]{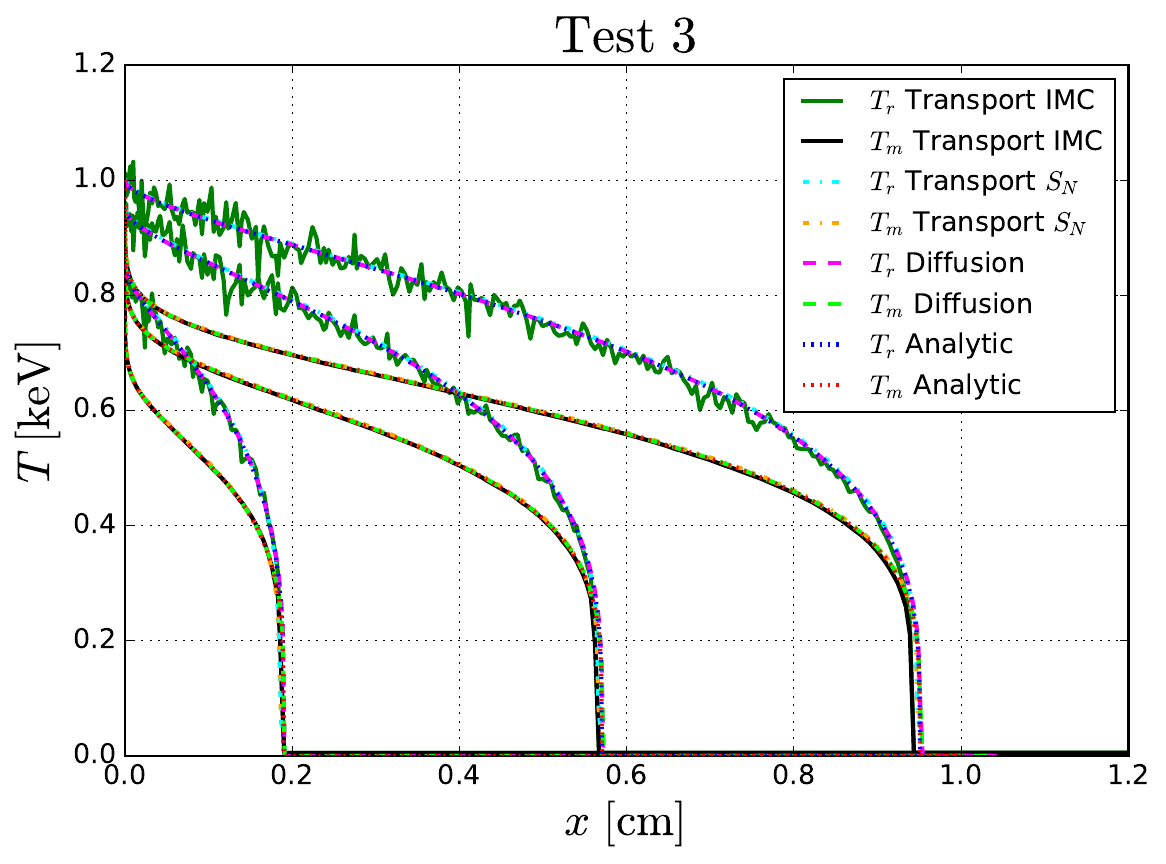}
\par\end{centering}
\caption{Radiation and material temperature profiles for Test 3. Results are
shown at times $t=0.2,\ 0.6$ and 1ns, as obtained from the gray diffusion
self-similar solution, a gray diffusion simulation and from Implicit-Monte-Carlo
(IMC) and discrete ordinates ($S_{N}$) transport simulations. \label{fig:simulation_3}}
\end{figure}

We now define a case with $\omega>0$. We take $\alpha=\alpha'=4.5$,
$\lambda=\lambda'=0.9$, $\mathcal{G}=0.5\frac{\text{cm}\left(\text{g}/\text{cm}^{3}\right)^{1.9}}{\text{keV}^{4.5}}$,
$\mathcal{G}'=10^{3}\frac{\text{cm\ensuremath{\left(\text{g}/\text{cm}^{3}\right)^{1.9}}}}{\text{keV}^{4.5}}$,
so that the total and absorption opacities are:
\begin{align*}
k_{t}\left(T,\rho\right) & =2\left(\frac{T}{\text{keV}}\right)^{-4.5}\left(\frac{\rho}{\text{g}/\text{cm}^{3}}\right)^{1.9}\ \text{cm}^{-1},\\
k_{a}\left(T,\rho\right) & =10^{-3}\left(\frac{T}{\text{keV}}\right)^{-4.5}\left(\frac{\rho}{\text{g}/\text{cm}^{3}}\right)^{1.9}\ \text{cm}^{-1}.
\end{align*}
For transport simulations, the scattering opacity is given by a power
law form as well: 
\begin{align*}
k_{s}\left(T,\rho\right) & =1.999\left(\frac{T}{\text{keV}}\right)^{-4.5}\left(\frac{\rho}{\text{g}/\text{cm}^{3}}\right)^{1.9}\ \text{cm}^{-1}.
\end{align*}
We also take $\mu=0.3$, so that $\beta_{c}=\beta_{c}'=2.3421$. We
set $\beta=6$ and $\mathcal{F}=10^{14}\frac{\text{keV}^{-6}}{\left(\text{g}/\text{cm}^{3}\right)^{0.7}}\frac{\text{erg}}{\text{cm}^{3}}$,
so that the material energy density is given by:

\[
u\left(T,\rho\right)=10^{14}\left(\frac{T}{\text{keV}}\right)^{6}\left(\frac{\rho}{\text{g}/\text{cm}^{3}}\right)^{0.7}\ \text{\ensuremath{\frac{\text{erg}}{\text{cm}^{3}}}}.
\]
For the exponents of this material model, we find $\tau=\frac{14}{139}\approx0.10072$
and $\omega=\frac{40}{139}\approx0.28777$, so that the surface temperature
and density profile are:
\[
T_{s}\left(t\right)=\left(\frac{t}{\text{ns}}\right)^{\frac{14}{139}}\ \text{keV},
\]

\[
\rho\left(x\right)=\left(\frac{x}{\text{cm}}\right)^{-\frac{40}{139}}\ \text{g}/\text{cm}^{3}.
\]
The dimensionless constants of the problem are $\mathcal{A}=0.0163665$
and $\mathcal{B}=0.0187098$, so we expect a significant deviation
from equilibrium. As in test 1, the similarity exponent is
$\delta=1$, since the absorption and total opacities have the same
exponents. The resulting numerical heat front similarity coordinate
is $\xi_{0}=0.31411518$, so that the heat front position is
\[
x_{F}\left(t\right)=0.95029077\left(\frac{t}{\text{ns}}\right)\ \text{cm}.
\]
The resulting numerical dimensionless flux at the origin is $\mathcal{S}\left(0\right)=0.20284$,
so that the bath temperature is {[}Eq. \eqref{eq:Tbath_marsh_bc}{]}:
\[
T_{\text{bath}}\left(t\right)=1.01008116\left(\frac{t}{\text{ns}}\right)^{\frac{14}{139}}\ \text{keV}.
\]
A comparison of the surface and bath temperatures as a function of
time is displayed in Fig. \ref{fig:simulation_3_Tbath}.

The radiation and material temperature profiles are given by the self-similar
solution {[}Eqs. \eqref{eq:Trss}-\eqref{eq:Tmss}{]}: 
\[
T_{r}\left(x,t\right)=\left(\frac{t}{\text{ns}}\right)^{\frac{14}{139}}f^{1/4}\left(\xi_{0}x/x_{F}\left(t\right)\right)\ \text{keV},
\]
\[
T\left(x,t\right)=\left(\frac{t}{\text{ns}}\right)^{\frac{14}{139}}g^{1/4}\left(\xi_{0}x/x_{F}\left(t\right)\right)\ \text{keV}.
\]
 The resulting numerical profiles are tabulated in table \ref{tab:tests_table}
and approximate closed form expressions are given in table \eqref{tab:test_fits}
as a function of $\xi/\xi_{0}=x/x_{F}\left(t\right)$. In this case,
since we have $\beta>4$, the material temperature is equal to the
radiation temperature at the origin.

The temperature profiles are displayed in Fig. \ref{fig:simulation_3},
showing again a great agreement between the various simulations and
analytic gray diffusion solution. 

\subsubsection*{TEST 4}

\begin{figure}[t]
\begin{centering}
\includegraphics[width=1\columnwidth]{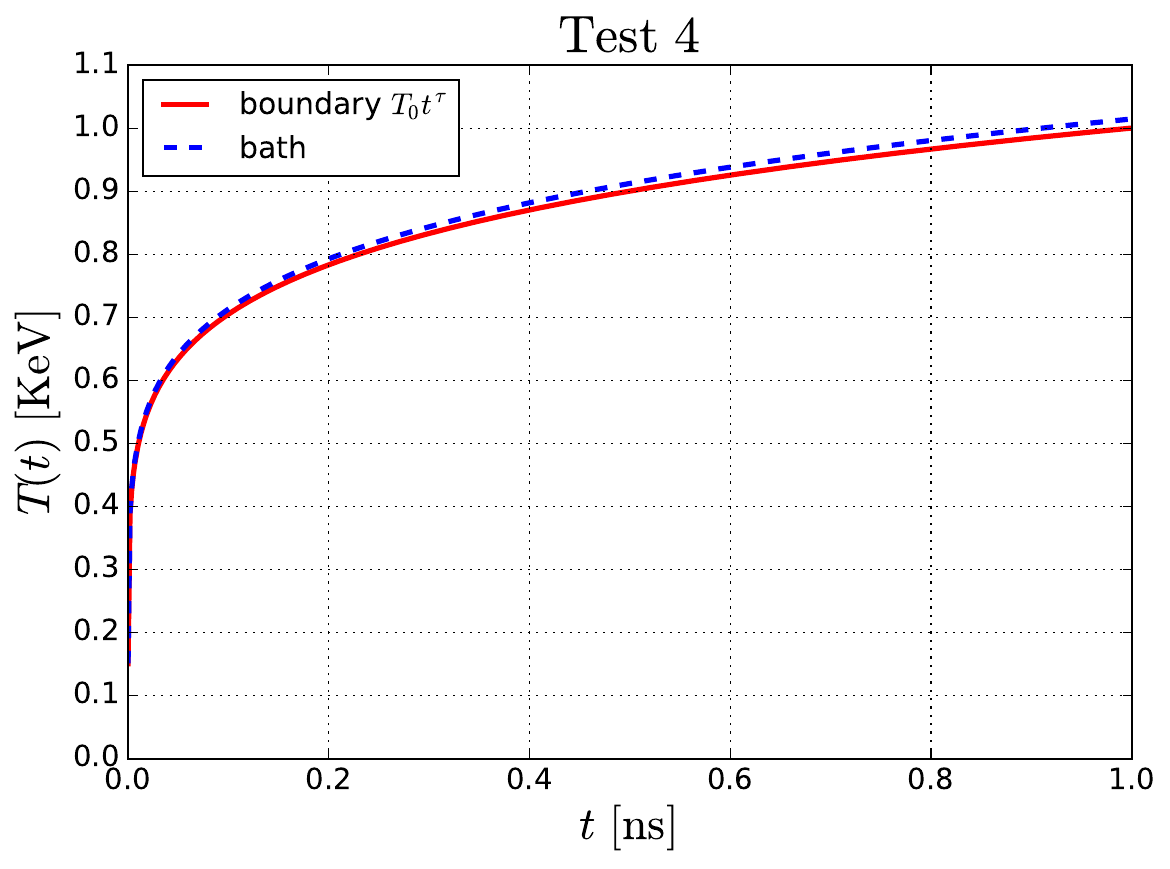}
\par\end{centering}
\caption{A comparison between the surface and bath drive temperatures for test
4. \label{fig:simulation_4_Tbath}}
\end{figure}

\begin{figure}[t]
\begin{centering}
\includegraphics[width=1\columnwidth]{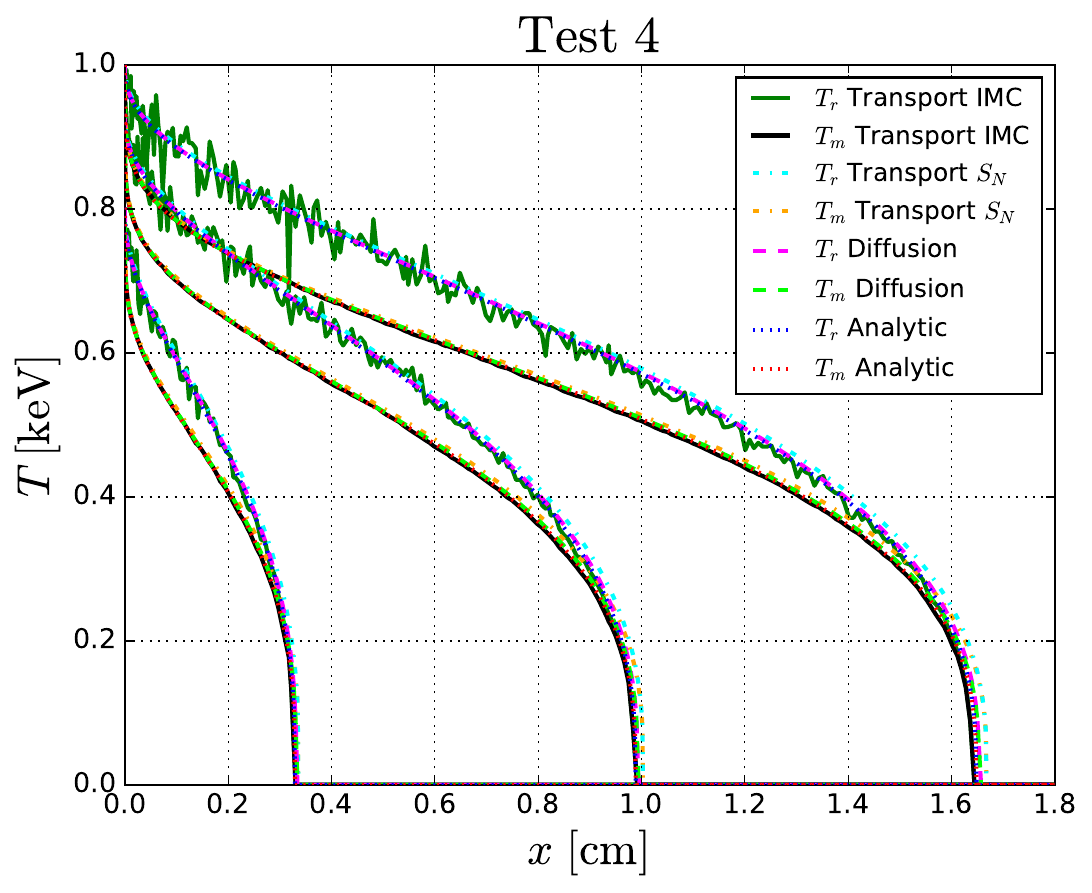}
\par\end{centering}
\caption{Radiation and material temperature profiles for Test 4. Results are
shown at times $t=0.233846,\ 0.630525$ and 1ns, as obtained from
the gray diffusion self-similar solution, a gray diffusion simulation
and from Implicit-Monte-Carlo (IMC) and discrete ordinates ($S_{N}$)
transport simulations.\label{fig:simulation_4}}
\end{figure}

We define, as in test 3, a test with $\omega>0$ but with a less steep
heat front. We take, as in test 2 the exponents $\alpha=3$, $\alpha'=2$,
$\lambda=0.2$, $\lambda'=0.1$ and $\mathcal{G}=0.5\frac{\text{cm}\left(\text{g}/\text{cm}^{3}\right)^{1.2}}{\text{keV}^{3}}$,
$\mathcal{G}'=10^{2}\frac{\text{cm\ensuremath{\left(\text{g}/\text{cm}^{3}\right)^{1.1}}}}{\text{keV}^{2}}$,
so that the total and absorption opacities are:
\begin{align*}
k_{t}\left(T,\rho\right) & =2\left(\frac{T}{\text{keV}}\right)^{-3}\left(\frac{\rho}{\text{g}/\text{cm}^{3}}\right)^{1.2}\ \text{cm}^{-1},\\
k_{a}\left(T,\rho\right) & =10^{-2}\left(\frac{T}{\text{keV}}\right)^{-2}\left(\frac{\rho}{\text{g}/\text{cm}^{3}}\right)^{1.1}\ \text{cm}^{-1}.
\end{align*}
For transport simulations, the scattering opacity {[}Eq. \eqref{eq:scatt}{]}
is given by: 
\begin{align*}
k_{s}\left(T,\rho\right) & =2\left(\frac{T}{\text{keV}}\right)^{-3}\left(\frac{\rho}{\text{g}/\text{cm}^{3}}\right)^{1.2}\\
 & -10^{-2}\left(\frac{T}{\text{keV}}\right)^{-2}\left(\frac{\rho}{\text{g}/\text{cm}^{3}}\right)^{1.1}\ \text{cm}^{-1}.
\end{align*}
We also take $\mu=0.4$, so that $\beta_{c}=2.909$ and $\beta_{c}'=2.69565$.
We set $\beta=6.5$ and $\mathcal{F}=10^{14}\frac{\text{keV}^{-6.5}}{\left(\text{g}/\text{cm}^{3}\right)^{0.6}}\frac{\text{erg}}{\text{cm}^{3}}$,
so that the material energy density is given by:

\[
u\left(T,\rho\right)=10^{14}\left(\frac{T}{\text{keV}}\right)^{6.5}\left(\frac{\rho}{\text{g}/\text{cm}^{3}}\right)^{0.6}\ \text{\ensuremath{\frac{\text{erg}}{\text{cm}^{3}}}}.
\]
For the exponents of this material model we get $\tau=\frac{12}{79}\approx0.151899$
and $\omega=\frac{4}{7}\approx0.571429$, so that the surface temperature
and density profile are:
\[
T_{s}\left(t\right)=\left(\frac{t}{\text{ns}}\right)^{\frac{12}{79}}\ \text{keV},
\]

\[
\rho\left(x\right)=\left(\frac{x}{\text{cm}}\right)^{-\frac{4}{7}}\ \text{g}/\text{cm}^{3}.
\]
The dimensionless constants of the problem are $\mathcal{A}=0.138891$
and $\mathcal{B}=0.178419$. As in the previous case, $\mathcal{B}$
is not large and we expect a significant deviation from equilibrium.
From Eq. \eqref{eq:delta_def} we find the similarity exponent $\delta=\frac{175}{158}\approx1.10759$,
so that the the wave front accelerates over time. The resulting numerical
heat front similarity coordinate is $\xi_{0}=0.48463864$ and the
heat front position is:
\[
x_{F}\left(t\right)=1.648216882\left(\frac{t}{\text{ns}}\right)^{\frac{175}{158}}\ \text{cm}.
\]
The resulting numerical dimensionless flux at the origin is $\mathcal{S}\left(0\right)=0.260125$,
so that the bath temperature is {[}Eq. \eqref{eq:Tbath_marsh_bc}{]}:
\[
T_{\text{bath}}\left(t\right)=\left(1+0.0590184\left(\frac{t}{\text{ns}}\right)^{\frac{17}{158}}\right)^{\frac{1}{4}}\left(\frac{t}{\text{ns}}\right)^{\frac{12}{79}}\ \text{keV}.
\]
A comparison of the surface and bath temperatures as a function of
time is displayed in Fig. \ref{fig:simulation_4_Tbath}.

The radiation and material temperature profiles are given by the self-similar
solution {[}Eqs. \eqref{eq:Trss}-\eqref{eq:Tmss}{]}: 
\[
T_{r}\left(x,t\right)=\left(\frac{t}{\text{ns}}\right)^{\frac{12}{79}}f^{1/4}\left(\xi_{0}x/x_{F}\left(t\right)\right)\ \text{keV},
\]
\[
T\left(x,t\right)=\left(\frac{t}{\text{ns}}\right)^{\frac{12}{79}}g^{1/4}\left(\xi_{0}x/x_{F}\left(t\right)\right)\ \text{keV}.
\]
 The resulting numerical profiles are tabulated in table \ref{tab:tests_table}
and approximate closed form expressions are given in table \eqref{tab:test_fits}
as a function of $\xi/\xi_{0}=x/x_{F}\left(t\right)$. As in the previous
case, since we have $\beta>4$, the material temperature is equal
to the radiation temperature at the origin.

The temperature profiles are displayed in Fig. \ref{fig:simulation_4},
showing again a great agreement between the various simulations and
analytic gray diffusion solution. 

\subsubsection*{TEST 5}

\begin{figure}[t]
\begin{centering}
\includegraphics[width=1\columnwidth]{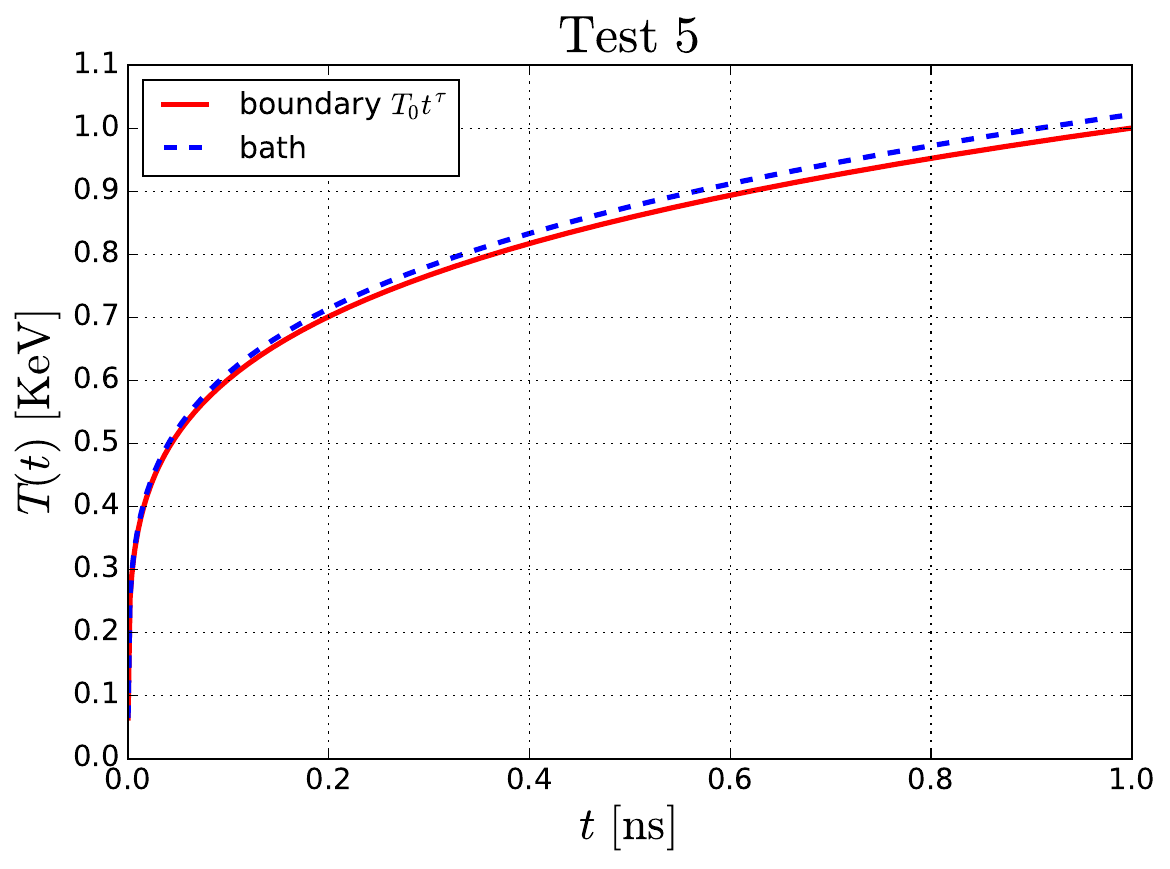}
\par\end{centering}
\caption{A comparison between the surface and bath drive temperatures for test
5. \label{fig:simulation_5_Tbath}}
\end{figure}

\begin{figure}[t]
\begin{centering}
\includegraphics[width=1\columnwidth]{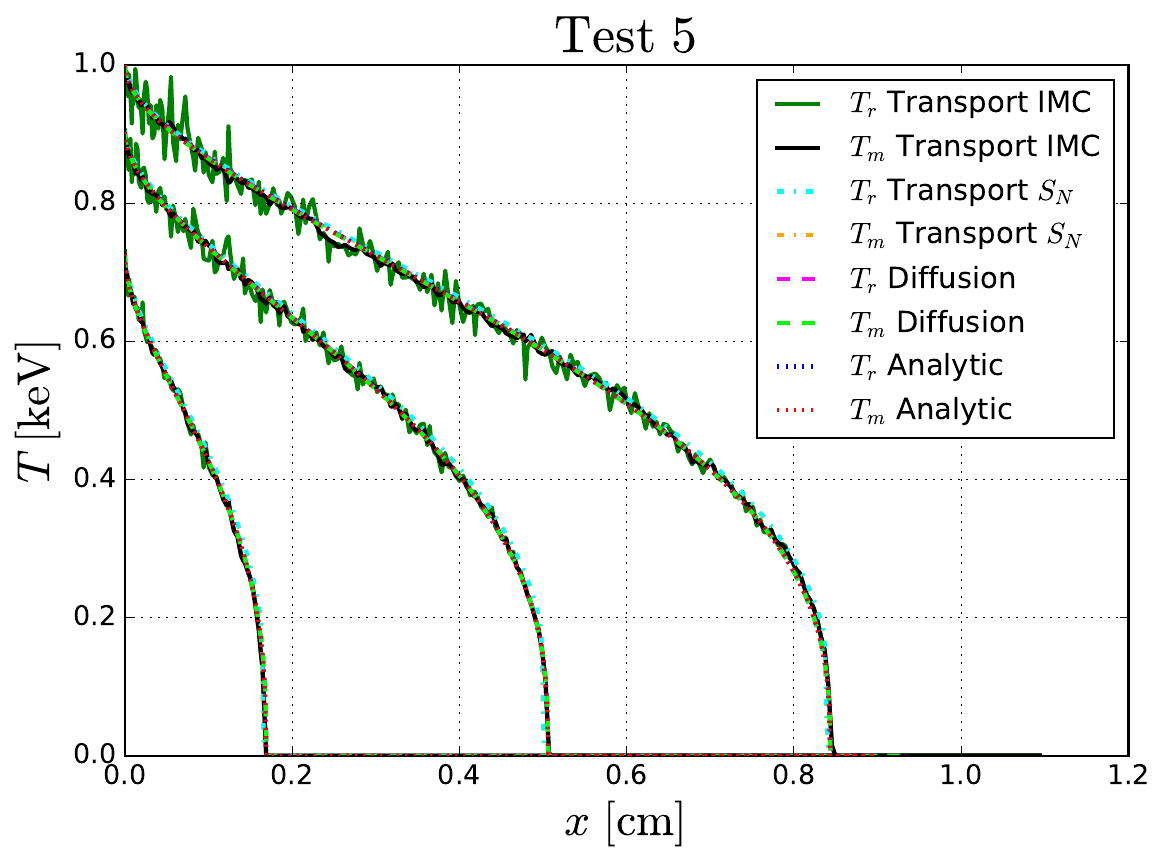}
\par\end{centering}
\caption{Radiation and material temperature profiles for Test 5. Results are
shown at times $t=0.2346891,\ 0.63124555$ and 1ns, as obtained from
the gray diffusion self-similar solution, a gray diffusion simulation
and from Implicit-Monte-Carlo (IMC) and discrete ordinates ($S_{N}$)
transport simulations.\label{fig:simulation_5}}
\end{figure}

\begin{figure}[t]
\begin{centering}
\includegraphics[width=1\columnwidth]{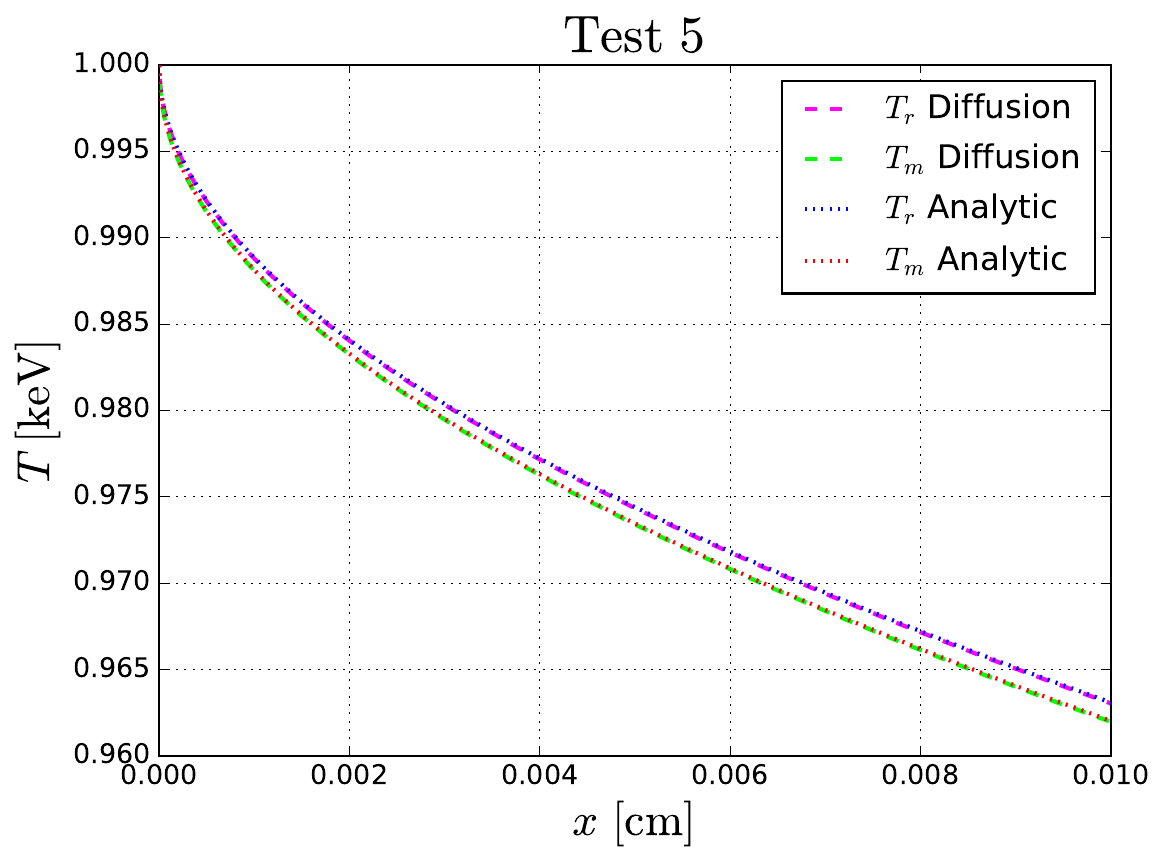}
\par\end{centering}
\caption{A close view near the origin of the temperature profiles at time $t=1\text{ns}$
(from the gray diffusion simulation and analytic solution), for Test
5 (see Fig. \ref{fig:simulation_5}).\label{fig:simulation_5_zoom}}
\end{figure}

We define another case with $\omega>0$, and such that $\mathcal{B}\gg1$,
so that the thermodynamic equilibrium limit is reached, unlike
the previous cases which were significantly out of equilibrium. We
take $\alpha=3$, $\alpha'=2$, $\lambda=\lambda'=0.35$, $\mathcal{G}=0.2\frac{\text{cm}\left(\text{g}/\text{cm}^{3}\right)^{1.35}}{\text{keV}^{3}}$,
$\mathcal{G}'=0.2\frac{\text{cm\ensuremath{\left(\text{g}/\text{cm}^{3}\right)^{1.35}}}}{\text{keV}^{2}}$,
so that the total and absorption opacities are:
\begin{align*}
k_{t}\left(T,\rho\right) & =5\left(\frac{T}{\text{keV}}\right)^{-3}\left(\frac{\rho}{\text{g}/\text{cm}^{3}}\right)^{1.35}\ \text{cm}^{-1},\\
k_{a}\left(T,\rho\right) & =5\left(\frac{T}{\text{keV}}\right)^{-2}\left(\frac{\rho}{\text{g}/\text{cm}^{3}}\right)^{1.35}\ \text{cm}^{-1}.
\end{align*}
For transport simulations, the scattering opacity {[}Eq. \eqref{eq:scatt}{]}
is given by: 
\begin{align*}
k_{s}\left(T,\rho\right) & =5\left(\frac{\rho}{\text{g}/\text{cm}^{3}}\right)^{1.35}\left[\left(\frac{T}{\text{keV}}\right)^{-3}-\left(\frac{T}{\text{keV}}\right)^{-2}\right]\ \text{cm}^{-1}.
\end{align*}
We also take $\mu=0.2$, so that $\beta_{c}=2.8148$ and $\beta_{c}'=2.5185$.
We set $\beta=5.5$ and $\mathcal{F}=2\times10^{14}\frac{\text{keV}^{-5.5}}{\left(\text{g}/\text{cm}^{3}\right)^{0.8}}\frac{\text{erg}}{\text{cm}^{3}}$,
so that the material energy density is given by:
\[
u\left(T,\rho\right)=2\times10^{14}\left(\frac{T}{\text{keV}}\right)^{5.5}\left(\frac{\rho}{\text{g}/\text{cm}^{3}}\right)^{0.8}\ \text{\ensuremath{\frac{\text{erg}}{\text{cm}^{3}}}}.
\]
For the exponents of this material model we get $\tau=\frac{32}{145}\approx0.22069$
and $\omega=\frac{60}{161}\approx0.372671$, so that the surface temperature
and density profile are:LTE
\[
T_{s}\left(t\right)=\left(\frac{t}{\text{ns}}\right)^{\frac{32}{145}}\ \text{keV},
\]

\[
\rho\left(x\right)=\left(\frac{x}{\text{cm}}\right)^{-\frac{60}{161}}\ \text{g}/\text{cm}^{3}.
\]
The dimensionless constants of the problem are $\mathcal{A}=118.772$
and $\mathcal{B}=68.0203$. Since $\mathcal{B}$ significantly larger
than unity, we expect the material and radiation temperatures to be
very close (and, as in tests 3-4, equal at the origin, since $\omega>0$).
From Eq. \eqref{eq:delta_def} we find the similarity exponent $\delta=\frac{161}{145}\approx1.11034$,
so that the the wave front accelerates over time. The resulting numerical
heat front similarity coordinate is $\xi_{0}=0.53073002$ and the
heat front position is:
\[
x_{F}\left(t\right)=0.8428997\left(\frac{t}{\text{ns}}\right)^{\frac{161}{145}}\ \text{cm}.
\]
The resulting numerical dimensionless flux at the origin is $\mathcal{S}\left(0\right)=0.840029$,
so that the bath temperature is {[}Eq. \eqref{eq:Tbath_marsh_bc}{]}:
\[
T_{\text{bath}}\left(t\right)=\left(1+0.0890032\left(\frac{t}{\text{ns}}\right)^{\frac{16}{145}}\right)^{\frac{1}{4}}\left(\frac{t}{\text{ns}}\right)^{\frac{32}{145}}\ \text{keV}.
\]
A comparison of the surface and bath temperatures as a function of
time is displayed in Fig. \ref{fig:simulation_5_Tbath}.

The radiation and material temperature profiles are given by the self-similar
solution {[}Eqs. \eqref{eq:Trss}-\eqref{eq:Tmss}{]}: 
\[
T_{r}\left(x,t\right)=\left(\frac{t}{\text{ns}}\right)^{\frac{32}{145}}f^{1/4}\left(\xi_{0}x/x_{F}\left(t\right)\right)\ \text{keV},
\]
\[
T\left(x,t\right)=\left(\frac{t}{\text{ns}}\right)^{\frac{32}{145}}g^{1/4}\left(\xi_{0}x/x_{F}\left(t\right)\right)\ \text{keV}.
\]
 The resulting numerical profiles are tabulated in table \ref{tab:tests_table}
and approximate closed form expressions are given in table \eqref{tab:test_fits}
as a function of $\xi/\xi_{0}=x/x_{F}\left(t\right)$.

The temperature profiles are displayed in Fig. \ref{fig:simulation_5},
showing again a great agreement between the various simulations and
analytic gray diffusion solution. It is evident that resulting
temperature profiles are very close to a state of equilibrium, as
the radiation and material temperatures differ by about $0.1\%$.
This difference, is evident in Fig. \ref{fig:simulation_5_zoom},
where we show a close up view near the origin. It is interesting to
see the great agreement of the gray diffusion simulations with the
analytic solution even on that scale.

\subsubsection*{TEST 6}

\begin{figure}[t]
\begin{centering}
\includegraphics[width=1\columnwidth]{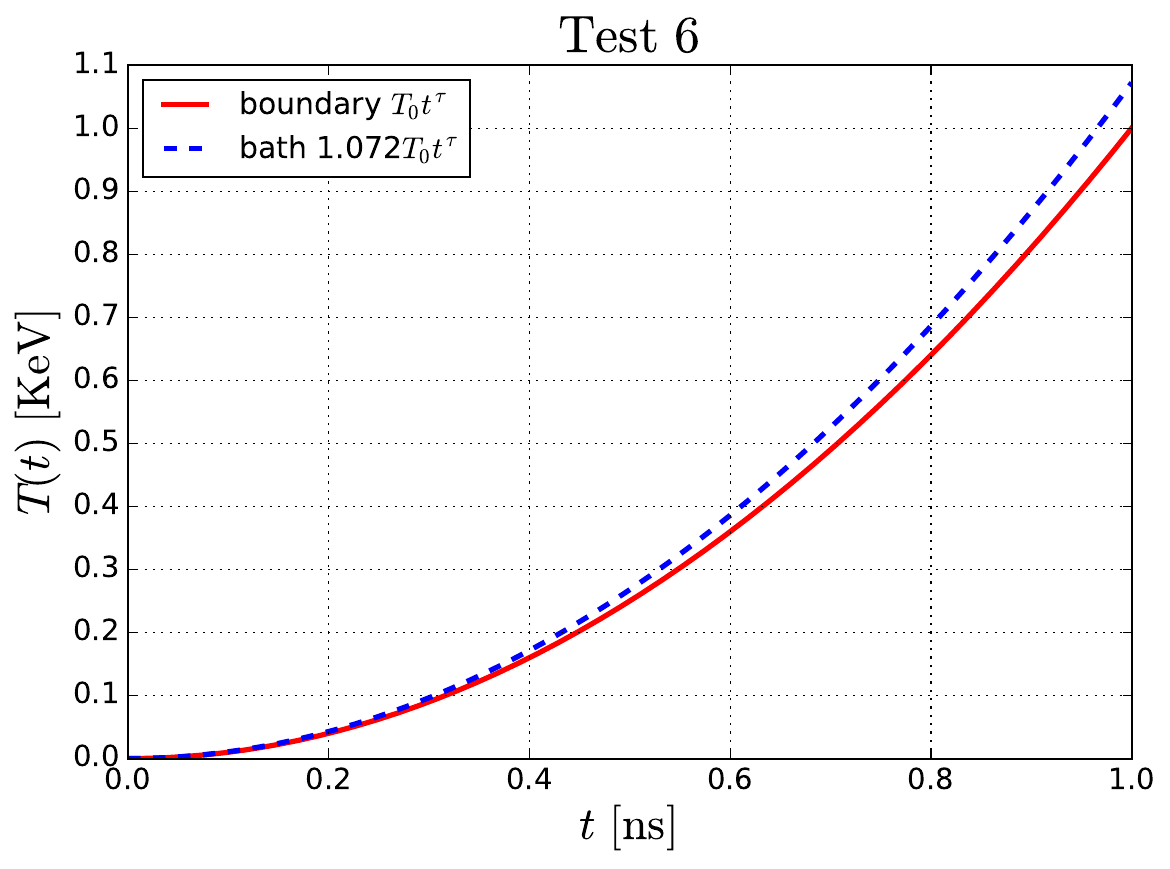}
\par\end{centering}
\caption{A comparison between the surface and bath drive temperatures for test
6. \label{fig:simulation_6_Tbath}}
\end{figure}

\begin{figure}[t]
\begin{centering}
\includegraphics[width=1\columnwidth]{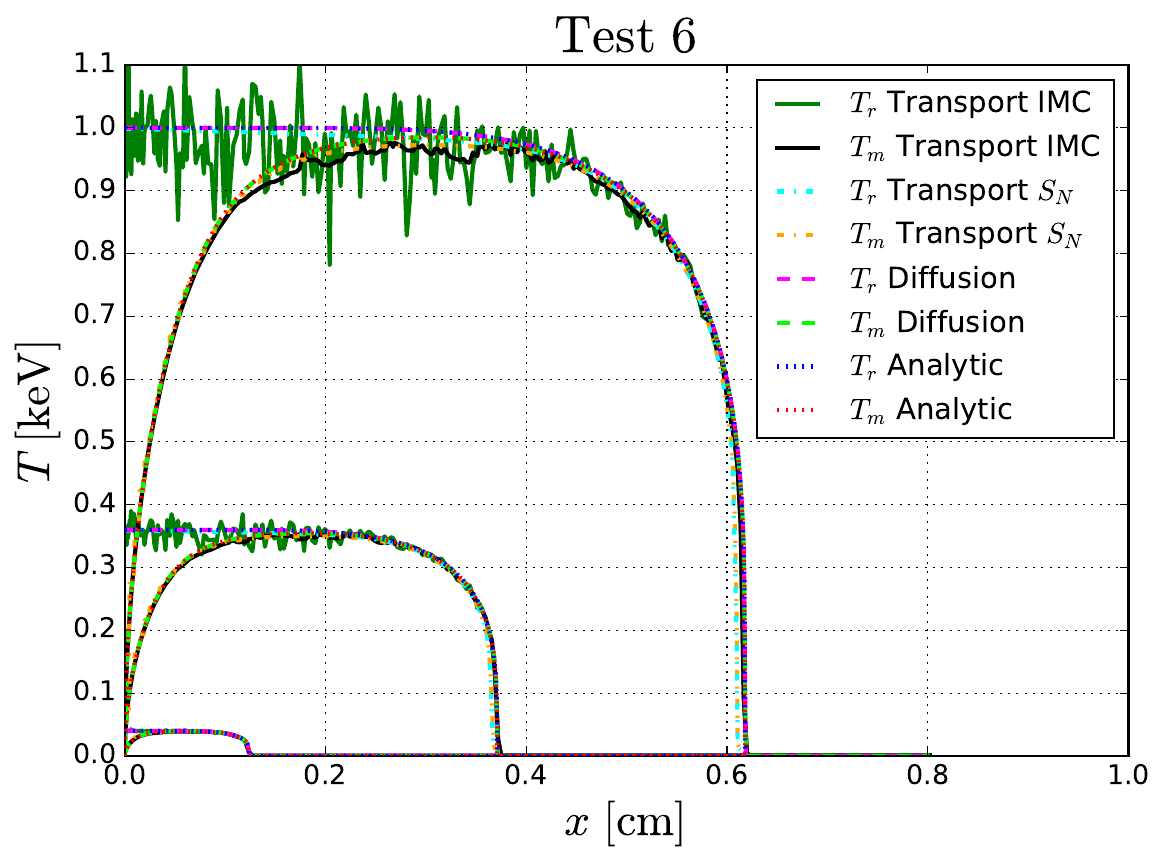}
\par\end{centering}
\caption{Radiation and material temperature profiles for Test 6. Results are
shown at times $t=0.2,\ 0.6$ and 1ns, as obtained from the gray diffusion
self-similar solution, a gray diffusion simulation and from Implicit-Monte-Carlo
(IMC) and discrete ordinates ($S_{N}$) transport simulations.\label{fig:simulation_6}}
\end{figure}

We define the last case to have $\omega=-3$, and $\mathcal{B}\gg1$,
so that the thermodynamic equilibrium limit should be reached.
We set a material with free-free like absorption, which depends on
$\rho^{2}$. Since $\rho\left(x\right)\propto x^{3}$, the density,
and as a result, the coupling between the material and radiation,
is expected to be very low in a wide region near the origin. This
results in a significant state of non-equilibrium ranging from the
origin and halfway towards the front (where the coupling is large
due to the lower material temperature). We take $\alpha=\alpha'=3.5$,
$\lambda=\lambda'=1$, $\mathcal{G}=0.0005\frac{\text{cm}\left(\text{g}/\text{cm}^{3}\right)^{2}}{\text{keV}^{3.5}}$,
$\mathcal{G}'=0.01\frac{\text{cm\ensuremath{\left(\text{g}/\text{cm}^{3}\right)^{2}}}}{\text{keV}^{3.5}}$,
so that the total and absorption opacities are:
\begin{align*}
k_{t}\left(T,\rho\right) & =2000\left(\frac{T}{\text{keV}}\right)^{-3.5}\left(\frac{\rho}{\text{g}/\text{cm}^{3}}\right)^{2}\ \text{cm}^{-1},\\
k_{a}\left(T,\rho\right) & =100\left(\frac{T}{\text{keV}}\right)^{-3.5}\left(\frac{\rho}{\text{g}/\text{cm}^{3}}\right)^{2}\ \text{cm}^{-1}.
\end{align*}
For transport simulations, the scattering opacity {[}Eq. \eqref{eq:scatt}{]}
is given by: 
\begin{align*}
k_{s}\left(T,\rho\right) & =1900\left(\frac{T}{\text{keV}}\right)^{-3.5}\left(\frac{\rho}{\text{g}/\text{cm}^{3}}\right)^{2}\ \text{cm}^{-1}.
\end{align*}
We take and ideal like energy density, $\mu=0$, so that $\beta_{c}=\beta_{c}'=2.25$.
We set $\beta=2.5$ and $\mathcal{F}=10^{14}\frac{\text{keV}^{-2.5}}{\text{g}/\text{cm}^{3}}\frac{\text{erg}}{\text{cm}^{3}}$,
so that the material energy density is given by:
\[
u\left(T,\rho\right)=10^{14}\left(\frac{T}{\text{keV}}\right)^{2.5}\left(\frac{\rho}{\text{g}/\text{cm}^{3}}\right)\ \text{\ensuremath{\frac{\text{erg}}{\text{cm}^{3}}}}.
\]
For the exponents of this material model we get $\tau=2$ and a $\omega=-3$,
so that the surface temperature and density profile are:
\[
T_{s}\left(t\right)=\left(\frac{t}{\text{ns}}\right)^{2}\ \text{keV},
\]

\[
\rho\left(x\right)=\left(\frac{x}{\text{cm}}\right)^{3}\ \text{g}/\text{cm}^{3}.
\]
The dimensionless constants of the problem are $\mathcal{A}=56.3408$
and $\mathcal{B}=902.197$. The resulting similarity exponent is $\delta=1$,
and the numerical heat front coordinate is $\xi_{0}=1.19867771$,
so that the heat front position is:
\[
x_{F}\left(t\right)=0.61806779\left(\frac{t}{\text{ns}}\right)\ \text{cm}.
\]
The resulting numerical dimensionless flux at the origin is $\mathcal{S}\left(0\right)=9.31253$,
so that the bath temperature is {[}Eq. \eqref{eq:Tbath_marsh_bc}{]}:
\[
T_{\text{bath}}\left(t\right)=1.0719423\left(\frac{t}{\text{ns}}\right)^{2}\ \text{keV}.
\]
A comparison of the surface and bath temperatures as a function of
time is displayed in Fig. \ref{fig:simulation_6_Tbath}.

The radiation and material temperature profiles are given by the self-similar
solution {[}Eqs. \eqref{eq:Trss}-\eqref{eq:Tmss}{]}: 
\[
T_{r}\left(x,t\right)=\left(\frac{t}{\text{ns}}\right)^{2}f^{1/4}\left(\xi_{0}x/x_{F}\left(t\right)\right)\ \text{keV},
\]
\[
T\left(x,t\right)=\left(\frac{t}{\text{ns}}\right)^{2}g^{1/4}\left(\xi_{0}x/x_{F}\left(t\right)\right)\ \text{keV}.
\]
 The resulting numerical profiles are tabulated in table \ref{tab:tests_table}
and approximate closed form expressions are given in table \eqref{tab:test_fits}
as a function of $\xi/\xi_{0}=x/x_{F}\left(t\right)$. The temperature
profiles are displayed in Fig. \ref{fig:simulation_6}, showing again
a great agreement between the various simulations and analytic gray
diffusion solution. We note the significant lack of equilibrium in
a wide range near the origin.

\section{Conclusion}

In this work we have developed new self-similar solutions to a nonlinear
non-equilibrium supersonic Marshak wave problem in non-homogeneous
media in the diffusion limit of radiative transfer. The solutions
exist under the assumptions of a temporal power-law surface radiation
temperature drive of the form $T_{s}\left(t\right)=T_{0}t^{\tau}$,
a spatial power law density profile $\rho\left(x\right)=\rho_{0}x^{-\omega}$,
and material model with power law temperature and density dependent
total and absorption opacities $k_{t}\left(T,\rho\right)=\frac{1}{g}T^{-\alpha}\rho^{\lambda+1}$
and $k_{a}\left(T,\rho\right)=\frac{1}{g'}T^{-\alpha'}\rho^{\lambda'+1}$,
and a material energy density $u\left(T,\rho\right)=\mathcal{F}T^{\beta}\rho^{1-\mu}$.
The solutions are a generalization of the recent work \cite{krief2024self},
were such non-linear non-equilibrium solutions were developed for
a homogeneous media ($\omega=0$), which required a material energy
density that is proportional to the radiation energy density ($\beta=4$).
It is shown that the generalized solutions exist for specific values
of the temporal drive exponent $\tau$ and the spatial density exponent
$\omega$, which are functions of the temperature and density material
model exponents $\alpha,\lambda,\alpha',\lambda',\beta,\mu$ (given
by Eqs. \eqref{eq:tau_ss}-\eqref{eq:omega_ss}). The properties of
the solutions were analyzed in detail, including the range of validity
and the thermodynamic equilibrium limit for which the radiation
and material temperature become very close. The behavior of the solutions
near the origin was analyzed, and it was shown that the material temperature
at the origin is always zero for $\omega<0$; can have any value between
0 and the radiation temperature for $\omega=0$; and is always equal
to the radiation temperature when $\omega>0$.

We constructed a set of non-equilibrium and non-homogeneous Marshak
wave benchmarks for supersonic radiative heat transfer. The numerical
solutions of the similarity profiles for these benchmarks were tabulated,
and approximate closed form analytic functions were also given for
convenience. The benchmarks were run using implicit Monte-Carlo and
discrete-ordinate radiation transport simulations as well gray diffusion
simulations. All benchmarks, which were defined to be optically thick,
resulted in a very good agreement with the similarity solutions of
the gray diffusion equation. We conclude that the solutions developed
in this work can be used as non-trivial but easy to implement code
verification test problems for non-equilibrium radiation heat transfer
computer simulations.

\subsection*{Availability of data}

The data that support the findings of this study are available from
the corresponding author upon reasonable request.

 \bibliographystyle{naturemag}
\bibliography{datab}

\pagebreak{}

\appendix

\section{Dimensional analysis\label{sec:Dimensional-analysis}}

\begin{table*}[t]
\begin{centering}
\begin{tabular}{|c|c|c|c|c|c|c|c|}
\hline 
$E$  & $U$  & $E_{0}$  & $K$  & $M$  & $P$  & $x$  & $t$\tabularnewline
\hline 
\hline 
$\left[E\right]$  & $\left[E\right]$  & $\frac{\left[E\right]}{\left[\text{time}\right]^{4\tau}}$  & $\frac{\left[\text{length}\right]^{2-\omega\left(1+\lambda\right)}}{\left[E\right]^{\frac{\alpha}{4}}\left[\text{time}\right]}$  & $\frac{\left[E\right]{}^{\frac{\alpha'}{4}}\left[\text{length}\right]^{\omega\left(1+\lambda'\right)}}{\left[\text{time}\right]}$  & $\frac{\left[E\right]^{\frac{\alpha'+\beta}{4}-1}\left[\text{length}\right]^{\omega\left(\lambda'+\mu\right)}}{\left[\text{time}\right]}$  & $\left[\text{length}\right]$  & $\left[\text{time}\right]$\tabularnewline
\hline 
\end{tabular}
\par\end{centering}
\caption{The dimensional quantities in the problem (upper line) and their dimensions
(lower line). $\left[E\right]$ denotes the dimensions of energy per
unit volume. \label{tab:The-dimensional-quantities}}
\end{table*}

In this appendix, we use the method of dimensional analysis to construct
a self-similar solution of the problem defined by the nonlinear gray
diffusion model in Eqs. \eqref{eq:Eform}-\eqref{eq:Uform}, with
the initial and boundary conditions \eqref{eq:init_cond}-\eqref{eq:bc}.
As detailed in table \ref{tab:The-dimensional-quantities}, the problem
is formulated in terms $8$ dimensional quantities, which are composed
of $3$ different units - time, length and energy density. According
the the Pi (Buckingham) theorem of dimensional analysis \cite{buckingham1914physically,zeldovich1967physics,barenblatt1996scaling},
the problem can be expressed in terms of $8-3=5$ dimensionless variables
$\Pi_{1},\Pi_{2},\Pi_{3},\Pi_{4},\Pi_{5}$ which are defined as products
of power laws in terms of the dimensional quantities. In order to
express the solution with a single similarity independent variable
$\Pi_{1}=\xi$ and two self-similar dependent variables for the radiation
$\Pi_{2}=f\left(\xi\right)\propto E$ and matter $\Pi_{3}=g\left(\xi\right)\propto U$,
we require that the remaining dimensionless quantities, which we denote
as $\mathcal{A}=\Pi_{4}$, $\mathcal{B}=\Pi_{5}$, would not depend
on $E,U,x,t$, that is, they should be dimensionless quantities that
depend on the dimensional \uline{constants} characterizing the
problem, which we write as: 
\begin{align}
\Pi_{4} & =\mathcal{A}=E_{0}^{n}K^{k}M^{m},\\
\Pi_{5} & =\mathcal{B}=E_{0}^{n'}K^{k'}P^{p}.
\end{align}
The requirement that $\mathcal{A}$ is dimensionless results in the
following homogeneous system of linear equations: 
\begin{align*}
 & n-\frac{\alpha}{4}k+\frac{\alpha'}{4}m=0,\\
 & k\left(2-\omega\left(1+\lambda\right)\right)+m\omega\left(1+\lambda'\right)=0,\\
 & 4\tau n+k+m=0,
\end{align*}
which has the solution: 
\begin{align*}
 & k\left(2-\omega\left(1+\lambda\right)-\omega\frac{1+\alpha\tau}{1-\alpha'\tau}\left(1+\lambda'\right)\right)=0,\\
 & n=-\frac{1}{4\tau}\left(k+m\right),\\
 & m=-\frac{1+\alpha\tau}{1-\alpha'\tau}k.
\end{align*}
A set of infinite non-trivial solutions exists only if: 
\begin{equation}
\frac{\omega\left(1+\lambda'\right)}{2-\omega\left(1+\lambda\right)}=\frac{1-\alpha'\tau}{1+\alpha\tau}.\label{eq:tauomega}
\end{equation}
By setting (without loss of generality) $m=1$, we have $n=-\frac{1}{1+\alpha\tau}\frac{\alpha+\alpha'}{4},\,k=-\frac{1-\alpha'\tau}{1+\alpha\tau}$
so that: 
\begin{equation}
\mathcal{A}=E_{0}^{-\frac{\alpha+\alpha'}{4\left(\alpha\tau+1\right)}}K^{\frac{\alpha'\tau-1}{\alpha\tau+1}}M.\label{eq:Aapp}
\end{equation}
Similarly, by the requirement that $\mathcal{B}$ is dimensionless,
we get the linear homogeneous system: 
\begin{align*}
 & n'-\frac{\alpha}{4}k'+\left(\frac{\alpha'+\beta}{4}-1\right)p=0,\\
 & k'\left(2-\omega\left(1+\lambda\right)\right)+\omega p\left(\lambda'+\mu\right)=0,\\
 & 4\tau n'+k'+p=0,
\end{align*}
which has the solution: 
\begin{align*}
 & \left[\left(2-\omega\left(1+\lambda\right)\right)-\frac{1+\alpha\tau}{1-\left(\alpha'+\beta-4\right)\tau}\omega\left(\lambda'+\mu\right)\right]k'=0,\\
 & p=-\frac{1+\alpha\tau}{1-\left(\alpha'+\beta-4\right)\tau}k',\\
 & n'=-\frac{1}{4\tau}\left(k'+p\right).
\end{align*}
As before, an infinite set of nontrivial solutions exists only if:

\begin{align}
\left(1-\left(\alpha'+\beta-4\right)\tau\right)\left(2-\omega\left(1+\lambda\right)\right)\nonumber \\
-\omega\left(1+\alpha\tau\right)\left(\lambda'+\mu\right) & =0\label{eq:tauomega2}
\end{align}
By solving Eqs. \eqref{eq:tauomega}, \eqref{eq:tauomega2} for $\tau$
and $\omega$, we obtain Eqs. \eqref{eq:tau_ss}-\eqref{eq:omega_ss}.
Therefore, given a material model which is defined by the exponents
$\alpha,$$\lambda,$$\alpha',$$\lambda',$$\beta,$$\mu$, a self-similar
solution exists only for a temporal exponent $\tau$ and spatial exponent
$\omega$ which obey the relations Eqs. \eqref{eq:tau_ss}-\eqref{eq:omega_ss},
respectively. By setting (without loss of generality) $p=1$, we get:

\begin{equation}
\mathcal{B}=E_{0}^{\frac{4-\beta-\alpha-\alpha'}{4\left(\alpha\tau+1\right)}}K^{\frac{\left(\alpha'+\beta-4\right)\tau-1}{\alpha\tau+1}}P.\label{eq:Bapp}
\end{equation}

We now write the dimensionless independent similarity coordinate as:
\begin{equation}
\Pi_{1}=\xi=xt^{-\delta}E_{0}^{n}K^{k}.\label{eq:xsiapp}
\end{equation}
The requirement that $\xi$ is dimensionless results in the following
(non-homogeneous) system of linear equations:

\begin{align*}
 & k\left(2-\omega\left(1+\lambda\right)\right)=-1,\\
 & n-\frac{\alpha}{4}k=0,\\
 & 4\tau n+k+\delta=0.
\end{align*}
which has the solution: 
\begin{align*}
 & \delta=\frac{1+\alpha\tau}{2-\omega\left(1+\lambda\right)},\\
 & n=-\frac{\alpha}{4\left(2-\omega\left(1+\lambda\right)\right)},\\
 & k=-\frac{1}{2-\omega\left(1+\lambda\right)},
\end{align*}
which proves Eq. \eqref{eq:xsi_def}. Finally, we can write the dimensionless
similarity profiles directly, since the radiation energy density at
the system boundary, $E_{0}t^{4\tau}$, has units of energy density:
\begin{equation}
\Pi_{2}=\frac{E}{E_{0}t^{4\tau}}=f\left(\xi\right),\label{eq:pi2}
\end{equation}
\begin{equation}
\Pi_{3}=\frac{U}{E_{0}t^{4\tau}}=g\left(\xi\right).\label{eq:pi3}
\end{equation}

We will now show explicitly, that by substituting the self-similar
ansatz Eqs. \eqref{eq:xsiapp},\eqref{eq:pi2}-\eqref{eq:pi3} in
the gray diffusion equations \eqref{eq:Eform}-\eqref{eq:Uform},
under the constrains Eqs. \eqref{eq:tau_ss}-\eqref{eq:omega_ss},
results in a dimensionless ODE system for the similarity profiles
$f,g$. By using the relation $\frac{\partial}{\partial x}=\frac{\xi}{x}\frac{\partial}{\partial\xi}$,
the flux term in Eq. \eqref{eq:Eform} reads: 
\begin{align}
\frac{\partial}{\partial x}\left(x^{\omega\left(1+\lambda\right)}U^{\frac{\alpha}{4}}\frac{\partial}{\partial x}E\right) & =x^{\omega\left(1+\lambda\right)-2}\xi t^{\left(\alpha+4\right)\tau}E_{0}^{\frac{\alpha}{4}+1}\times\nonumber \\
\Bigg(\omega\left(1+\lambda\right)g^{\frac{\alpha}{4}}f'+\xi & g^{\frac{\alpha}{4}-1}\left[\frac{\alpha}{4}f'g'+gf''\right]\Bigg).\label{eq:fluxderivative}
\end{align}
Similarly, by using the relation $\frac{\partial}{\partial t}=-\delta\frac{\xi}{t}\frac{\partial}{\partial\xi}$
the left hand side of Eq. \eqref{eq:Eform} reads: 
\begin{align}
\frac{\partial E}{\partial t} & =E_{0}t^{4\tau-1}\left(4\tau f-\delta\xi f'\right).\label{eq:energytimederivative}
\end{align}
By substituting Eqs. \eqref{eq:fluxderivative}-\eqref{eq:energytimederivative}
in Eq. \eqref{eq:Eform}, we find: 
\begin{align*}
 & E_{0}t^{4\tau-1}\left(4\tau f-\delta\xi f'\right)=E_{0}t^{4\tau-1}\xi^{\omega\left(1+\lambda\right)}\times\\
 & \left(\frac{1}{\xi}\omega\left(1+\lambda\right)g^{\frac{\alpha}{4}}f'+g^{\frac{\alpha}{4}-1}\left[\frac{\alpha}{4}f'g'+gf''\right]\right)\\
 & +ME_{0}^{-\frac{\alpha'}{4}}\xi^{-\omega\left(1+\lambda'\right)}\left(\frac{1}{t^{1+\alpha\tau}E_{0}^{\frac{\alpha}{4}}K}\right)^{\frac{\omega\left(1+\lambda'\right)}{2-\omega\left(1+\lambda\right)}}\times\\
 & E_{0}g^{-\frac{\alpha'}{4}}t^{\left(4-\alpha'\right)\tau}\left(g-f\right).
\end{align*}
By using Eq. \eqref{eq:Aapp}, the dimensional quantity $M$ can be
written in terms of the dimensionless $\mathcal{A}$, and by using
Eq. \eqref{eq:tauomega}, we find: 
\begin{align*}
E_{0}t^{4\tau-1}\left(4\tau f-\delta\xi f'\right) & =E_{0}t^{4\tau-1}\xi^{\omega\left(1+\lambda\right)}\times\\
\Bigg(\frac{1}{\xi}\omega\left(1+\lambda\right)g^{\frac{\alpha}{4}}f' & +g^{\frac{\alpha}{4}-1}\left[\frac{\alpha}{4}f'g'+gf''\right]\Bigg)\\
+E_{0}t^{4\tau-1}\mathcal{A} & \xi^{-\omega\left(1+\lambda'\right)}g^{-\frac{\alpha'}{4}}\left(g-f\right),
\end{align*}
which upon factoring out the common dimensional $E_{0}t^{4\tau-1}$
term, results in the dimensionless similarity ODE \eqref{eq:f_ode}.
Similarly, the material equation \eqref{eq:Uform} can be written
as: 
\begin{align*}
E_{0}t^{4\tau-1}\left(4\tau f-\delta\xi f'\right) & =P\xi^{-\omega\left(\mu+\lambda'\right)}\left(\frac{1}{t^{1+\alpha\tau}E_{0}^{\frac{\alpha}{4}}K}\right)^{-\frac{\omega\left(\mu+\lambda'\right)}{2-\omega\left(1+\lambda\right)}}\times\\
 & \left(E_{0}t^{4\tau}g\right)^{1-\frac{\alpha'+\beta}{4}}E_{0}t^{4\tau}\left(f-g\right).
\end{align*}
By using Eq. \eqref{eq:Bapp}, the dimensional quantity $P$ can be
written in terms of the dimensionless $\mathcal{B}$, and by using
Eq. \eqref{eq:tauomega2}, which gives 
\[
\frac{\omega\left(\mu+\lambda'\right)}{2-\omega\left(1+\lambda\right)}=\frac{1-\left(\alpha'+\beta-4\right)\tau}{1+\alpha\tau},
\]
we obtain: 
\[
E_{0}t^{4\tau-1}\left(4\tau g-\delta\xi g'\right)=E_{0}t^{4\tau-1}\mathcal{B}\xi^{-\omega\left(\mu+\lambda'\right)}g^{1-\frac{\alpha'+\beta}{4}}\left(f-g\right),
\]
which upon factoring out the common dimensional $E_{0}t^{4\tau-1}$
term, results in the dimensionless similarity ODE \eqref{eq:g_ode}. 
\end{document}